\documentclass[preprint,sort&compress]{elsarticle}

\usepackage{hyperref}
\usepackage{amsmath, amsthm, amssymb}
\usepackage{algorithm}
\usepackage[normalem]{ulem}
\usepackage{algorithmic}
\graphicspath{{fig/}}
\usepackage{stfloats}
\usepackage[caption=false,font=footnotesize]{subfig}
\usepackage{float}
\usepackage[autostyle]{csquotes}
\usepackage[british]{babel}
\journal{Elsevier}
\usepackage[dvipsnames]{xcolor}
\colorlet{LightRubineRed}{RubineRed!70!}
\colorlet{Mycolor1}{green!10!orange!90!}
\definecolor{Mycolor2}{HTML}{00F9DE}
\usepackage{mathtools}

\DeclarePairedDelimiter\floor{\lfloor}{\rfloor}
\DeclareMathOperator*{\argmin}{arg\,min}
\DeclareMathOperator*{\argmax}{arg\,max}

\begin{document}

\begin{frontmatter}

\title{Detection and Prediction of Equilibrium States in Kinetic
Plasma Simulations via Mode Tracking using Reduced-Order Dynamic Mode Decomposition}

\author[address1]{Indranil Nayak}
\cortext[mycorrespondingauthor]{Corresponding author}
\ead{nayak.77@osu.edu}

\author[address2]{Mrinal Kumar}

\author[address1]{Fernando L. Teixeira}

\address[address1]{ElectroScience Laboratory and Department of Electrical and Computer Engineering, The Ohio State University, Columbus, Ohio 43212, USA}
\address[address2]{Laboratory for Autonomy in Data-Driven and Complex Systems, Department of Mechanical and Aerospace Engineering, The Ohio State University, Columbus, Ohio 43210, USA}

\begin{abstract}
A dynamic mode decomposition (DMD) based reduced-order model (ROM) is developed for tracking, detection, and prediction of kinetic plasma behavior. DMD is applied to the high-fidelity kinetic plasma model based on the electromagnetic particle-in-cell (EMPIC) algorithm to extract the underlying dynamics and key features of the model. In particular, the ability of DMD to reconstruct the spatial pattern of the self electric field from high-fidelity data and the effect of DMD extrapolated self-fields on charged particle dynamics are investigated. An in-line sliding-window DMD method is presented for identifying the transition from transient to equilibrium state based on the loci of DMD eigenvalues in the complex plane. The in-line detection of equilibrium state combined with time extrapolation ability of DMD has the potential to effectively expedite the simulation.
Case studies involving electron beams and plasma ball are presented to assess the strengths and limitations of the proposed method.
\end{abstract}


\end{frontmatter}

%\linenumbers

\section{Introduction}
Kinetic plasma simulations are important for a wide range of applications, including but not limited to the design and analysis of high-power microwave sources, particle accelerators, laser ignited devices, and ionosphere and magnetosphere problems~\citep{gold1997review, booske2008plasma, benford2015high, lapenta2006kinetic,nayak2019progress,karimabadi2011petascale,chen2020magnetohydrodynamic}. Electromagnetic particle-in-cell (EMPIC) algorithms are typically used for simulating kinetic collisionless plasmas governed by Maxwell-Vlasov equations. EMPIC algorithms compute the electromagnetic field on the spatial mesh based on a discretized form of Maxwell's equations while simultaneously updating, via a kinetic model based on the Lorentz force equation, the velocity and position of computational superparticles that effect a coarse-graining of the phase space of charged particles in the plasma~\citep{birdsall2004plasma, bettencourt2008performance, wang2010three, moon2015exact, meierbachtol2015conformal}.
The inherent nonlinearity and multi-scale nature of the problem
make the interpretation of the underlying physics often difficult and serve as one of the motivations for a reduced-order model that can characterize, with sufficient accuracy, the plasma system using a small number of degrees of freedom. Reduced-order models may also facilitate the possible use of model-based control methods such as model predictive control (MPC) \cite{allgower1999nonlinear, kaiser2018sparse}. 
Several recent studies \cite{pandya2016low, van2014use, byrne2017study, kaptanoglu2020physics} in the plasma physics community have indicated the practicality of adopting a lower dimensional feature space that can model the system through a small set of spatio-temporal coherent structures. A variety of model-order reduction techniques, such as proper orthogonal decomposition (POD) \citep{beyer2000proper, nicolini2019model,kaptanoglu2020physics},  bi-orthogonal decomposition (BOD) \citep{de1995enhancement,dudok1994biorthogonal}, principal component analysis (PCA) \cite{bellemans2017reduced} have been proposed in the past. These methods are invariably limited in their ability to resolve the time dynamical properties using low rank modelling. Dynamic mode decomposition (DMD) \cite{schmid2010dynamic,schmid2011applications,tu2013dynamic} helps to overcome this difficulty. In particular, it was recently shown in \citep{taylor2018dynamic,kaptanoglu2020characterizing,sasaki2019using} that DMD can efficiently extract the underlying characteristic features of 
(fluid-model) magnetohydrodynamics based plasma simulations with reasonable accuracy. Our preliminary study \citep{nayak2020prediction} shows promise of DMD in reconstructing self electric fields from (kinetic-model) EMPIC plasma simulations. However, a detailed analysis regarding ability of DMD to capture relevant plasma dynamics from the particle-in-cell simulation is yet to be explored.

Another challenge of particle-in-cell (PIC) based algorithms is the large computational load \citep{hockney1988computer}. Several improvements have been proposed in the literature to speed up PIC simulations, ranging from computational architecture to the underlying algorithmic structure, e.g. see \citep{werner2018speeding,decyk2014particle,WOLF2016342}. Here, we address the issue also from a reduced-order model perspective. In order to minimize the computational cost, ideally one would like to perform reduced-order modeling such as DMD using data from high-fidelity simulations based on relatively short time windows and extrapolate the results in future time. However, as is shown in this work, accurate prediction of the equilibrium dynamics using data-driven methods such as DMD requires sufficient data harvesting near equilibrium. As a result, a related important question to be addressed is how to leverage DMD to optimally predict the equilibrium state. The question becomes particularly crucial for timely termination of the high-fidelity simulations such as those based on EMPIC algorithms.

In order to exploit the time extrapolation ability of DMD for reducing computation cost of high-fidelity simulations, it is important to identify the transition from transient to equilibrium state of a dynamical system in an in-line fashion. Several past works \cite{van2013interaction, feng2014deep, chekroun2014rough, tantet2015early} deal with identification of state transition in high-dimensional physical systems. Some recent publications \citep{gottwald2019detecting,alessandri2019dynamic} highlight the importance of DMD in identifying such regime transitions. The authors in \citep{gottwald2019detecting} rely on the DMD reconstruction error difference between transient and equilibrium states of a dynamical system to identify such transitions. However, one of the key assumptions in \citep{gottwald2019detecting} is the fast relaxation of the dynamical system in transience, i.e. a faster time scale of the transient dynamics compared to equilibrium dynamics. The present work does not rely on the fast relaxation assumption since the transience is characterized by temporal variations in the amplitude and changing frequency content.  %since the transient state is identified on the basis of growing/decaying nature of solution and the changing frequency content.} 
Rather, we compute the \textit{residue} based on the relative position of dominant DMD eigenvalues with respect to the unit circle. While Ref. \citep{alessandri2019dynamic} also performs identification of regime transition, it does so by observing the variation of a DMD-based least-squares residual term as the DMD window is gradually increased to span the spatial domain.  %also deals with identifying regime transition using DMD, but in the spatial domain. \citep{alessandri2019dynamic} uses a least-square residual term from DMD algorithm and observe it's variation as DMD window width is increased gradually to cover the spatial region. 
In contrast, the %Our 
residual term in this work is based on the loci of DMD eigenvalues in the complex plane. We keep track of the residual term as a \textit{fixed-width} DMD window is moved forward in time.  %and we move the DMD window in time keeping the width fixed. 
Finally, in \citep{alessandri2019dynamic}, the change in the slope of the residual term is detected by fitting two straight lines. This work employs instead a rolling average to detect non-negative slopes that is suitable for %makes our algorithm applicable for
 in-line application. This work addresses all these issues in the context of kinetic plasma simulations from a modal analysis perspective. The main contributions of the present work can be summarized as follows:
\begin{enumerate}
    \item As mentioned above, while DMD has been recently applied to fluid-based plasma simulations, it has not yet been studied for kinetic plasma simulations. In this work we study the performance of DMD in reconstructing the self electric fields and its effect on the superparticle dynamics, for several test cases.
    \item  We propose an algorithm for in-line detection of the onset of the equilibrium state of a dynamical system using a sliding-window DMD approach. This advancement has the potential to speed up EMPIC simulations for long term predictions when combined with the time-extrapolation ability of DMD. We propose a sliding window approach that tracks the position of DMD eigenvalues relative to the unit circle on the complex plane for detecting the equilibrium state. We analyze the prediction error in self-field pattern, as well as the superparticle dynamics, produced by the reduced-order model extrapolated solution.
    \item We perform a first-of-its-kind analysis to investigate the convergence in DMD mode shapes and shifting of DMD eigenvalues as the DMD window slides from transient to the equilibrium state. We do so by in-line tracking of the DMD modes and eigenvalues as a part of equilibrium detection algorithm. Such analysis can provide insight on how hidden features in the transient state can manifests itself as the system approaches equilibrium.
\end{enumerate}
%\par

\begin{figure}[t]
    \centering
	\includegraphics[width=1\linewidth]{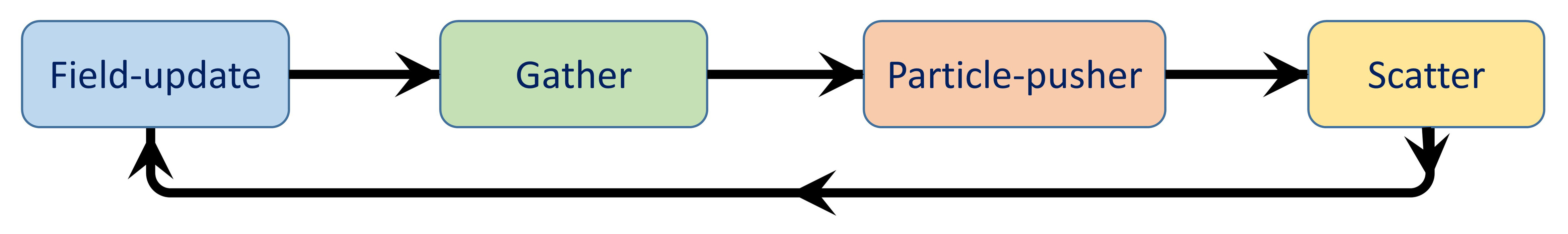}
    \caption{\small{Main cyclic steps in the EMPIC algorithm.}}
    \label{fig:empic_cycle} 
\end{figure}

\section{DMD Applied to EMPIC Kinetic Plasma Simulations}

\subsection{EMPIC Algorithm} \label{empic}
The EMPIC algorithm \citep{moon2015exact,kim2011parallel,EVSTATIEV2013376,doi:10.1063/1.4742985,doi:10.1063/1.4976849,kraus-kormann,jianyuan2018structure}  generates the high-fidelity data for the DMD reduced-order model. It executes a marching-on-time procedure in four stages (Fig.~\ref{fig:empic_cycle}) during each timestep: field-update, gather, particle-pusher and scatter. 

For the field update, time-dependent Maxwell's equations are discretized on simplicial (triangular or tetrahedral) meshes using finite elements based on discrete exterior calculus~\cite{flanders1989, teixeira1999lattice, kotiuga2004, he2007differential, deschamps1981electromagnetics, he2006sparse, donderici2008mixed, teixeira2013differential}.
The electric $\mathcal{E}\left(t, \vec{r}\right)$ and magnetic (flux) $\mathcal{B}\left(t, \vec{r}\right)$ fields are expanded as a sum of Whitney forms (natural interpolants for discrete differential forms) as explained in~\cite{moon2015exact,nicolini2019model, kim2011parallel,na2016local},
\begin{align} 
\mathcal{E}\left(t, \vec{r}\right) &= \sum_{i=1}^{N_{1}} e_{i}\left(t\right) w^{(1)}_{i} (\vec{r}), \label{eq:EDoF}\\
\mathcal{B}\left(t, \vec{r}\right) &= \sum_{j=1}^{N_{2}} b_{j}\left(t\right) w^{(2)}_{j} (\vec{r})\label{eq:BDoF}. 
\end{align}
The functions $w^{(1)}_{i} (\vec{r})$ and $w^{(2)}_{j} (\vec{r})$ represent Whitney 1-forms (edge-based functions) and Whitney 2-forms (facet-based functions), respectively
\cite{moon2015exact, teixeira2014lattice, he2006geometric}.
These functions 
have a biunivocal association to the edges and facets of the finite element mesh, respectively, with $N_1$ denoting the number of edges and $N_2$ the number of facets.
A detailed description of the discrete field update equations can be found in~\citep{na2016local,moon2015exact,kim2011parallel,nicolini2019model}.
The discrete degrees of freedom (DoF) for the electric field and magnetic {flux} can be represented by column vectors comprising the set of 
time-dependent coefficients in \eqref{eq:EDoF},\eqref{eq:BDoF}, i.e. 
 $\mathbf{e}(t) = [e_{1}(t) \ e_{2}(t) \ \ldots \ e_{N_1}(t)]^{\text{T}}$ and  $\mathbf{b}(t) = [b_{1}(t) \ b_{2}(t) \ \ldots \ b_{N_2}(t)]^{\text{T}}$
 where \textquote{$^\text{T}$} denotes transpose. Their time-discrete counterparts at the $n^{th}$ timestep produced by the EMPIC algorithm, denoted as
   $\mathbf{e}^{(n)} = [e_{1}^{(n)}  \ e_{2}^{(n)} \ \ldots \ e_{N_1}^{(n)}]^{\text{T}}$ and {  $\mathbf{b}^{(n+1/2)} = [b_{1}^{(n+1/2)}  \ b_{2}^{(n+1/2)} \ \ldots \ b_{N_2}^{(n+1/2)}]^{\text{T}}$,}
are used as input to the DMD, as described {in the following section}.

In the gather step, the fields are interpolated at each superparticle position based on the same Whitney forms expansion as above. Then, in the particle-pusher step, the position and velocity of the superparticles are updated using Newton's law of motion (with relativistic corrections if necessary) and the Lorentz force equation.
Finally, the scatter step maps the electric current density and the
electric charge distribution produced by the updated velocities and positions of the superparticles back onto the mesh edges and nodes, respectively, while ensuring charge conservation~\cite{moon2015exact}.

\subsection{Koopman Operator}
\label{sec:koopman}
DMD derives its ability to model nonlinear dynamics from its close relation to the Koopman operator. Indeed, DMD can be viewed as a finite dimensional approximation of the infinite dimensional Koopman operator \cite{mezic2013analysis,rowley2009spectral}. The infinite dimensional linear Koopman operator is associated with evolution of a nonlinear dynamical system on an $N$-dimensional manifold \cite{kutz2016dynamic}, where $N$ is the dimensionality of the state-space.  \par
Let us consider a discrete-time dynamical system
\begin{align}
    \mathbf{x}^{(n+1)}=F(\mathbf{x}^{(n)}),
\end{align}
where $\mathbf{x}$ is the state of the system belonging to an $N$-dimensional manifold $\mathcal{M}$ ($\mathbf{x}\in\mathcal{M}$) and $F$ is the flow map, $F:\mathcal{M}\mapsto \mathcal{M}$. 
In the present application,  $\mathbf{x} = \mathbf{e} = [e_{1} \ e_{2} \ \ldots \ e_{N_1}]^{\text{T}}$ from the expansion in \eqref{eq:EDoF}.

The discrete time Koopman operator denoted by $\mathcal{K}$ operates on $g(\mathbf{x})$ ($g: \mathcal{M}\mapsto \mathbb{C}$), the so-called ``observables of the state'' as follows
\begin{align}
    \mathcal{K}g(\mathbf{x}^{(n)})=g(F(\mathbf{x}^{(n)}))=g(\mathbf{x}^{(n+1)}).
\end{align}
Suppose the eigenfunctions and eigenvalues of the operator $\mathcal{K}$ are represented as $\phi_i: \mathcal{M}\mapsto \mathbb{C}$ and $\lambda_i \in \mathbb{C}$ respectively, ie. $    \mathcal{K}\phi_j(\mathbf{x})=\lambda_j\phi_j(\mathbf{x})~,~ j=1,2, \ldots$. % "j" used as complex number
We can represent a vector valued observable $\mathbf{g(x)}=[g_1(\mathbf{x}) \ g_2(\mathbf{x}) \ \ldots \ g_P(\mathbf{x})]^{\text{T}} $ using Koopman modes $\mathbf{v}_j$ and Koopman eigenfunctions $\phi_j$, so long as the eigenfunctions span each observable, $g_i$, $i=1,2, \ldots,P$, as $\mathbf{g(x)} = \sum_{j=1}^\infty \phi_j(\mathbf{x})\mathbf{v}_j$ \cite{rowley2009spectral, mezic2005spectral}. For the $n^{th}$ time instant,
\begin{align}
    \mathbf{g}(\mathbf{x}^{(n)})=\sum_{j=1}^\infty \mathcal{K}^n\phi_j(\mathbf{x}^{(0)})\mathbf{v}_j=\sum_{j=1}^\infty\lambda_j^n\phi_j(\mathbf{x}^{(0)})\mathbf{v}_j\label{koop_recon},
\end{align}
$\mathbf{x}^{(0)}$ being the initial state. Eq. \eqref{koop_recon} is the basis for DMD which is a finite dimensional approximation of infinite dimensional Koopman operator \cite{rowley2009spectral}. \par

\subsection{DMD Algorithm}\label{DMD_algo}
In classical DMD, the state, $\mathbf{x}$, itself serves as the set of observables, i.e. $\mathbf{g(x)} = \mathbf{x}$. DMD is commonly employed to retrieve dominant spatio-temporal patterns of a dynamical system by harvesting time snapshots of the state. %wherein time snapshots of the state vector (spatial) are harvested and used as input to the DMD algorithm.}. 
It produces a set of DMD modes $\Phi$, corresponding DMD frequencies $\omega$ and a set of scaling factors $\vartheta$. The DMD modes $\Phi$ capture spatial variation while temporal variation is captured by the term $e^{\omega t}$. A linear combination of properly scaled modes multiplied by $e^{\omega t}$ reconstructs the original data \cite{schmid2010dynamic,tu2013dynamic,kutz2016dynamic} as shown in \eqref{DMD_recon}. Consider a harvesting window of $(l+1)$ snapshots, starting at $t_0=n_0\Delta_t$ and ending at $(n_0+l\Delta n)\Delta_t$, where $\Delta n$ is the number of timesteps between two consecutive snapshots and $\Delta_t$ is the timestep interval. The snapshot matrix $X$ and the shifted snapshot matrix $X'$ are given as
\begin{gather}
 X
 =
  \begin{bmatrix}\label{snapshot_eq1}
  \mathbf{x}^{(n_0)}& \mathbf{x}^{(n_0+\Delta n)} & . & . & . & \mathbf{x}^{(n_0+(l-1)\Delta n)}  \\
   \end{bmatrix},
\\
X'
=
  \begin{bmatrix}\label{snapshot_eq2}
  \mathbf{x}^{(n_0+\Delta n)}& \mathbf{x}^{(n_0+2\Delta n)} & . & . & . & \mathbf{x}^{(n_0+l\Delta n)}  \\
   \end{bmatrix}.
\end{gather}
DMD assumes $X'\approx A\cdot X$ and proceeds to extract the eigenvalues and eigenvectors of $A$ in an efficient manner, where
$A=X'X^\dagger$ (`$\dagger$' is the Moore-Penrose pseudo inverse). 
The first step towards low-dimensional representation of $A$ involves performing singular value decomposition (SVD) of the snapshot matrix $X$, resulting in the $U$, $\Sigma$, and $V$ matrices as follows
\begin{align} \label{SVD_eq}
    X=U\Sigma V^*,
\end{align}
where `$^*$' denotes complex-conjugate transpose. Next, rank reduction is performed by retaining only the first $r$ columns ($r<l$) of $U, V$ as $U_r$ and $V_r$ respectively, as well as the first $r$ columns and rows of $\Sigma$, as $\Sigma_r$. Typically, the value of $r$ is chosen based on a hard energy threshold or through optimal hard thresholding, {as discussed in \cite{gavish2014optimal,opt_thr_code}}. {In this work we choose an optimal hard thresholding based on the nearest odd value of $r$. An odd value of $r$ ensures at least one DMD eigenvalue on the real axis and thus facilitates tracking of DMD eigenvalues and modes.} The Moore-Penrose pseudo inverse of $X$ is then approximated by $X^\dagger\approx V_r\Sigma_r^{-1}U_r^*$ and $A$ by
\begin{align}\label{A_formula}
    A\approx X'V_r\Sigma_r^{-1}U_r^*.
\end{align}
Spectral decomposition of $A$ is invariably computationally expensive due to its high dimensionality. An acceptable compromise is to  project $A$ onto the columns of $U_r$ (its POD basis), resulting in $\tilde{A} = U_r^*AU_r = U_r^*X'V_r\Sigma_r^{-1}$. The spectral decomposition of $\tilde{A}$ is given by $\tilde{A}W=W\Lambda$, where the diagonal matrix $\Lambda$ contains eigenvalues $\lambda_i$, $i = 1, 2, \ldots, r$ that are an adequate approximation of eigenvalues of $A$.  
Exact DMD modes can be constructed as the columns of $\mathbf{\Phi} = X'V_r\Sigma_r^{-1}W$ \cite{tu2013dynamic}, resulting in the DMD reconstruction ($\hat{\mathbf{x}}$) of the state for $t\geq t_0$, i.e.
\begin{align}\label{DMD_recon}
   \mathbf{x}(t)\approx\hat{\mathbf{x}}(t)=\sum_{i=1}^r \vartheta_i\Phi_i e^{\omega_i(t-t_0)},
\end{align} 
where $\omega_i=ln(\lambda_i)/\Delta t$, $\Delta t$ being the time interval between two consecutive snapshots ($\Delta t=\Delta n\Delta_t$). The scaling factor $\vartheta_i$ can be calculated by solving an optimization problem as described in \cite{jovanovic2014sparsity}. This paper employs stacked snapshot matrices for better accuracy. Further details can be found in \cite{schmid2010dynamic, tu2013dynamic, kutz2016dynamic}. In practical applications DMD is performed on real signal, generating complex conjugate pairs of DMD modes with corresponding complex conjugate pairs of frequencies and scaling factors. So, we can re-write \eqref{DMD_recon} in terms of $M$  complex-conjugate ( $\overline{(.)}$ ) pairs of DMD modes as
\begin{align}\label{DMD_recon_conj}
    \hat{\mathbf{x}}(t)=\sum_{m=1}^M (\vartheta_m\Phi_m e^{\omega_m(t-t_0)}+\overline{\vartheta}_m\overline{\Phi}_m e^{\overline{\omega}_m(t-t_0)}).
\end{align}
For purely real modes, two terms in \eqref{DMD_recon_conj} collapse to single term $(2M\geq r)$. To accurately capture the periodic behavior of limit-cycle oscillations, the DMD harvesting window should cover multiple cycles. Note that the inter-snapshot sampling interval is dictated by the Nyquist criterion and noise frequency. \par

\section{\label{idet_eq} Equilibrium State Identification}
Detection of onset of the ``equilibrium state'' is motivated by the need to identify the ideal data-harvesting window for data-driven reduced order methods, as well as for control applications. Accurate long term prediction of equilibrium behavior requires the DMD harvesting region to include the equilibrium region. One may terminate high-fidelity simulations once the system has reached equilibrium, ensuring enough quality data for the DMD to work with. Therefore, in-line detection (i.e. concomitantly with the ongoing simulation) of the equilibrium state is highly desirable. In this work, we introduce a sliding-window DMD approach for identification of the equilibrium state. This is particularly useful while characterizing highly nonlinear physical systems \cite{PhysRevE.99.063311}, as the sliding-window DMD approximates the evolution of a nonlinear system through piecewise linear dynamic systems supported by the windowed data \cite{costa2019adaptive,10.3389/fncom.2019.00075,taylor2018dynamic,hemati2014dynamic, zhang2019online, alfatlawi2019incremental}. Next, we present the algorithm to track DMD modes followed by detection of the equilibrium phase.

\subsection{Tracking DMD Modes}
DMD captures key features of a dynamical system within the data-harvesting time window. For a sufficiently ``well-behaved'' dynamic system, an infinitesimal shift in the DMD window is not expected to produce a drastic change in its constituent spatio-temporal features. We aim to track each DMD eigenvalue-mode pair $(\lambda, \Phi)$ from one DMD window to next, because doing so provides insights into how constitutive features of the dynamic system evolve. More importantly, it also helps identify if certain $(\lambda, \Phi)$ pairs become ``sufficiently'' stationary over several windows, indicating the  onset of equilibrium. Mode tracking is an evolving field of study \cite{alden1985eigenvalue, beaverstock2015automatic, safin2016advanced, raines2012wideband}. Generally, the tracking of eigenvectors is preferred over eigenvalues due to convergence issues caused by repeated (or nearly equal) eigenvalues \cite{alden1985eigenvalue}. However, in DMD theoretical framework we work with the assumption that DMD eigenvalues are distinct \cite{schmid2010dynamic, tu2013dynamic,hirsh2019centering}.\par
In DMD, the effect of a sliding window can be viewed as a perturbation in  {the} snapshot matrices. {Let} $X_k, X_k'$ (from \eqref{snapshot_eq1}, \eqref{snapshot_eq2}) be the snapshot matrices for the $k^{th}$ window and $X_{k+1}, X_{k+1}'$ for the $(k+1)^{th}$ window, with $k^{th}$ and $(k+1)^{th}$ window usually multiple snapshots apart. One can write $X_{k+1} = X_k + \delta_1$ and $X_{k+1}'=X_k' + \delta_2$. The amount of perturbation ($\delta_1, \delta_2$) depends on how fast the system changes between two consecutive DMD windows. Through the arguments presented below, we first point out that  infinitesimal perturbations in the snapshot matrix will result in only infinitesimal changes in DMD modes and eigenvalues. The following arguments concerning \eqref{snapshot_eq1}-\eqref{DMD_recon} support this claim:
\begin{enumerate}
    \item DMD elements $A$, $\tilde{A}$, $\Phi$, as well as the reconstruction in \eqref{DMD_recon} are linear transformations whose continuity ensures small change in output with small change in input.
    \item Continuity is less obvious for \eqref{SVD_eq} and the spectral decomposition of $\tilde{A}$.
    However, the perturbation bounds for singular values and singular vectors are well documented \cite{stewart1973error, stewart1977perturbation, li1993performance, chen2020asymmetry}, ensuring infinitesimal change in output given infinitesimal change in input for \eqref{SVD_eq}. Regarding the eigendecomposition step, continuity of the roots of a polynomial ensures that eigenvalues of $\tilde{A}$ (roots of its characteristic polynomial) do not experience discontinuities under small perturbations.
   Similarly, perturbation bounds for eigenvectors of simple eigenvalues \cite{greenbaum2019first}
   assures an infinitesimal change in $W$, thus an infinitesimal change in DMD modes with infinitesimal change in $\tilde{A}$.
\end{enumerate}
Following the above arguments, a gradual shift in the DMD window is expected to lead to a gradual change in DMD eigenvalues and mode shapes. An exception arises at bifurcation points, which we address in the tracking algorithm described below.
\par
 The tracking algorithm refers to each DMD mode and corresponding eigenvalue as the pair $(\lambda, \Phi)$. In other words, both the position of $\lambda$ in the complex plane as well as information on the spatial distribution of $\Phi$ are employed for mode tracking. Define ($\lambda_i^{(k)},\Phi_i^{(k)}$) as the DMD eigenvalue-mode pair in the $k^{th}$ window, where $1 \leq i \leq p$. The aim of the tracking algorithm is to assign $(\lambda_j^{(k+1)},\Phi_j^{(k+1)}),~(1\leq j\leq q)$ from $(k+1)^{th}$ window to ($\lambda_i^{(k)},\Phi_i^{(k)}$) as its successor. Assuming $p$ and $q$ to be the number of DMD modes in the $k^{th}$ and $(k+1)^{th}$ window respectively, there can be broadly three scenarios, 
\begin{enumerate} [i.)]
    \item $p = q$: In this case, each DMD eigenvalue-mode pair in the $k^{th}$ window is associated with exactly one pair in the $(k+1)^{th}$ window.
    \item $p > q$: The algorithm must terminate the tracking of some pairs ($\lambda_i^{(k)},\Phi_i^{(k)}$) to which no successors can be assigned.
    \item $p < q$: The algorithm must initiate the tracking of newly identified pairs starting at the $(k+1)^{th}$ window after all pairs from the $k^{th}$ window have been assigned unique successors.
\end{enumerate} 
The primary condition for successor assignment is given in terms of the placement of DMD eigenvalues. In other words, the first candidate for mode-matching of ($\lambda_i^{(k)},\Phi_i^{(k)}$) is
\begin{align}
    \label{matchconditionA} \tilde{j} = \argmin_{j = 1, \ldots q} \| \lambda_i^{(k)} - \lambda_j^{(k+1)} \|
\end{align}
If a conflict arises, resulting in assignment of the same $\tilde{j}$ for multiple $i$, mode-shape matching is invoked as the secondary criterion for tracking. The modal assurance criterion (MAC) is a popular metric used for comparing mode shapes  \cite{beaverstock2015automatic}, given by,
\begin{align}
    \text{MAC}(\Phi_i,\Phi_j)=\frac{\big| \Phi_i^\text{T}  \ \overline{\Phi}_j\big|^2 }{(\Phi_i^\text{T} \  \overline{\Phi}_i)\cdot(\Phi_j^\text{T} \ \overline{\Phi}_j) }.
\end{align}
This work uses the absolute value of MAC, defined as $\rho(\Phi_i, \Phi_j) = |\text{MAC} (\Phi_i, \Phi_j)|$. The maximum value $\rho$ can attain is $1$, denoting an exact configuration match, while $\rho = 0$ indicates no match at all. The tracking algorithm is described in Algorithm 1.

 \begin{algorithm}[tbh!] \label{tracking_algo}
 \caption{Algorithm for tracking DMD eigenvalue-mode pair $(\lambda, \Phi)$.}
 \begin{algorithmic}[1]
 \renewcommand{\algorithmicrequire}{\textbf{Input: }}
 \renewcommand{\algorithmicensure}{\textbf{Output:}}
 \REQUIRE  DMD eigenvalue-mode pair ($\lambda_i^{(k)},\Phi_i^{(k)}$) from $k^{th}$ window, $i=1,2,\ldots,p$ and ($\lambda_j^{(k+1)},\Phi_j^{(k+1)}$) from $(k+1)^{th}$ window, $j=1,2,\ldots,q$. 
 \ENSURE Successor of ($\lambda_{i}^{(k)},\Phi_i^{(k)}$).
 \FOR {i = 1 to p}
 \STATE {Find  $\tilde{j} =\argmin_{j = 1, \ldots q}d(i,j)= \argmin_{j = 1, \ldots q} \| \lambda_i^{(k)} - \lambda_j^{(k+1)} \|$.}
 \ENDFOR
 \IF{All $i$ are associated with distinct $\tilde{j}$} \label{repeat_step}
 \RETURN{($\lambda_{\tilde{j}}^{(k+1)},\Phi_{\tilde{j}}^{(k+1)}$) as successor of respective ($\lambda_i^{(k)},\Phi_i^{(k)}$).}
 \ELSE
 \STATE{Identify the set of indices $i$ which share a common $\tilde{j}$. Let $I$ be the set of $i$ ($i\in I$) which have common $\tilde{j}=\tilde{j}_I$.}
   \STATE{ Identify $ \hat{i}= \argmax_{i\in I} \rho(\Phi_i^{(k)},\Phi_{\tilde{j_I}}^{(k+1)})$ }\label{repeat1}
   \STATE{Identify the second closest eigenvalue to $\lambda_{\hat{i}}^{(k)}$ from $(k+1)^{th}$ window after $\lambda_{\tilde{j}_{I}}^{(k+1)}$. Let the index of second closest eigenvalue be $\tilde{j}_{2I}$ .}
  \IF{$\big(\rho(\Phi_{\hat{i}}^{(k)},\Phi_{\tilde{j}_I}^{(k+1)})\geq\rho(\Phi_{\hat{i}}^{(k)},\Phi_{\tilde{j}_{2I}}^{(k+1)})\big)$}
   \RETURN{($\lambda_{\tilde{j_I}}^{(k+1)},\Phi_{\tilde{j_I}}^{(k+1)}$) as successor of $(\lambda_{\hat{i}}^{(k)},\Phi_{\hat{i}}^{(k)})$.}
   \STATE{ For, $\forall{i}\in I-\{\hat{i}\}$, replace the closest eigenvalue index $\tilde{j}_I$ by next closest eigenvalue index $\tilde{j}_{2I}$ from $(k+1)^{th}$ window. If there is no next closest eigenvalue, the eigenvalue-mode pair corresponding to $i\in I-\{\hat{i}\}$ is not tracked further. Repeat from step \ref{repeat_step} for rest of the eigenvalues. }
  \ELSE
  \STATE{Delete $\hat{i}$ from $I$, so that $\hat{i}\not\in I$. Then repeat from step \ref{repeat1}. }
  \ENDIF
  \IF{$\big(\rho(\Phi_{i}^{(k)},\Phi_{\tilde{j}_I}^{(k+1)})<\rho(\Phi_{i}^{(k)},\Phi_{\tilde{j}_{2I}}^{(k+1)})\big)$, $\forall{i}\in I$}
  \STATE{Identify the $ \tilde{i}= \argmax_{i\in I} |d(i,\tilde{j}_I)-d(i,\tilde{j}_{2I})|$.}
 \RETURN{($\lambda_{\tilde{j}_I}^{(k+1)},\Phi_{\tilde{j}_I}^{(k+1)}$) as successor of ($\lambda_{\tilde{i}}^{(k)},\Phi_{\tilde{i}}^{(k)}$).} 
  \STATE{ For, $\forall{i}\in I-\{\tilde{i}\}$, replace the closest eigenvalue index $\tilde{j}_I$ by next closest eigenvalue index $\tilde{j}_{2I}$ from $(k+1)^{th}$ window. If there is no next closest eigenvalue, the eigenvalue-mode pair corresponding to $i\in I-\{\tilde{i}\}$ is not tracked further. Repeat from step \ref{repeat_step} for rest of the eigenvalues. }
 \ENDIF
 \ENDIF
 \end{algorithmic}
 \end{algorithm}
{At the bifurcation point, broadly two scenarios are possible. First, one complex conjugate pair of DMD eigenvalues generates two real eigenvalues after encountering the real axis. Second, two real DMD eigenvalues merge and become a complex conjugate pair of eigenvalues. Since for real data the DMD eigenvalues are mirrored with respect to the real axis, the tracking algorithm concentrates only on the upper half complex plane including the real axis. The first scenario leads to $p<q$, where the algorithm starts tracking the newly generated DMD eigenvalues from that particular window. For the second case $p>q$, the algorithm stops tracking some eigenvalues from the previous window. }
\subsection{State Transition to Equilibrium}\label{state_trans}
DMD analysis of a dynamical system focuses on low-dimensional modeling of the equilibrium state, usually ignoring transient phenomena \cite{bagheri2013koopman, page2019koopman, pascarella2019analysis}. Regardless of the ROM employed, knowing when the transient phase comes to an end is useful for terminating the high-fidelity simulation in a timely fashion so that future solution can be predicted with the ROM (sec. \ref{DMD_algo}). In the literature, several methods are available for detecting state transition of high-dimensional dynamical systems \cite{van2013interaction, feng2014deep, chekroun2014rough, tantet2015early}. The authors in \cite{gottwald2019detecting} have presented a method exploiting DMD reconstruction error for identifying such transitions. The current paper takes advantage of the temporal variation in position of the DMD eigenvalues in the complex plane with respect to the unit circle. The algorithm presented here can be exploited for in-line applications, given some a priori knowledge about the timescale of the problem. This will be discussed in details later. A preliminary version of this state transition algorithm was described in our work \cite{nayak2021detecting}. \par
\begin{figure} [t]
    \centering
  \subfloat[\label{fig:abs_move} ]{%
       \includegraphics[width=0.5\linewidth]{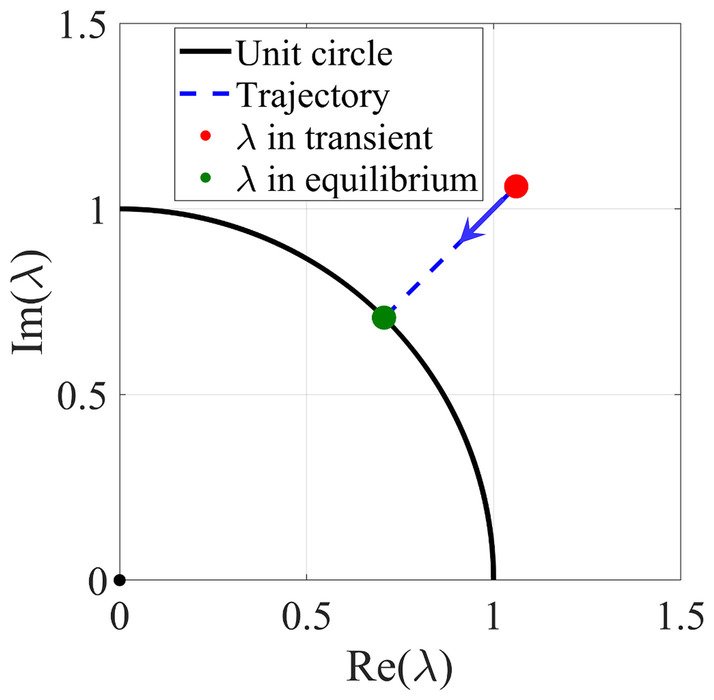}}
  \subfloat[\label{fig:ang_move} ]{%
        \includegraphics[width=0.5\linewidth]{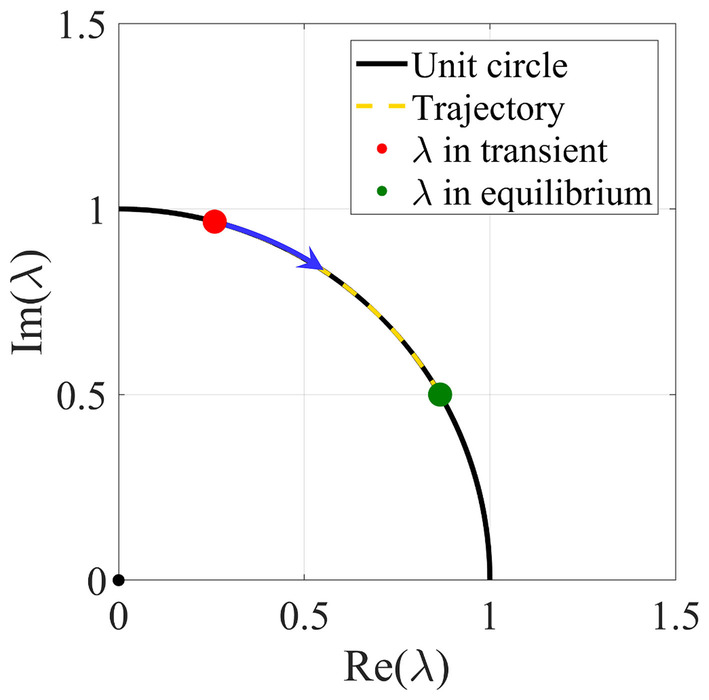}}
    \hfill
  \caption{\small{ Schematic representation of DMD eigenvalue ($\lambda$) migration. In practical cases, the trajectory can be complex combination of both the trajectories shown. (a) Radial movement of eigenvalue towards unit circle as the DMD window moves from transient to equilibrium state. (b)  Eigenvalue movement along unit circle to the equilibrium position (green) as the DMD window moves towards equilibrium from transient.}   }\label{fig:general_eigen_move}
\end{figure}
The fundamental idea behind the proposed approach is that given a sufficiently wide data harvesting region within the equilibrium state, it follows that (a) the dominant DMD eigenvalues lie on the unit circle \cite{schmid2010dynamic, tu2013dynamic, sasaki2019using} and (b) the mode shapes and corresponding frequencies associated with the dominant DMD modes remain invariant. Intuitively, the latter makes sense because in equilibrium the dynamics of the system remain unchanged irrespective of the observation window as long as that window is sufficiently wide. The Maxwell-Vlasov equations (governing equations in kinetic plasma simulations) are autonomous in nature. In the equilibrium state, the number of particles entering the solution domain remains same as the number of particles leaving, ensuring that the governing dynamics are autonomous.
Note that the solution of the well-posed DMD is unique \cite{hirsh2019centering}, whereby the extracted dominant DMD modes and corresponding eigenvalues remain unchanged as we slide the window within equilibrium region. Of course, conditions (a) and (b) are not necessarily true in the transient state as indicated by the presence of dominant DMD eigenvalues away from the unit circle and continuously changing dynamics. \par

In practical scenarios, the data obtained is not free from noise {(in a general sense, either from finite machine precision and discretization errors present in a simulation or from ambient and instrument noise present in a measurement)}. As a result, conditions (a) and (b) are not exactly satisfied. 
Therefore, we emphasize that invariance of characteristics applies to only dominant DMD modes (i.e. with physically meaningful character) in the equilibrium stage. DMD modes corresponding to the  noise space of the data do not follow such observations. Here, we adopt a $5\%$ error criterion, so the first few high energy DMD modes corresponding to $\geq 95\%$ of the reconstructed amplitude are defined as the dominant modes. However, the modal amplitude, defined as $A_m(t) = \| \vartheta_m\Phi_m e^{\omega_m(t-t_0)}+\overline{\vartheta}_m\overline{\Phi}_m e^{\overline{\omega}_m(t-t_0)} \|_2$ ($\|.\|_2$ denotes the Frobenius norm) varies with time, so the measurements are performed at the end of the DMD harvesting window. As the harvesting window approaches the equilibrium state, two key parameters are tracked. The $\alpha$ parameter measures the relative error in the reconstructed data, assuming exponential growth or decay to be the only source of error due to non-zero distance of the dominant DMD eigenvalues from the unit circle. The $\beta$ parameter represents the error in reconstructed data considering the error only due to fluctuation in phase of dominant DMD eigenvalues. The expressions for $\alpha$ and $\beta$ are derived next. Recall the DMD reconstruction formula{,}
\begin{subequations}
\begin{align}
    \hat{\mathbf{x}}(t)&\approx\sum_{m=1}^{M_d} (\vartheta_m\Phi_m e^{\omega_m(t-t_0)}+\overline{\vartheta}_m\overline{\Phi}_m e^{\overline{\omega}_m(t-t_0)})\\ 
     & =\sum_{m=1}^{M_d} e^{\omega_{mR}(t-t_0)}(\vartheta_m\Phi_m e^{j\omega_{mI}(t-t_0)}+\overline{\vartheta}_m\overline{\Phi}_m e^{-j\omega_{mI}(t-t_0)})\nonumber\\
        & =\sum_{m=1}^{M_d} e^{\omega_{mR}(t-t_0)}\psi_m,
         \label{DMD_recon_alpha}
\end{align}
\end{subequations}
where complex frequency $\omega_m=\omega_{mR}+j\omega_{mI}$. $\psi_m=(\vartheta_m\Phi_m e^{j\omega_{mI}(t-t_0)}+\overline{\vartheta}_m\overline{\Phi}_m e^{-j\omega_{mI}(t-t_0)})$ 
is the oscillating part of the solution and $M_d$ is the number of dominant DMD modes. In equilibrium, the DMD solution must not include exponentially growing or decaying factors, i.e. $\omega_{mR}= 0$, $m = 1, 2, \ldots M_d$. Assuming non-zero $\omega_{mR}$ to be the only source of error, the ideal solution ${\mathbf{x}}(t)$ must be
\begin{align}\label{DMD_recon_conj2_org}
          {\mathbf{x}}(t)\approx\sum_{m=1}^{M_d} \psi_m.
\end{align}
Using $\omega_m = ln(\lambda_m)/\Delta t$, $\lambda_m = |\lambda_m| e^{j\theta_m}$, one can write $\omega_{m} = \frac{ln|\lambda_m|}{\Delta t} + \frac{j\theta_m}{\Delta t} = \omega_{mR} + j\omega_{mI}$, giving $\omega_{mR} = \frac{ln|\lambda_m|}{\Delta t}$. From \eqref{DMD_recon_alpha},
\begin{align}\label{DMD_recon_conj2}
          \hat{\mathbf{x}}(t)\approx\sum_{m=1}^{M_d} |\lambda_m|^{\frac{(t-t_0)}{\Delta t}} \psi_m.
\end{align}
The relative 2-norm error, $\delta(t)$, is given below, under the assumption that error is only due to exponential growth/decay. This results in the definition of the parameter $\alpha$:
\begin{align}\label{DMD_err_alpha}
          \delta(t)=\frac{||\hat{\mathbf{x}}(t)-{\mathbf{x}}(t)||_2}{||{\mathbf{x}}(t)||_2}
       &=\frac{||\sum_{m=1}^{M_d} |\lambda_m|^{\frac{(t-t_0)}{\Delta t}} \psi_m-\sum_{m=1}^{M_d} \psi_m||_2}{||\sum_{m=1}^{M_d} \psi_m||_2}\\
    &=\frac{||\sum_{m=1}^{M_d} (|\lambda_m|^{\frac{(t-t_0)}{\Delta t}}-1) \psi_m||_2}{||\sum_{m=1}^{M_d} \psi_m||_2}\nonumber\\
    \end{align}
 and hence,  from the triangle inequality, it follows 
 \begin{align}\label{DMD_err_alpha_2}
    \delta(t)&\leq\frac{\sum_{m=1}^{M_d} ||(|\lambda_m|^{\frac{(t-t_0)}{\Delta t}}-1) \psi_m||_2}{||\sum_{m=1}^{M_d} \psi_m||_2}=\alpha(t-t_0)=\alpha(\tilde{t}).
\end{align}
Note that $\psi_m$ is a function of the time difference between target time $t$ and the reference initial time of that particular DMD window $t_0$, which we denote as $\tilde{t}$. As a result, we can write $\alpha(t-t_0)$=$\alpha(\tilde{t})$ in the above. As the DMD window moves towards equilibrium, the dominant DMD eigenvalues move closer towards unit circle (Fig. \ref{fig:general_eigen_move}), thus decreasing $\alpha(\tilde{t})$, for a fixed $\tilde{t}$. We define the convergence in $\alpha$ as the termination of its secular decay. However, a dynamical system can continue to be in transience even after all DMD eigenvalues have moved to the unit circle (Fig. \ref{fig:ang_move}). This happens when the transient state involves variation of frequency content instead of amplitude. Thus the presented approach also validates that the dominant DMD eigenvalues do not move along the unit circle. Doing so ensures that the error due to shift in phase ($\Delta \theta_m$) over successive windows is less than some predetermined threshold $\beta_{thr}$.
\begin{align}
    \hat{\mathbf{x}}(t)& \approx \sum_{m=1}^{M_d} (\vartheta_m\Phi_m e^{\tilde{\omega}_m(t-t_0)}+\overline{\vartheta}_m\overline{\Phi}_m e^{\overline{\tilde{\omega}}_m(t-t_0)})\\
    &=\sum_{m=1}^{M_d} (\vartheta_m\Phi_m e^{(\omega_{mR}+j(\omega_{mI}+\Delta\omega_{mI}))(t-t_0)}+\overline{\vartheta}_m\overline{\Phi}_m e^{(\omega_{mR}-j(\omega_{mI}+\Delta\omega_{mI}))(t-t_0)})\nonumber\\
     &=\sum_{m=1}^{M_d} (\chi_m e^{j\Delta\omega_{mI}(t-t_0)}+\overline{\chi}_m e^{-j\Delta\omega_{mI}(t-t_0)})\nonumber
    \end{align}
    \begin{align}
     &=\sum_{m=1}^{M_d} (\chi_m e^{j\frac{\Delta\theta_{m}}{\Delta t}(t-t_0)}+\overline{\chi}_m e^{-j\frac{\Delta\theta_{m}}{\Delta t}(t-t_0)})\nonumber\\
     &=\sum_{m=1}^{M_d} 2\text{Re}\{\chi_m e^{j\frac{\Delta\theta_{m}}{\Delta t}(t-t_0)}\},
\end{align}
where, $\tilde{\omega}_m=\omega_m+j\Delta\omega_{mI}$ and $\chi_m=\vartheta_m\Phi_m e^{\omega_m(t-t_0)}$, a function of $(t-t_0)=\tilde{t}$. As above, we compute the relative 2-norm error $\delta(t)$ under the assumption that error is only due to $\Delta\omega_{mI}$ with ${\mathbf{x}}(t)\approx\sum_{m=1}^{M_d} 2\text{Re}\{\chi_m\}$ as the ideal {solution. This results} in the definition of the parameter $\beta$:
\begin{align}\label{DMD_err_beta}
          \delta(t)=\frac{||\hat{\mathbf{x}}(t)-{\mathbf{x}}(t)||_2}{||{\mathbf{x}}(t)||_2}&=\frac{||\sum_{m=1}^{M_d} 2\text{Re}\{\chi_me^{j\frac{\Delta\theta_m}{\Delta t}(t-t_0)}\}-\sum_{m=1}^{M_d} 2\text{Re}\{\chi_m\}||_2}{||\sum_{m=1}^{M_d} 2\text{Re}\{\chi_m\}||_2}\nonumber\\
    &=\frac{||\sum_{m=1}^{M_d} 2\text{Re}\{\chi_m(e^{j\frac{\Delta\theta_m}{\Delta t}(t-t_0)}-1)\}||_2}{||\sum_{m=1}^{M_d} 2\text{Re}\{\chi_m\}||_2}\nonumber\\
    \end{align}
and hence     
\begin{align}\label{DMD_err_beta_2}
       \delta(t) \leq \frac{\sum_{m=1}^{M_d}|| \text{Re}\{\chi_m(e^{j\frac{\Delta\theta_m}{\Delta t}(t-t_0)}-1)\}||_2}{||\sum_{m=1}^{M_d} \text{Re}\{\chi_m\}||_2}
    =\beta(t-t_0)=\beta(\tilde{t})
\end{align}

Parameters $\alpha$ and $\beta$ are computed for dominant DMD modes only. Our goal is to detect the ``knee'' or ``elbow'' region in the graph of $\alpha$ against $k$ (window index), denoting transition to equilibrium state. However, in-line detection of the knee region is challenging, especially when data is noisy. We thus examine the rolling average of $\alpha$ over $W$ successive windows and search for a non-negative slope in the averaged graph, hinting convergence in $\alpha$. Once convergence in $\alpha$ is detected, the focus shifts to the parameter $\beta$ to ensure that the error due to phase shift of the dominant eigenvalues over $W$ windows is within an acceptable bound $(\leq \beta_{thr})$. We will illustrate the in-line algorithm (Algorithm 2) assuming we have some a priori knowledge about the timescale of the problem to inform the selection of appropriate window width $T$. \par

\begin{algorithm}[hbt!]\label{algo_equi}
 \caption{Algorithm for detecting onset of equilibrium state}
 \begin{algorithmic}[1]
 \renewcommand{\algorithmicrequire}{\textbf{Input: }}
 \renewcommand{\algorithmicensure}{\textbf{Output:}}
 \REQUIRE Data from high-fidelity simulation.
 \ENSURE  Window index indicating onset of equilibrium.
 \\ \textit{Initialization} : For first window (k=1), calculate $r$ as in \eqref{A_formula} using optimal hard thresholding and use it for rest of the algorithm. 
   \STATE At current ($k^{th}$) window, say $D$ denotes the set of dominant DMD eigenvalues. Identify the $i$ for which $\lambda_i^{(k)}\in D$, where $1\leq i \leq p$, and $p$ is number of DMD modes ($M$ from \eqref{DMD_recon_conj}) in the $k^{th}$ window. \label{first_step}
  \STATE Calculate $\alpha{(\tilde{t})}$ for $k^{th}$ window at $\tilde{t}=T$ denoted by $\alpha(T)^{(k)}$, where $T$ is the DMD window width. \label{second_step}
  \STATE For $k\geq Wh$, perform averaging of $\alpha$ over $W$ windows to get $<\alpha>_W^{(h)}=[\alpha(T)^{(s+1)}+\alpha(T)^{(s+2)}+\ldots+\alpha(T)^{(s+W)}]/W$, where $s=W(h-1)$, $h=1,2,\ldots$.
  \IF {(log$(<\alpha>_W^{(h)})\geq$ log$(<\alpha>_W^{(h-1)})$)}\label{check_step}
  \STATE From the tracking Algorithm 1, identify the predecessors of $\lambda_i^{(k)}\in D$ for previous $W$ windows, $\lambda_{(i)}^{(k-1)},\lambda_{(i)}^{(k-2)},~.~.~.,\lambda_{(i)}^{(k-W)}$, calculate
  $\Delta\theta_{a,i}^{(k)}=|\text{Arg}(\lambda_{(i)}^{(k-a)})-\text{Arg}(\lambda_{i}^{(k)})|$, where $a=1,2,\ldots,W$.
  \STATE Calculate $\beta(T)$ wrt. $\Delta\theta_{a,i}^{(k)}$ at $k^{th}$ window for $W$ predecessors, $\beta(T)_1^{(k)},\beta(T)_2^{(k)},~.~.~.,\beta(T)_W^{(k)}$.
  \IF {($\beta(T)_1^{(k)},\beta(T)_2^{(k)},~.~.~.,\beta(T)_W^{(k)}\leq \beta_{thr}$)}
  \STATE Stop harvesting.
   \RETURN $k$
   \ELSE 
   \STATE Continue harvesting, move to the $(k+1)^{th}$ window and return to the step \ref{first_step}.
  \ENDIF
  \ELSE 
  \STATE Continue harvesting, move to the $(k+1)^{th}$ window and return to the step \ref{first_step}.
  \ENDIF 
 \end{algorithmic}
 \end{algorithm}

It is important to note that the performance of Algorithm 2 depends on the choice of  parameters $T$, $W$, $\beta_{thr}$ and $\Delta_k$, where $\Delta_k$ is the shift between two successive sliding DMD windows. These parameters must be selected beforehand and do not adapt during the run. We make the following observations:

\begin{itemize}
    \item The parameter $T$ denotes the width of each sliding window. Prior knowledge about the time-scale of the problem helps to make sure that $T$ covers multiple oscillation cycles (if any) in the equilibrium state. If the window width is not sufficient to capture the dynamics of the equilibrium, temporal variation in the parameter $\alpha$ might not elicit convergence even as the window slides towards equilibrium. For offline applications, one of many simple algorithms such as zero crossing detection or peak detection can be used to approximate the period of limit cycle oscillations. 
    \item The shift $\Delta_k$ generally spans an integer multiple of snapshots. For in-line processes, the natural choice is to shift by one snapshot, in which case DMD is performed when a new snapshot becomes available. In our test cases, we shift by two snapshots as it provides enough headroom to play with varying snapshot intervals, keeping the shift $\Delta_k$ constant. Ideally, overlap between successive windows must be avoided to minimize the computation cost. In practice however, intersection between consecutive sliding windows is employed for the following reasons: (i.) window overlap implies smaller perturbation in the snapshot matrix, which helps with tracking DMD eigenvalue-mode pair, and, (ii.) overlap helps determine $W$, the number of windows over which $\alpha$ is averaged. 
    \item $W$ is chosen such that it is the minimum number of shifts for there to be no overlap between $k^{th}$ and $(k+W)^{th}$ window. In other words, $W=\floor{T/\Delta_k}$. Too small a value for $W$ can result in premature, erroneous detection of equilibrium, especially for highly noisy $\alpha$ variation. While a large $W$ overcomes this difficulty, it is at the cost of delayed detection of equilibrium. Delayed detection of equilibrium does not pose risk of increased error in DMD extrapolation, only inefficiency. 
    \item The threshold $\beta_{thr}$ is based on the acceptable error limit for each application. In this work we follow a $5\%$ error criterion, but from experience, the error due to fluctuations in phase is much smaller than errors due to exponential growth/decay. Therefore, we set $\beta_{thr}=0.01$ $(1\% ~\text{error})$. As shown in \eqref{DMD_err_beta_2}, $\beta$ is calculated based on the difference in phase of dominant eigenvalue at the $k^{th}$ window and it's predecessors at previous windows. It aims to detect slow unidirectional movement (Fig. \ref{fig:ang_move}) of eigenvalues along the unit circle. As a rule of thumb, we check for the error due to shift in phase over $W$ previous windows. If there is extremely slow movement of DMD eigenvalues along the unit circle in the transient state, it might go undetected for small $W$ value. Of course, very slow phase variation might not be of  interest, as long as the reconstruction accuracy stays within acceptable limits.
\end{itemize}

\section{Results}
In this section we apply the tracking and equilibrium detection algorithm for several plasma examples. The effectiveness of DMD in the modeling and prediction of self electric fields as well as its effect on the particle dynamics is demonstrated. For application of the proposed equilibrium detection algorithm to a classic textbook use-case, the reader is referred to \ref{lorenz_96}, wherein
the well-known Lorenz'96 oscillator is studied. This section presents three test cases. The first two examples consider {a} two dimensional (2-D) plasma ball expansion and {an} oscillating electron beam respectively. We establish the effectiveness of DMD in extracting low dimensional key features from self electric field data $\mathbf{e}(t)$ of EMPIC kinetic plasma simulations and reconstruct the data with good accuracy. Sliding window DMD technique is then employed for in-line identification of the equilibrium state of each system. Finally, we investigate the extrapolation accuracy beyond the detected equilibrium point for both {the} predicted fields as well as {the} particle dynamics. 
 Convergence of dominant DMD mode shapes and movement of corresponding eigenvalues in the complex plane is presented.
The final example deals with virtual cathode formation, where the main focus is on the accuracy of predicted particle dynamics.
{We treat the data generated from high-fidelity EMPIC simulation as the ``ground truth" to evaluate DMD performance. Nevertheless, 
for long-term predictions, we should keep in mind that long simulation runtimes might introduce numerical noise in high-fidelity data queried at later time due to ``numerical heating'' effects \citep{PhysRevE.95.043302}.}{ Due to this and other sources of numerical error mentioned earlier, some of the dominant DMD eigenvalues might not lie exactly on the unit circle. After detecting equilibrium and before extrapolation, we adjust the dominant DMD eigenvalues in the radial direction so that they are exactly on the unit circle.}

\subsection{Plasma ball expansion}
\begin{figure} [t]
    \centering
  \subfloat[ \label{fig:ball_ss_snap} ]{%
       \includegraphics[width=0.475\linewidth]{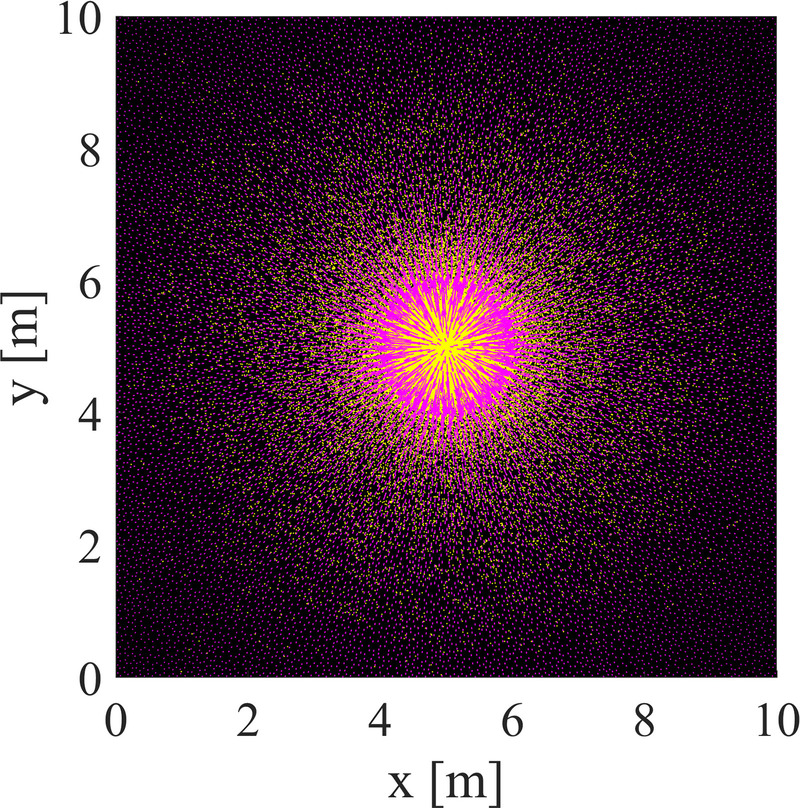}}
       \hspace{0.0cm}
  \subfloat[ \label{fig:ball_ss_sing} ]{%
        \includegraphics[width=0.51\linewidth]{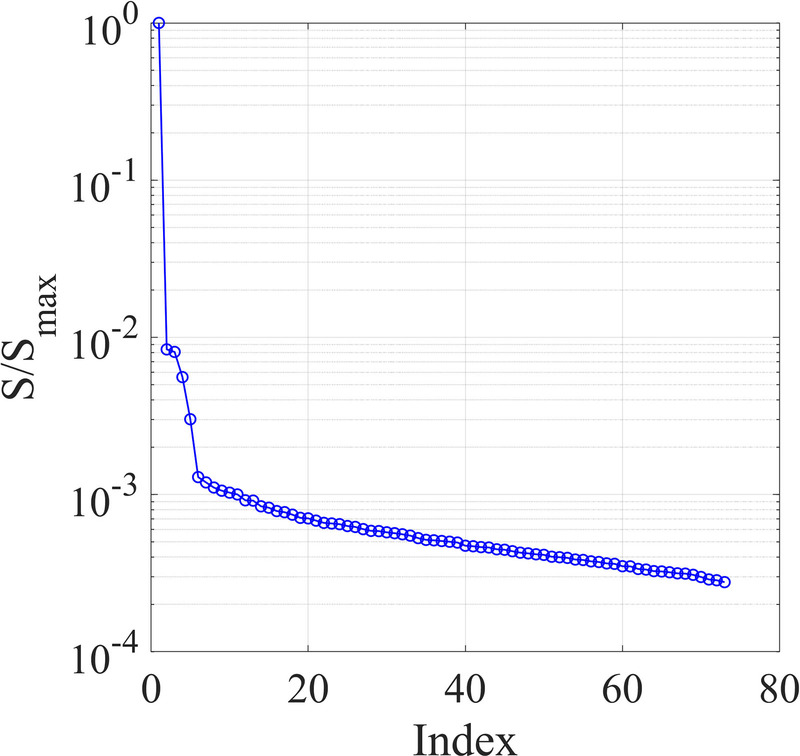}}
    \hfill
  \caption{\small{ (a) Snapshot of plasma ball expansion at $n=400000$ in a square cavity. The yellow dots represent superparticles and majenta arrows show the self electric field quiver plot. (b) Normalized singular values from SVD of snapshot matrix in equilibrium state. }  }
\end{figure}

The solution domain is a $L \times L$ square two-dimensional cavity ($L=10$ m: see Fig.~\ref{fig:ball_ss_snap}). It is discretized using an irregular triangular mesh with  $N_0=8037$ nodes, $N_1 = 23797$ edges and $N_2=15761$ triangles. Superparticles are initially placed at the center of the cavity within a circle of radius $0.5$ m. The plasma ball is initially assumed to be neutral as each electron-ion pair is initially located at the exact same position. All four sides of the cavity are assumed to be perfect magnetic conductors (PMC). Superparticles are given an initial radial velocity with Maxwellian distribution.  The timestep interval is $0.1$ ns and each superparticle represents $2\times 10^5$ electrons. Superparticles are absorbed as they hit the boundary. We sample the data every $\Delta_n=500$ timesteps until $n = 500000$.\par

\subsubsection{Self Electric Field Reconstruction}\label{ball_in_eq}
\begin{figure} [t]
    \centering
  \subfloat[ \label{fig:ball_ss_eig} ]{%
       \includegraphics[width=0.49\linewidth]{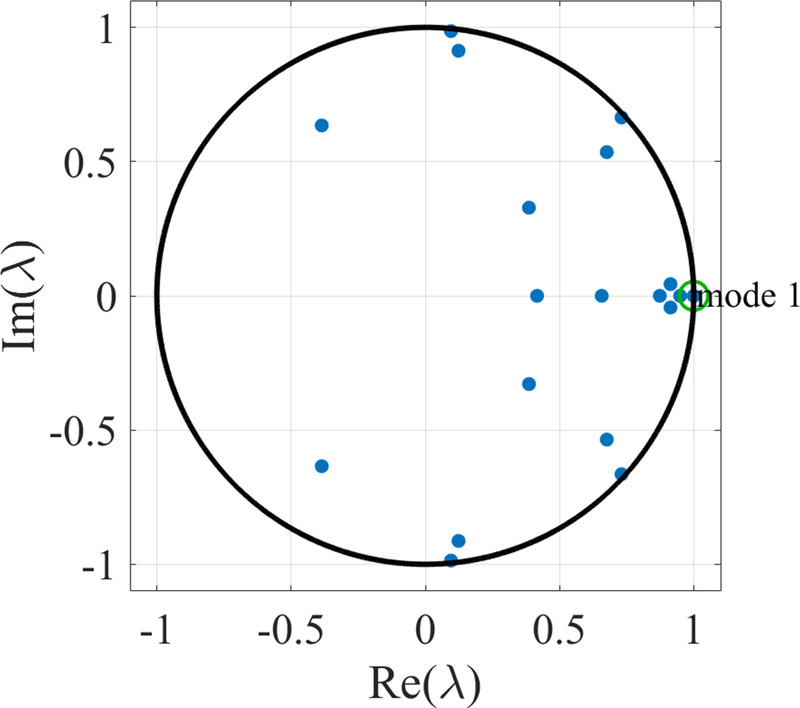}}
       \hspace{0.0cm}
  \subfloat[ $\Phi_1^{(ss)}$\label{fig:ball_ss_mode1} ]{%
        \includegraphics[width=0.5\linewidth]{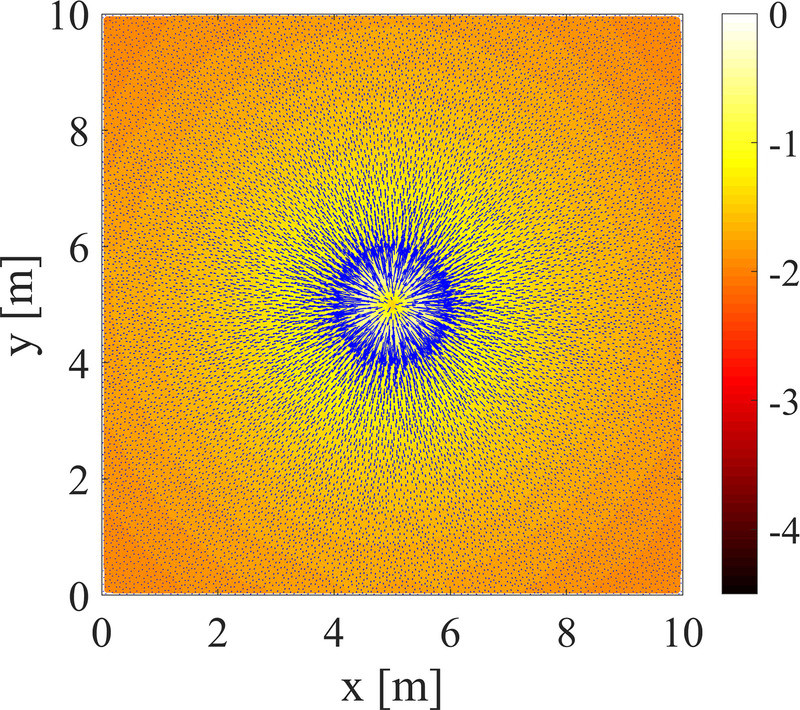}}
    \hfill
  \caption{\small{(a) DMD eigenvalues in the complex plane. The green circle denotes the dominant mode and black curve indicates the unit circle. (b) Dominant mode $\Phi_1^{(ss)}$. The blue arrows show the self electric field quiver plot. The colormap indicates logarithm (base 10) of amplitude. }  }\label{fig:ball_ss_eig_mode1}
\end{figure}

\begin{figure} [t]
\centering
  \subfloat[$\Phi_2^{(ss)}$ \label{fig:ball_ss_mode2} ]{%
       \includegraphics[width=0.33\linewidth]{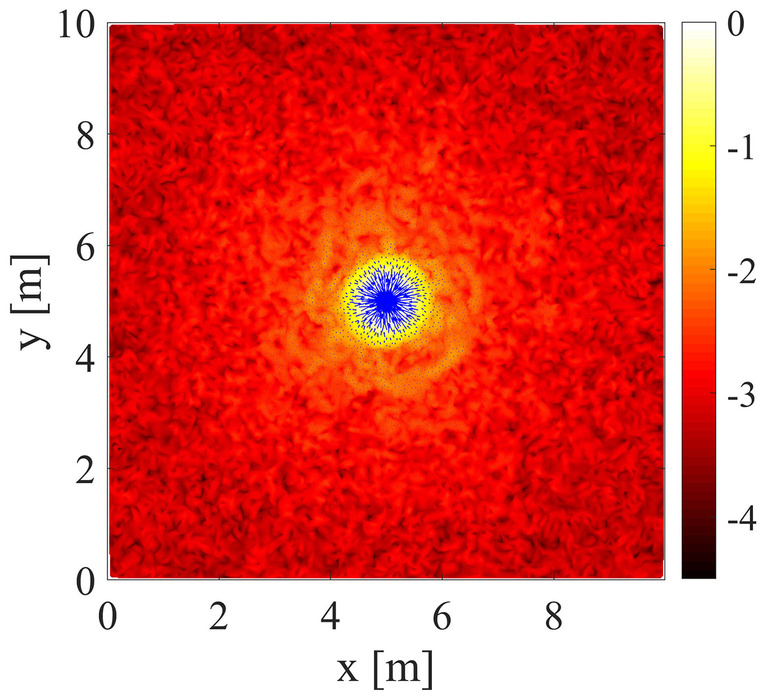}}
  \subfloat[$\Phi_3^{(ss)}$ \label{fig:ball_ss_mode3} ]{%
        \includegraphics[width=0.33\linewidth]{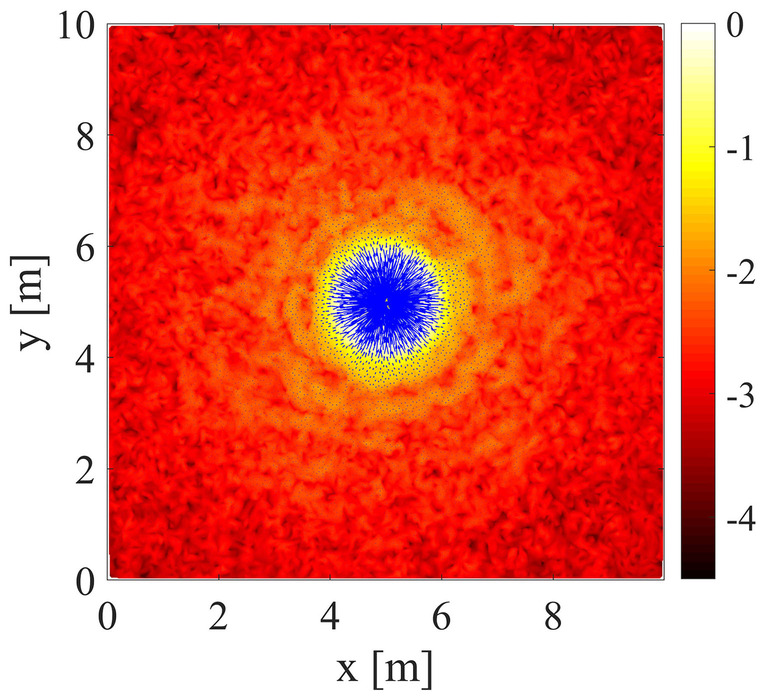}}
  \subfloat[$\Phi_4^{(ss)}$ \label{fig:ball_ss_mode4} ]{%
        \includegraphics[width=0.33\linewidth]{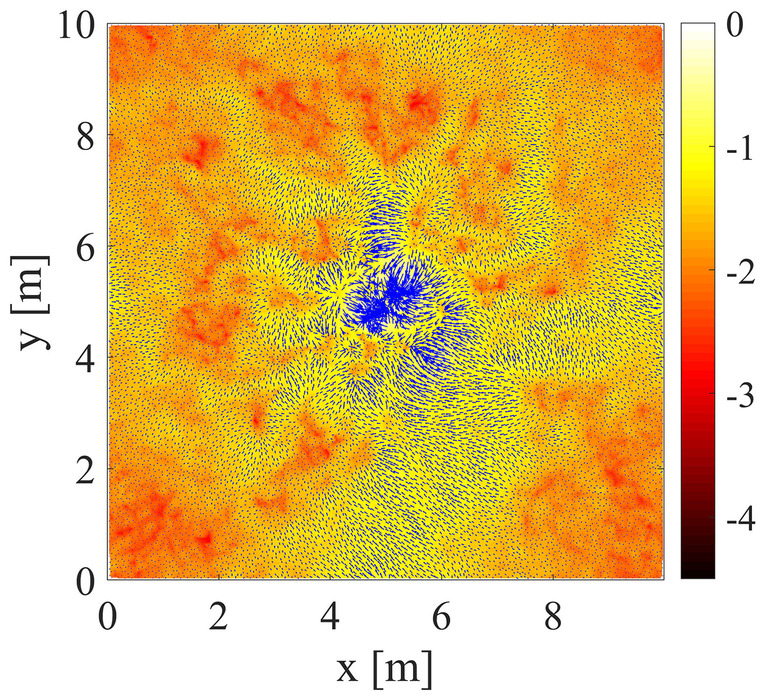}}\\
 \subfloat[$\Phi_5^{(ss)}$ \label{fig:ball_ss_mode5} ]{%
        \includegraphics[width=0.33\linewidth]{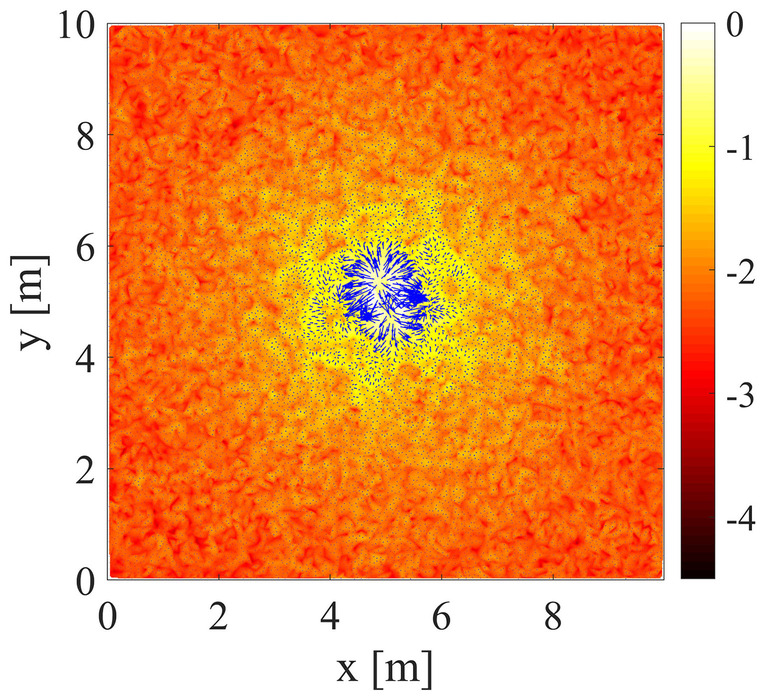}}
\subfloat[$\Phi_6^{(ss)}$ \label{fig:ball_ss_mode6} ]{%
        \includegraphics[width=0.33\linewidth]{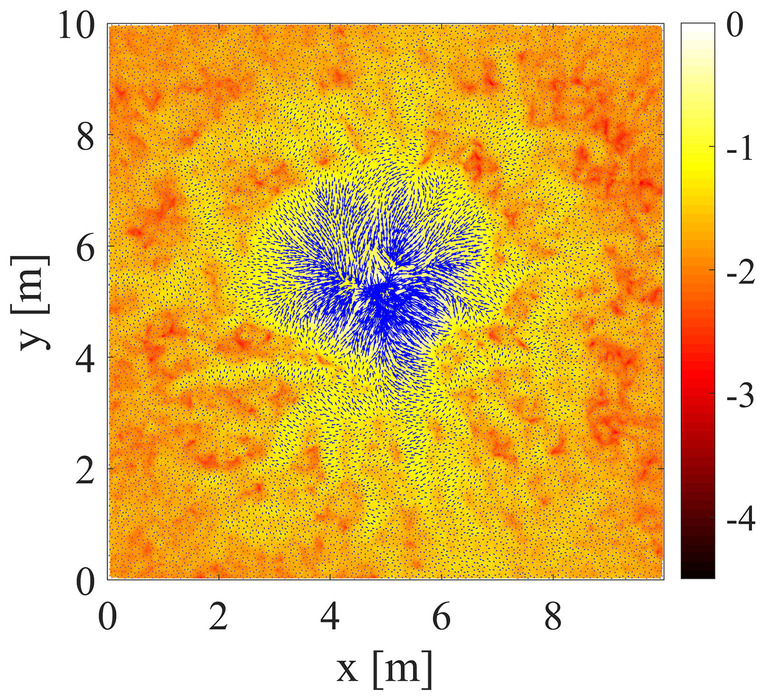}}
 \subfloat[$\Phi_7^{(ss)}$ \label{fig:ball_ss_mode7} ]{%
        \includegraphics[width=0.33\linewidth]{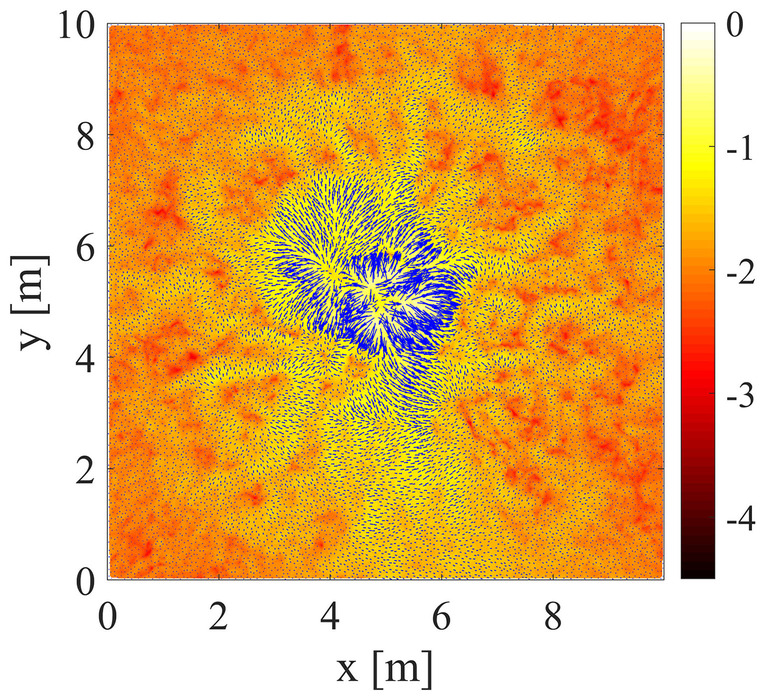}}
  \caption{\small{ First six recessive DMD modes for plasma ball in equilibrium.}
  \label{fig:ball_ss_rec}}
\end{figure}

In equilibrium, the self electric field attains a steady state with a constant spatial configuration (Fig.~\ref{fig:ball_ss_snap}). For extracting low-dimensional features in equilibrium through DMD, we harvest data from $n=200500$ to $n=275000$ with interval $\Delta t = 100$ ns between consecutive snapshots. A selection of $r = 19$ leads to $12$ DMD modes, effectively reducing the degrees of freedom from $23797$ to only $12$. Fig. \ref{fig:ball_ss_sing} shows the exponential decay of singular values, revealing the dominance of a single mode. The DMD eigenvalue distribution in the complex plane and dominant stationary mode ($\Phi_1^{(ss)}$) field configuration are shown in Fig. \ref{fig:ball_ss_eig_mode1}. The modes are numbered according to their energy content ($|A_m|^2$), with $\Phi_1^{(ss)}$ being the most energetic mode. With increasing mode indices (decreasing energy), the field configuration becomes more random, as can be observed in the recessive modes (Fig.~\ref{fig:ball_ss_rec}).

\begin{figure} [hbt!]
    \centering
  \subfloat[ Transient state DMD modes\label{fig:ball_trans_self_corr} ]{%
       \includegraphics[width=0.5\linewidth]{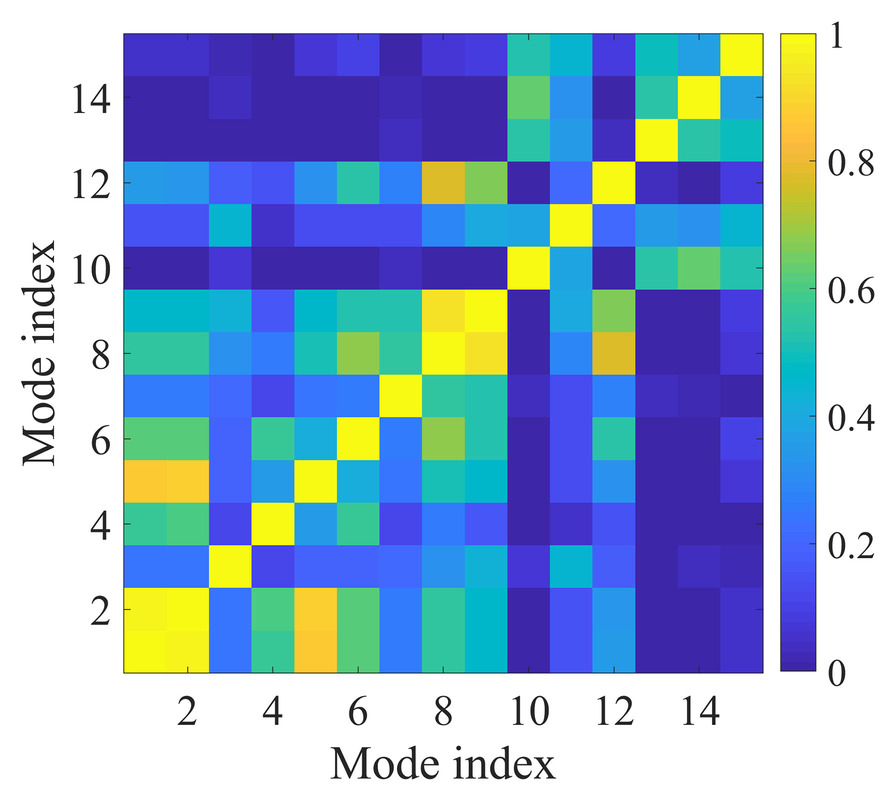}}
  \subfloat[ Equilibrium state DMD modes\label{fig:ball_ss_self_corr} ]{%
        \includegraphics[width=0.5\linewidth]{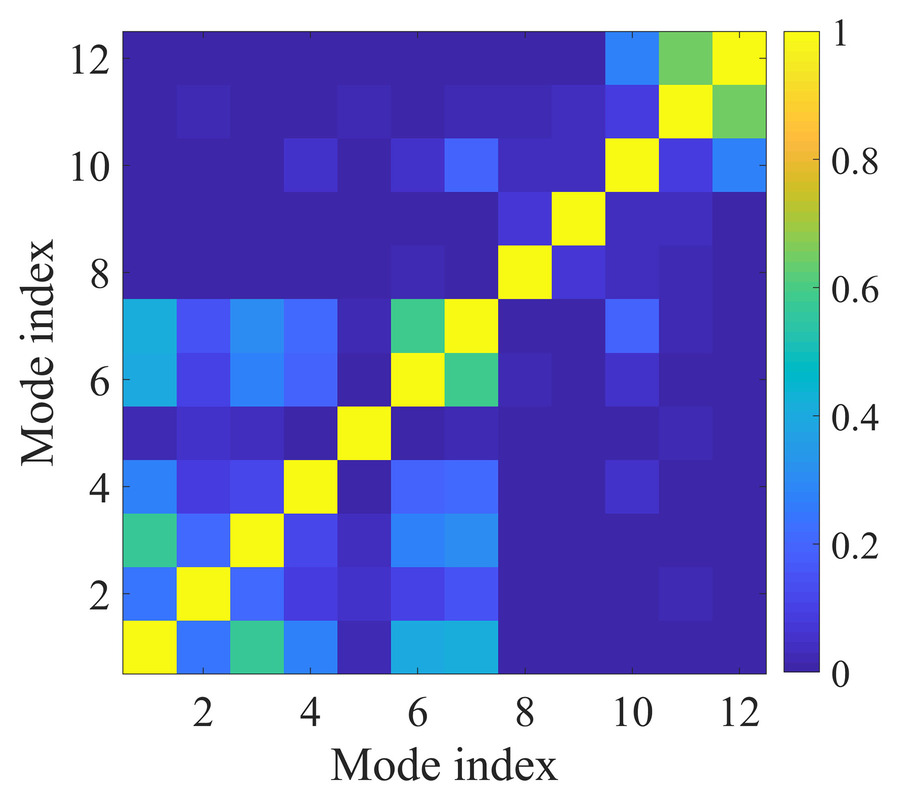}}
    \hfill
  \caption{\small{ (a) Absolute value of MAC coefficient $\rho$ between transient state DMD modes. (b) Coefficient $\rho$ between equilibrium state DMD modes. }  }\label{fig:ball_ss_trans_self_corr}
\end{figure}

\begin{figure} [t]
    \centering
  \subfloat[Transient state DMD\label{fig:ball_trans_err} ]{%
       \includegraphics[width=0.5\linewidth]{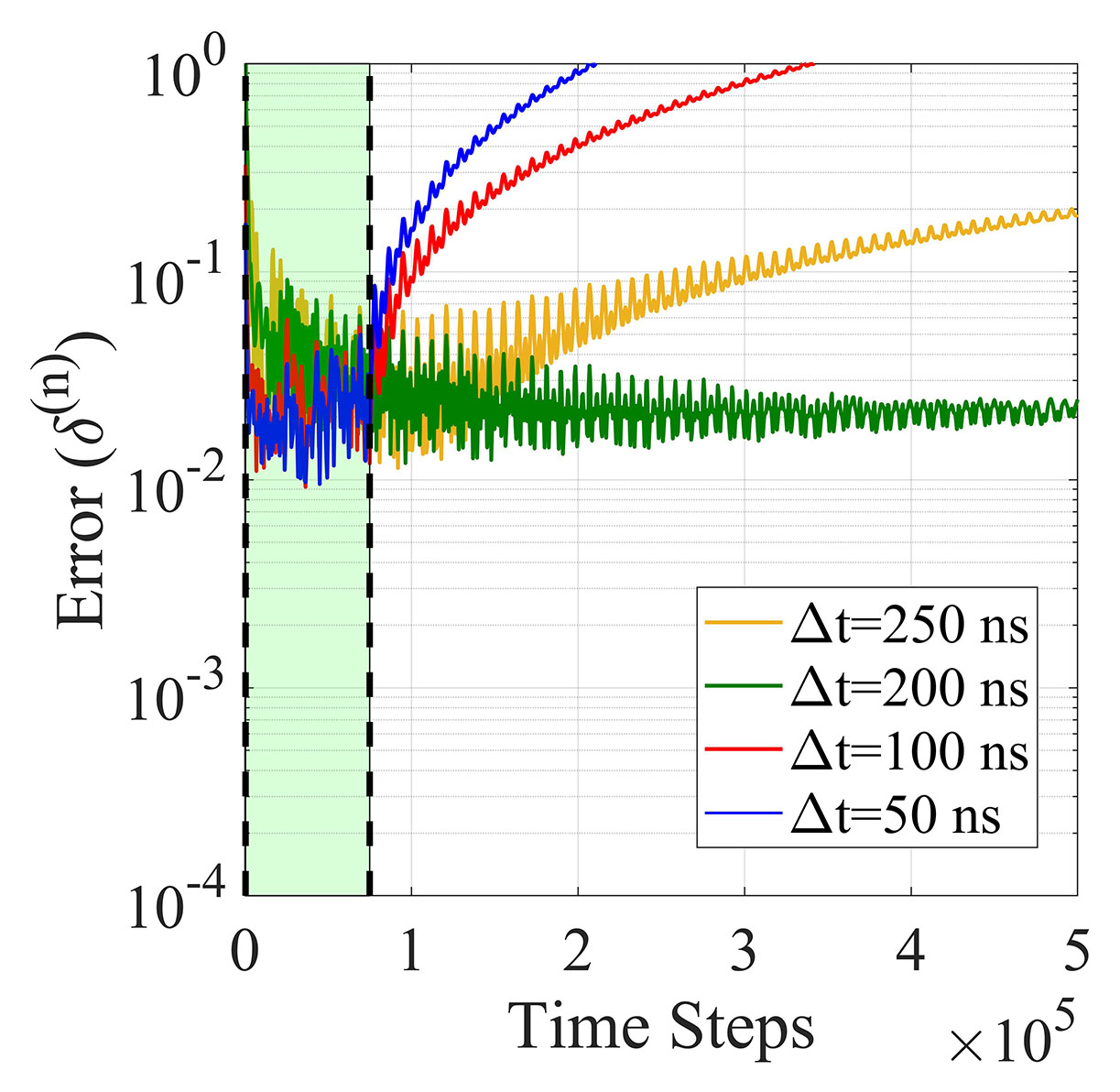}}
  \subfloat[Equilibrium state DMD\label{fig:ball_ss_err} ]{%
        \includegraphics[width=0.5\linewidth]{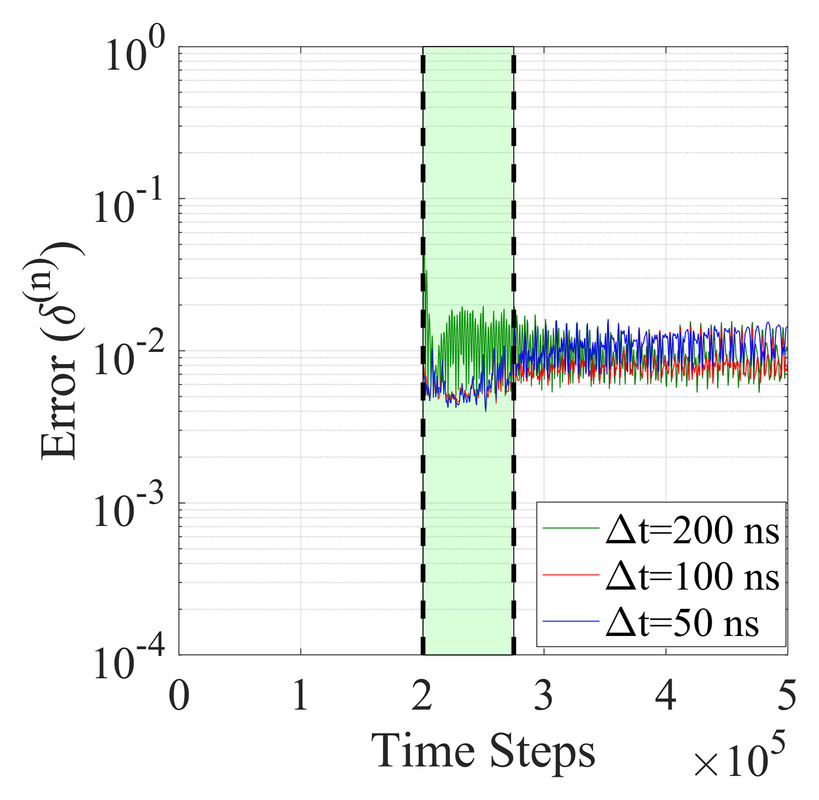}}
    \hfill
  \caption{\small{(a) 2-norm relative error when the DMD window (green shaded area) is in the transient region.} (b) 2-norm relative error when the DMD window (green shaded area) is in equilibrium region. }  \label{fig:ball_ss_trans_err}
\end{figure}

For comparison, we also perform DMD in the transient state with the harvesting region spanning from $n=500$ to $n=75000$, snapshots $\Delta t=100$ ns apart. We choose $r=27$, giving us $15$ DMD modes. For comparing the spatial configuration of DMD modes, we plot the absolute value of MAC ($\rho$) in a matrix form in Fig. \ref{fig:ball_ss_trans_self_corr}. Unlike other projection-based reduced order model techniques such as the proper orthogonal decomposition (POD), DMD does not enforce orthogonality of modes in the spatial domain, which explains the presence of nontrivial off-diagonal elements. At the same time, Fig.~\ref{fig:ball_ss_trans_self_corr}  reveals a clear distinction among various equilibrium DMD modes. 
On the other hand, transient-state DMD modes are less distinguishable from each other due to more complex dynamics.  The 2-norm relative error in reconstructed self electric field is shown in Fig.~\ref{fig:ball_ss_trans_err} for different sampling rates. 
The 2-norm relative error in DMD reconstruction ($\hat{\mathbf{e}}$) compared to the full-order solution (${\mathbf{e}}$) at $n^{th}$ timestep ($\delta^{(n)}$) is given by,
\begin{align}\label{error}
    \delta^{(n)}=\frac{||\hat{\mathbf{e}}^{(n)}-\mathbf{e}^{(n)}||_2}{||\mathbf{e}^{(n)}||_2}.
\end{align}
As expected, decreasing sampling interval ensures better accuracy, but the solution diverges more rapidly in the extrapolation region. Note that the surprisingly good performance for $\Delta t=200$ ns in Fig. \ref{fig:ball_trans_err} can be attributed to the ``aliasing" like effect for this particular sampling interval. The $\Delta t=200$ ns case is an anomaly for which the DMD frequencies are such that it produces a stable solution with relative error oscillating around a fixed value. This is further confirmed by the fact that the $\Delta t=250$ ns case continues to follow the trend as shown by the $\Delta t=50$ and $100$ ns cases. Higher error at the beginning of the simulation can be attributed simply to the very low field magnitudes then, causing a spike in the relative error. The extrapolation error is higher for transient state DMD compared to DMD in the equilibrium state, which further evokes the need to correctly determine the equilibrium state for good prediction accuracy. 

\subsubsection{Sliding-Window DMD}
In the plasma ball expansion case, the self-fields attain a non-oscillatory steady state in equilibrium.  This makes the prediction task trivial once equilibrium is detected and presents an opportunity to verify the accuracy of the equilibrium detection algorithm. First, we discuss the robustness of our algorithm with respect to the sampling interval and sliding window width. Then, the convergence of dominant mode shapes and accuracy of predicted particle dynamics is presented.

\begin{figure}[t]
    \centering
	\includegraphics[width=1\linewidth]{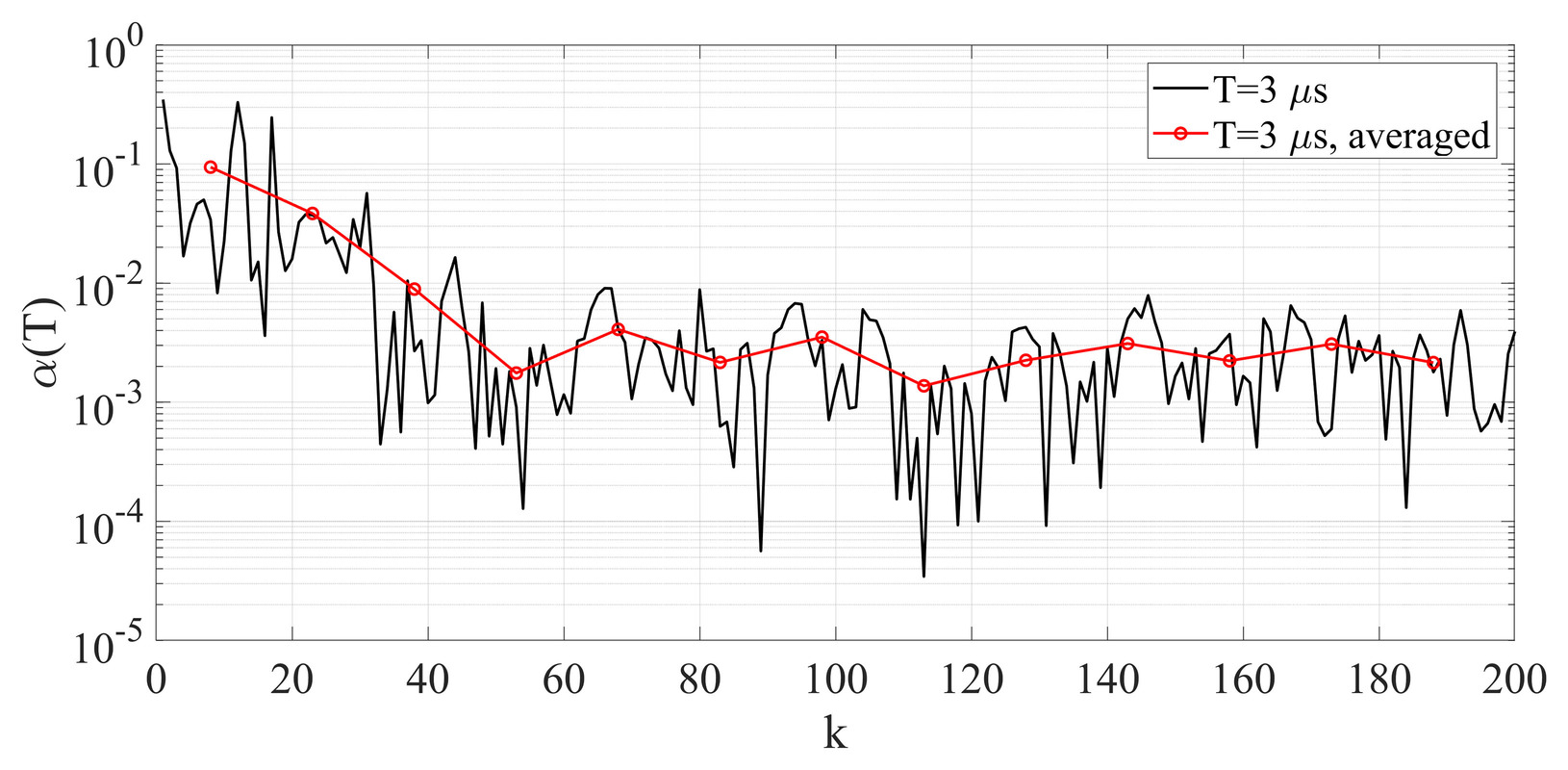}
    \caption{\small{Variation in $\alpha(T)$ as the window slides towards the equilibrium state for $\Delta t=100$ ns.}}
    \label{fig:ball_sld_alpha_var} 
\end{figure}

\begin{figure} [hbt!]
    \centering
  \subfloat[\label{fig:ball_sld_alpha_comp1} ]{%
       \includegraphics[width=0.49\linewidth]{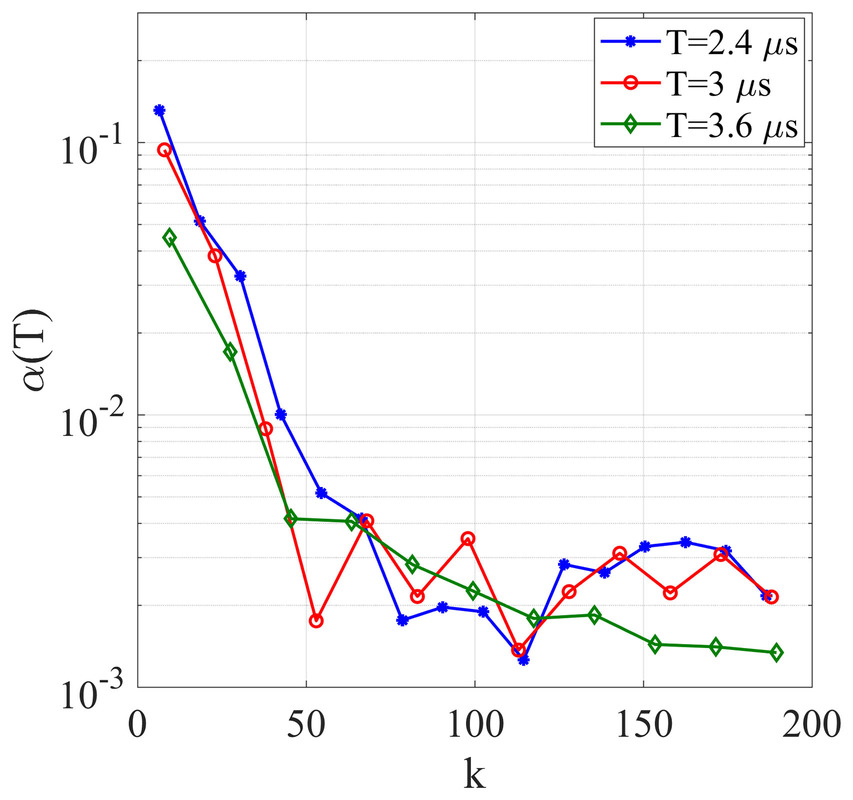}}
  \subfloat[\label{fig:ball_sld_alpha_comp2} ]{%
        \includegraphics[width=0.5\linewidth]{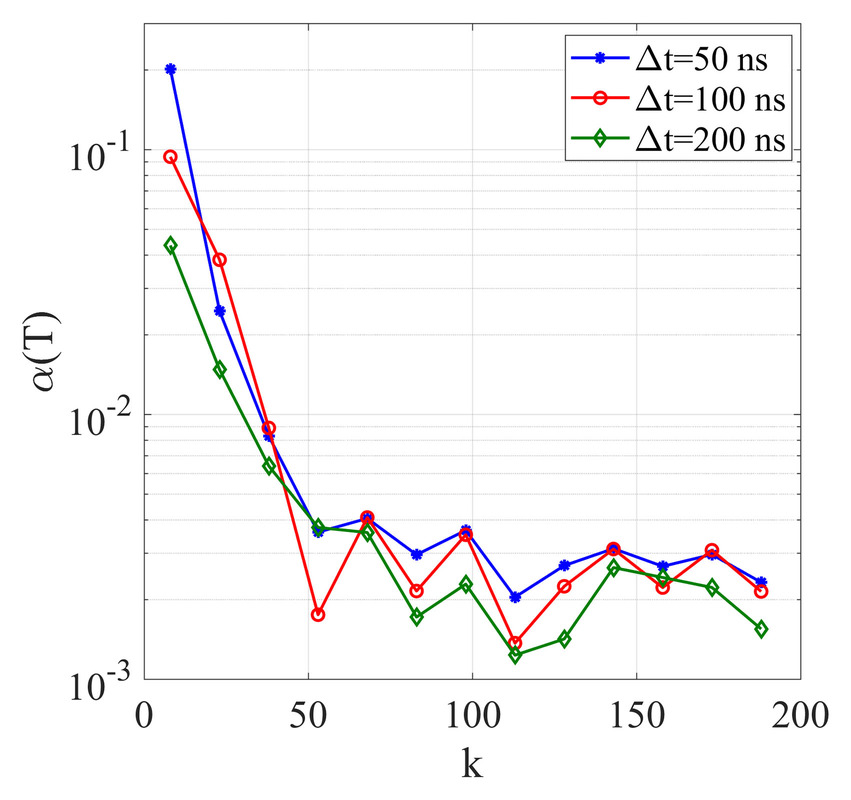}}
    \hfill
  \caption{\small{ (a) Sensitivity of Algorithm 2 towards window width $T~(\pm20\%)$, keeping fixed $\Delta t=100$ ns. (b) Sensitivity of Algorithm 2 towards sampling interval $\Delta t$ }, keeping fixed $T=3~\mu$s. }\label{fig:ball_sld_alpha_comp}
\end{figure}

\paragraph{Equilibrium Detection}\label{ball_eq_det}
Algorithm 2 is used for identifying the onset of equilibrium state with $\beta_{thr}=0.01$ and $\Delta_k=200$ ns. In this case we know that the fields will eventually attain steady state without limit-cycle oscillations, whereby the selection of $T$ is not a critical factor. We choose $T=3$ $\mu$s with 30 snapshots inside the harvesting window. Starting and ending points of the $k^{th}$ DMD window are given by $n_{st}(k)=\Delta_n+(k-1)\times n_{\Delta_{k}}$ and $n_{en}(k)=(\Delta_n+n_T)+(k-1)\times n_{\Delta_k}$ respectively, where $n_{\Delta_{k}}$ is the window shift in terms of timesteps ($\Delta_k=n_{\Delta_{k}}\Delta_t$) and $n_T$ denotes the number of timesteps forming the DMD window ($T=n_T\Delta_t$). \par

As seen in Fig. \ref{fig:ball_sld_alpha_var}, $\alpha(T)$ decreases initially with increasing $k$ and eventually converges, the knee/elbow region marking the transition from transient to steady-state. The algorithm detects the steady-state at $k=75$ ($n_{st}(75)=148500$). The sensitivity of $\alpha(T)$ towards variation in $T$ and $\Delta t$ is shown in Figs.~\ref{fig:ball_sld_alpha_comp1} and \ref{fig:ball_sld_alpha_comp2} respectively. For better comparison, we set $\Delta_k=200$ ns for all the three cases in Fig.~\ref{fig:ball_sld_alpha_comp2}. Using the non-negative slope criterion, the algorithm stops at $k=96,75$ and $144$ for $T=2.4~\mu$s, $3~\mu$s and $3.6~\mu$s respectively. As explained in \ref{lorenz_96},  non-negative slope is employed to indicate a ``knee'', which can potentially delay the detection of equilibrium. For $T=3.6~\mu$s in Fig. \ref{fig:ball_sld_alpha_comp1}, $\alpha(T)$ encounters non-negative slope at a much later time compared to the other two cases, even though the actual knee region appears earlier. For $\Delta t=50$ ns, $100$ ns and $200$ ns, the algorithm detects equilibrium at $k=75,75$ and $105$ respectively.

\begin{figure}[hbt!]
    \centering
	\includegraphics[width=1\linewidth]{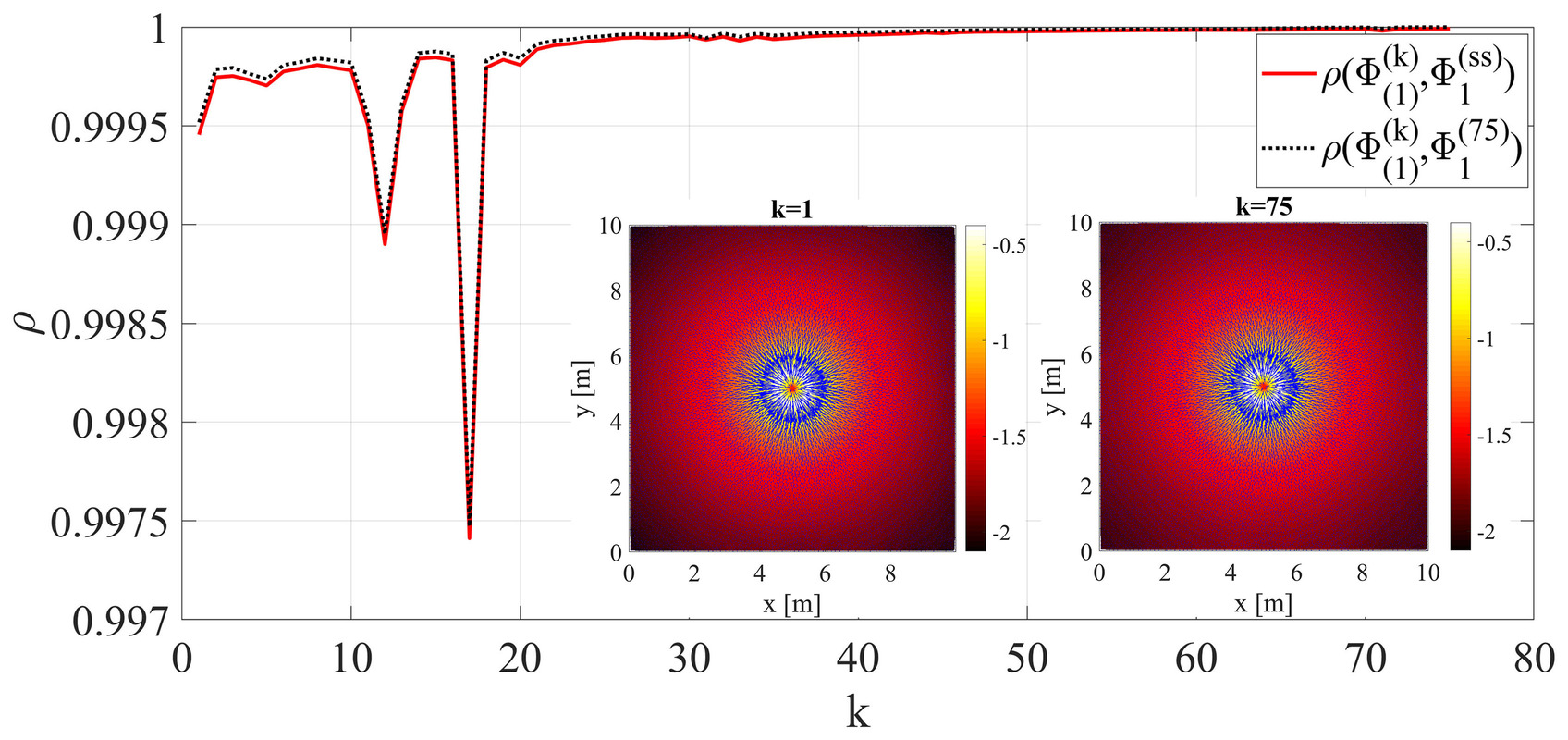}
    \caption{\small{Correlation coefficient ($\rho$) of $\Phi_1^{(75)}$ with its predecessors (black dotted curve). $\rho$ between $\Phi_1^{(ss)}$ and predecessors of $\Phi_1^{(75)}$ (red curve). Inset: $\Phi_1^{(75)}$ and its predecessor $\Phi_1^{(1)}$ at $k=1$.}}
    \label{fig:ball_mode1_conv} 
\end{figure}
\begin{figure}[hbt!]
    \centering
  \subfloat[\label{fig:ball_eig_move_mode1} ]{%
       \includegraphics[width=0.48\linewidth]{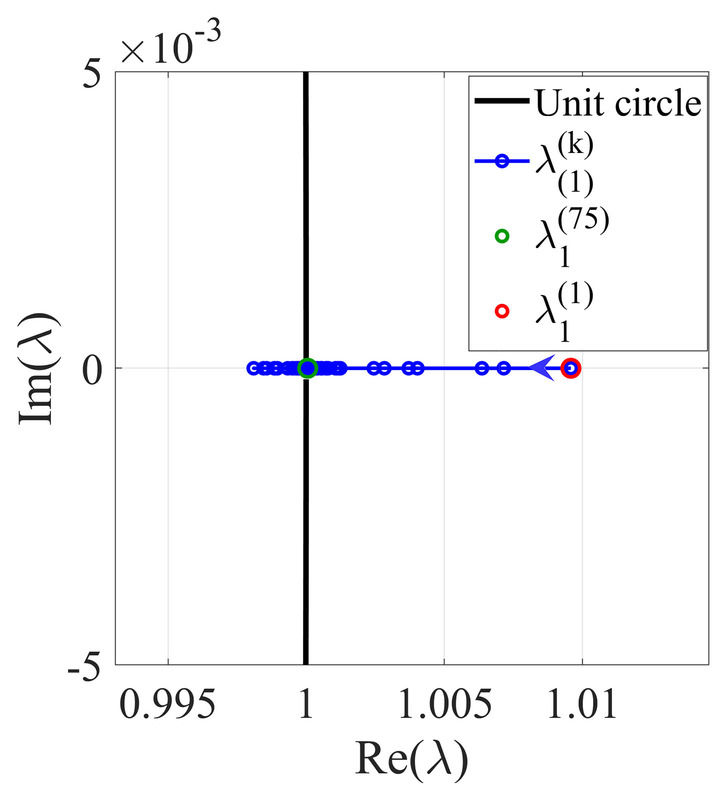}}
  \subfloat[\label{fig:ball_sld_E_err} ]{%
        \includegraphics[width=0.52\linewidth]{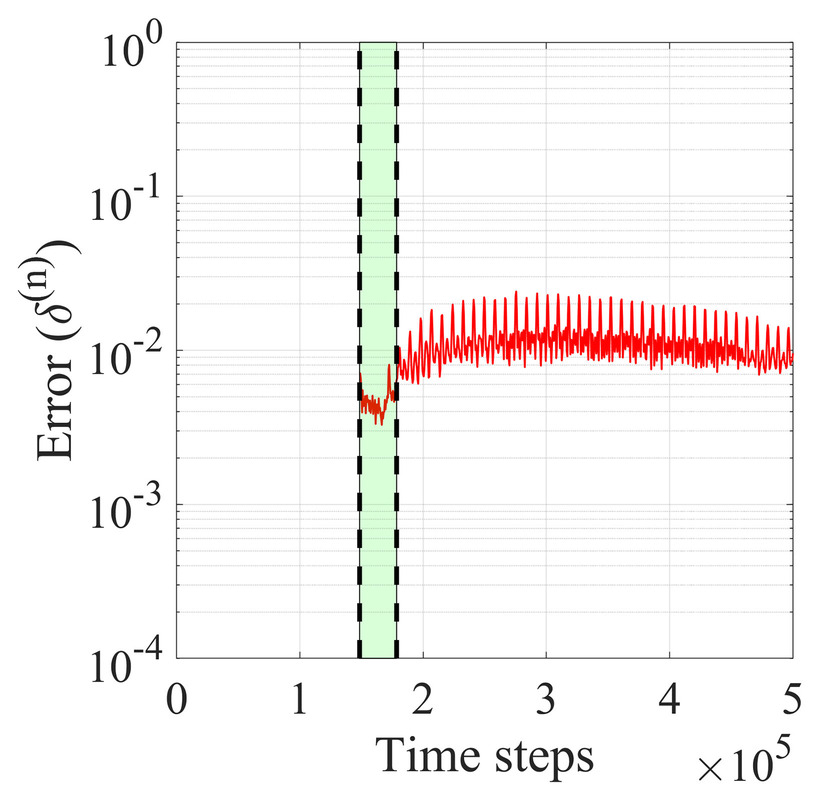}}
    \hfill
  \caption{\small{ (a) Movement of predecessors of the eigenvalue corresponding to $(\lambda_1^{(75)},\Phi_1^{(75)})$. (b) 2-norm relative error in self electric field reconstruction. The green shaded area denotes the DMD window corresponding to $k=75$.} }\label{fig:ball_eig_move_err}
\end{figure}
\paragraph{Convergence in DMD Mode Shape}
Algorithm 1 helps track the evolution of DMD mode shapes through the parameter $\rho$. We correlate the steady-state mode $\Phi_1^{(ss)}$ (from sec. \ref{ball_in_eq}) and dominant DMD mode in the last window $\Phi_1^{(75)}$ with its predecessors $\Phi_{(1)}^{(k)}$ ($k=1,2, \ldots, 75$, where $\Phi_{(1)}^{(75)}=\Phi_{1}^{(75)}$) . As can be seen in Fig. \ref{fig:ball_mode1_conv}, the high value of $\rho$ indicates that the dominant mode shape remains almost time invariant. Close proximity of red and black curves (Fig. \ref{fig:ball_mode1_conv}) further confirms that the equilibrium is attained at $k=75$ as $\Phi_1^{(ss)}$ and $\Phi_1^{(75)}$ are almost identical. Fig. \ref{fig:ball_eig_move_mode1} shows convergent movement of the dominant eigenvalue towards the unit circle. 
\paragraph{Prediction of Self-Fields and Particle Dynamics}
The high-fidelity simulation is stopped after detecting the equilibrium state. The final data harvesting window ($k=75$) is then used for extrapolation (Fig.~\ref{fig:ball_sld_E_err}).  Again, as mentioned earlier, extrapolating the self-fields in this particular example is trivial because of the non-oscillatory steady-state nature of the solution. \par

\begin{figure}[t]
    \centering
	\includegraphics[width=1\linewidth]{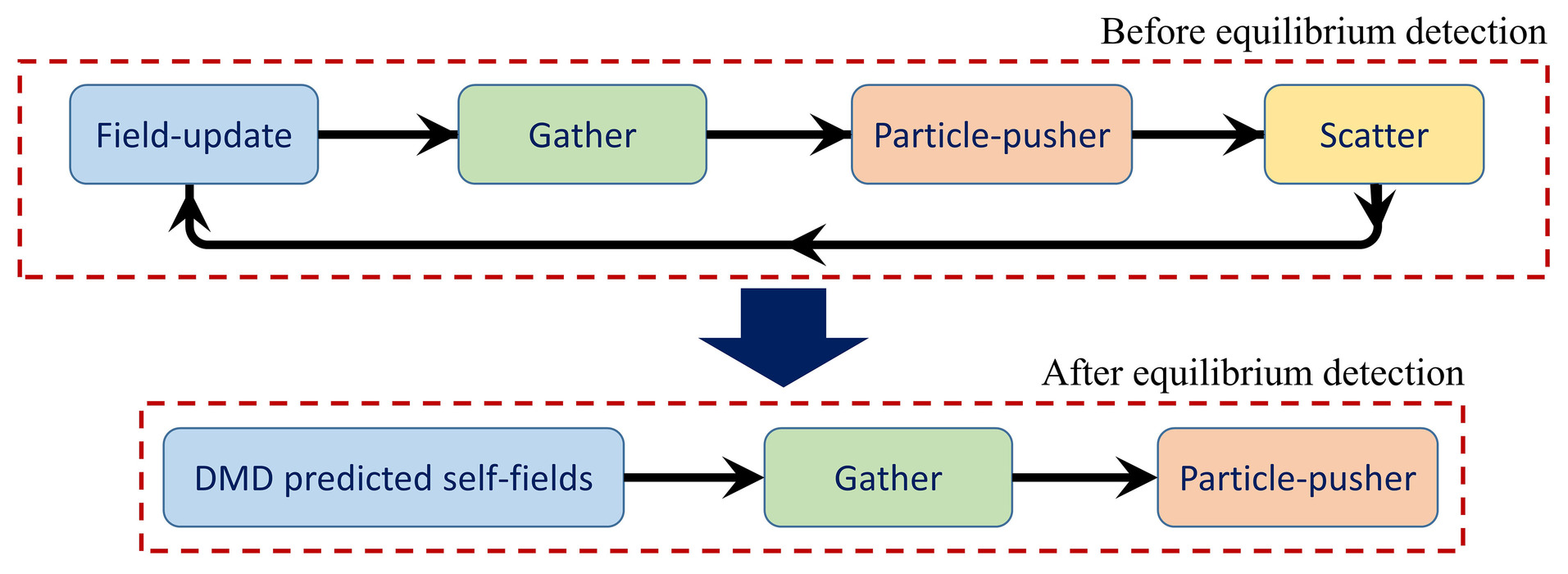}
    \caption{\small{Schematic representation of EMPIC algorithm with DMD predicted self-fields. Prior to detection of equilibrium state, the EMPIC algorithm consists of usual four stages. After the equilibrium is detected, we perform DMD to extrapolate self-field values and utilize those values bypassing field-update stage for future time. To observe the effect of predicted self-fields on particle behavior, we also perform the gather stage and particle pusher stage. }}
    \label{fig:ball_empic_DMD} 
\end{figure}

 As we are interested in how this predicted self electric field affects the (predicted) particle dynamics, we will substitute it in place of the self electric field generated by  original EMPIC algorithm. However, we can entirely bypass the field solver (update) stage of the EMPIC algorithm as illustrated in Fig. \ref{fig:ball_empic_DMD} by performing DMD on the self magnetic {flux} $\mathbf{b}(t)$ as well. In this work we identify the equilibrium performing sliding-window DMD on electric field dataset and extrapolate both the self electric and magnetic field from the last DMD window. However, the DMD extrapolated self-fields do not ensure energy conservation in the extrapolated region. To the extent that the extrapolated fields remain close to the original solution, the energy is approximately conserved in the extrapolation region given that the high-fidelity algorithm itself is energy-conserving. \par

\begin{figure}[hbt!]
    \centering
	\includegraphics[width=1\linewidth]{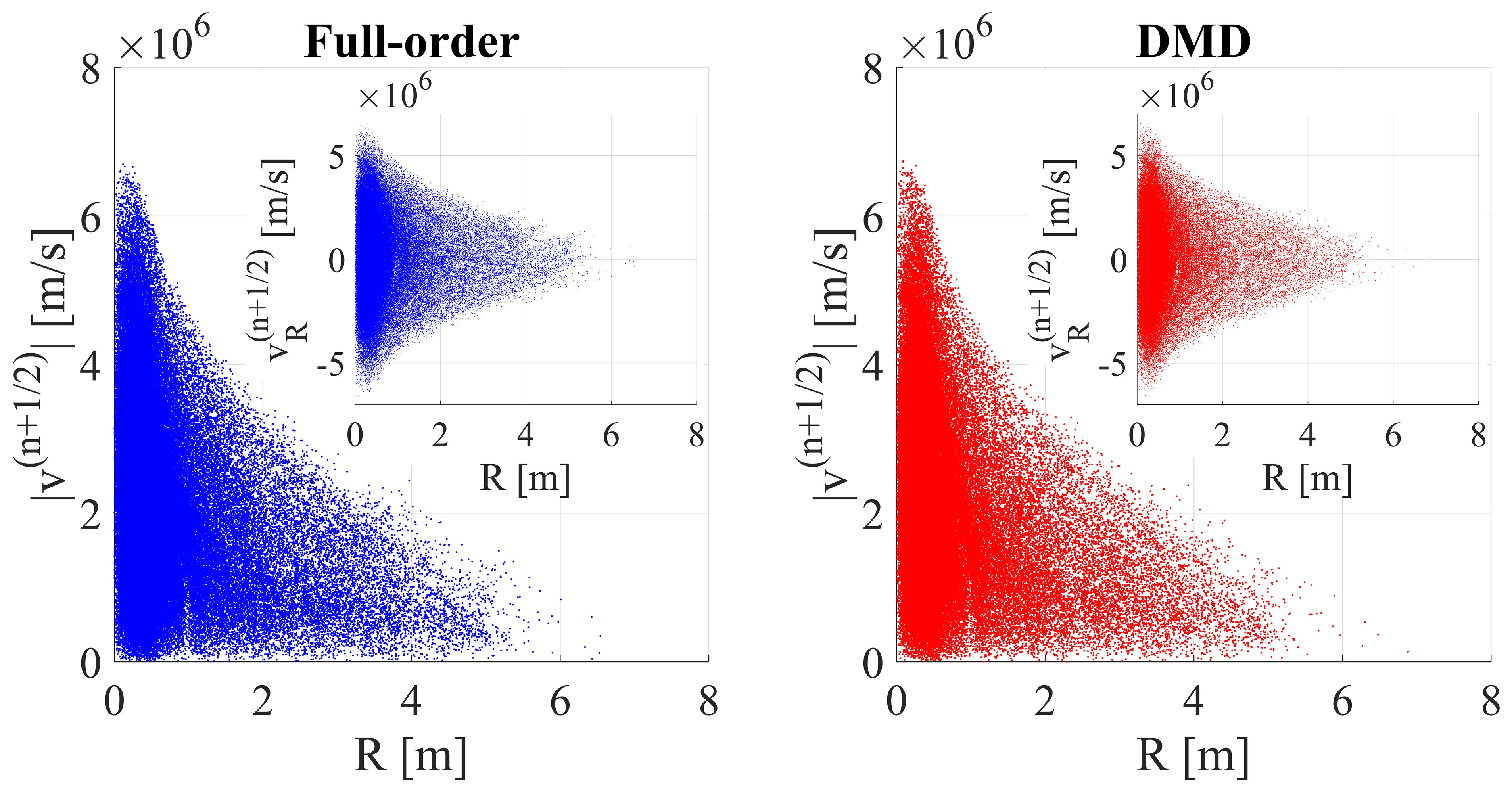}
    \caption{\small{Phase-space plot comparison between finite-element full-order EMPIC simulation (blue) and reduced-order DMD (red) in extrapolation region ($n=225000$). Phase-space plot for absolute velocity and radial distance $(R)$ from center of mesh (5,5). Inset: Phase-space plot corresponding to radial velocity and radial distance.}}
    \label{fig:ball_phase_space_450} 
\end{figure}
\begin{figure} [hbt!]
    \centering
  \subfloat[\label{fig:ball_av_rad_vel} ]{%
       \includegraphics[width=0.49\linewidth]{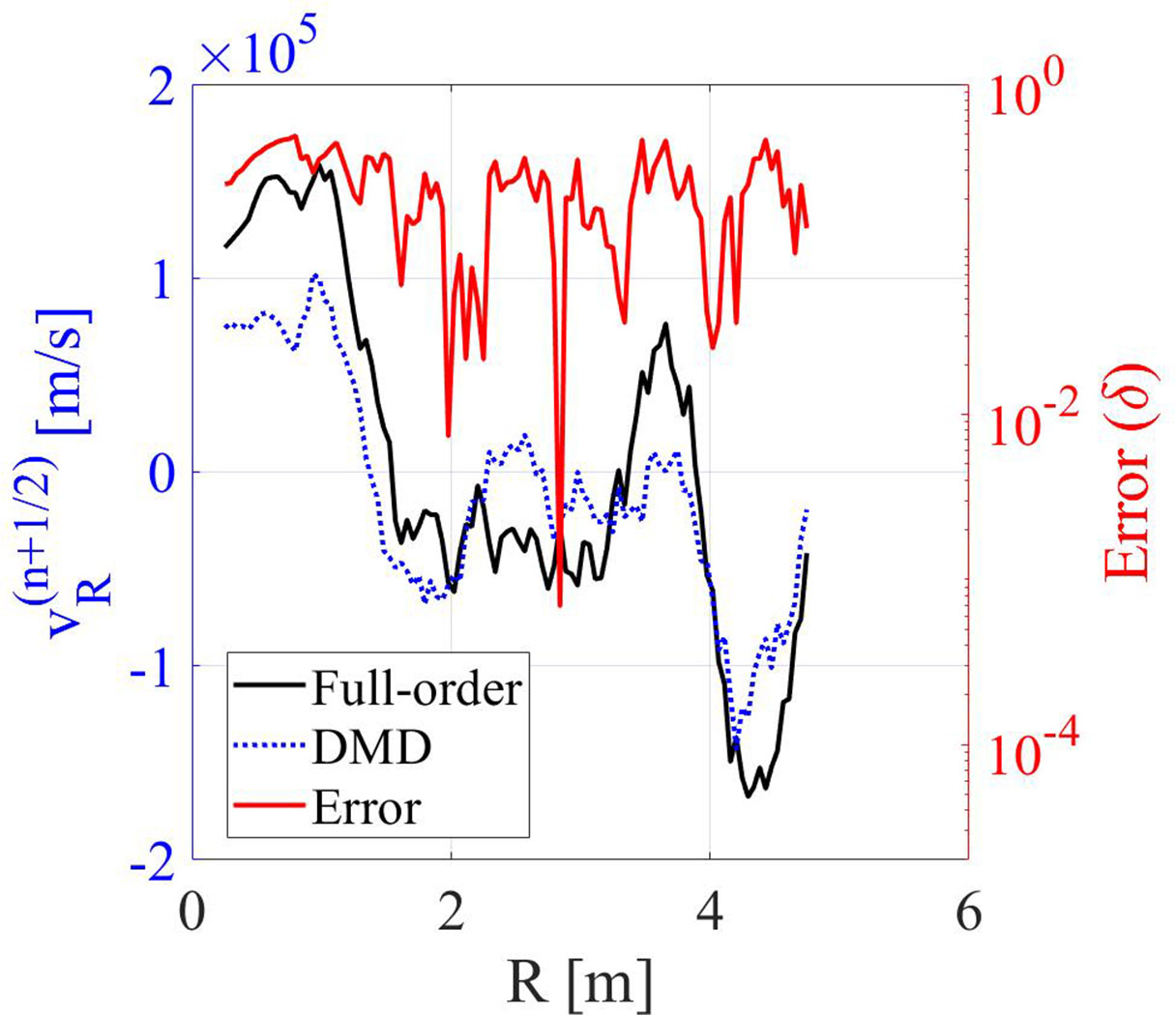}}
  \subfloat[\label{fig:ball_av_part_den} ]{%
        \includegraphics[width=0.51\linewidth]{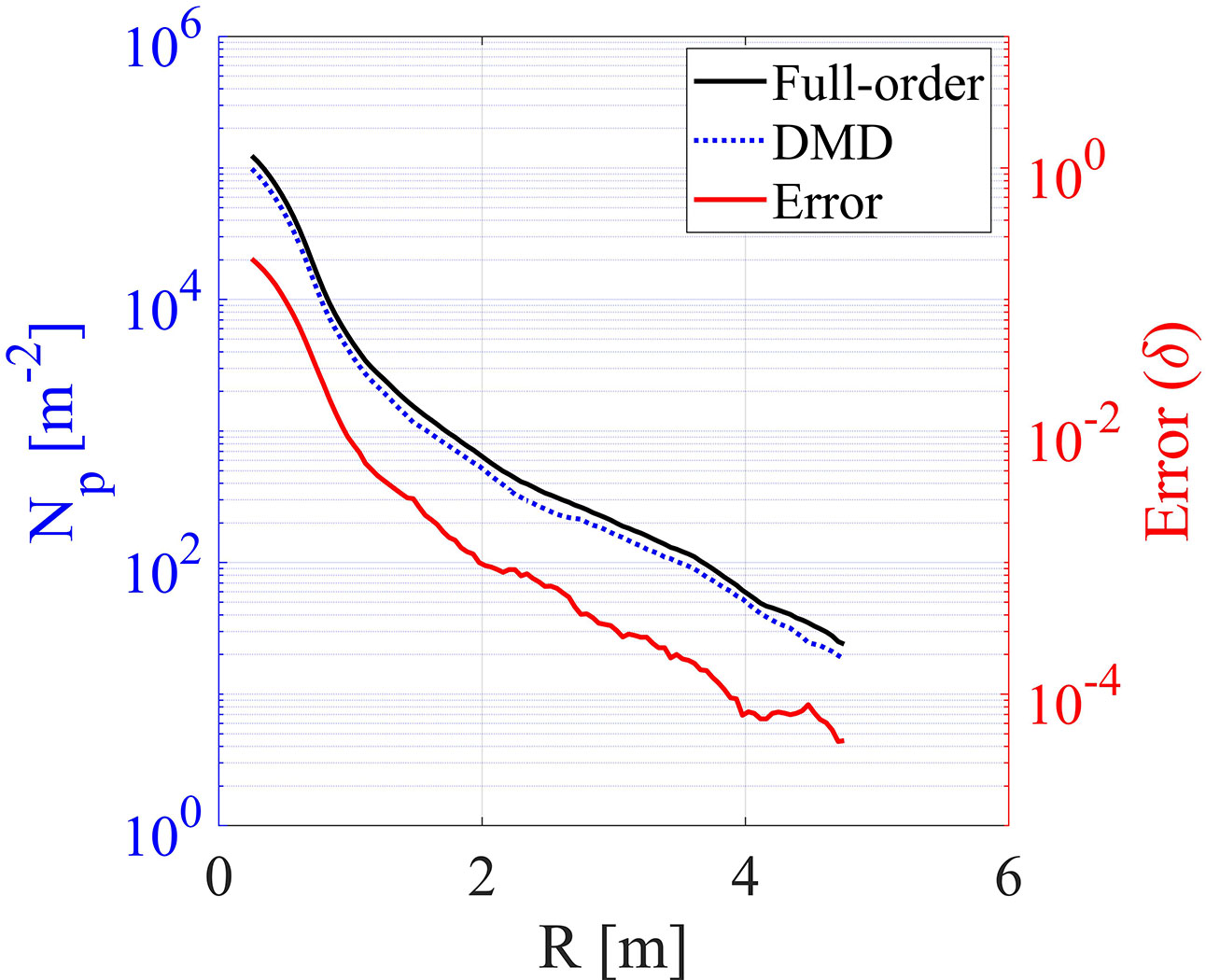}}
    \hfill
  \caption{\small{ Particle dynamics comparison at $n=225000$. (a) Radial variation of average radial velocity of particles. (b) Radial variation of particle density. For both cases, relative error is defined as $\delta=|\hat{\mathcal{X}}(R)-\mathcal{X}(R)|/\max|\mathcal{X}(R)|$, where $\mathcal{X}$ represents either $v_R^{(n+1/2)}$ or $N_p$ and ``hat'' denotes DMD approximation.}  }\label{fig:ball_av_vel_part_den}
\end{figure}

We next compare the particle dynamics generated from the full-order and reduced-order DMD in the extrapolation region at $n=225000$, beyond the final snapshot ($n=178500$) of the last window. Fig.~\ref{fig:ball_phase_space_450} shows a good match between the phase space plots of the full-order and reduced-order models in the radial direction ($R$).  Fig.~\ref{fig:ball_av_vel_part_den} compares the average radial velocity and particle density as a function of radial distance from the center of the plasma ball. For calculating the average radial particle velocity and particle density at $R$, we consider a thin annular region with outer radius $R+L/40$ and inner radius $R-L/40$ and perform the averaging for all the particles present inside that annular region. It is clear that the predicted fields produce good prediction of the particle dynamics and thus have the potential to speed-up EMPIC simulations for long term predictions.

\subsection{Oscillating Electron Beam}\label{wavy_beam}
Consider the case of a 2-D electron beam propagation along the positive $y$ direction in the $xy$ plane, under the influence of an external oscillating transverse magnetic {flux} (Fig. \ref{fig:wavy_beam_ss_snap}). The solution domain is a square cavity of size $1~\text{m}\times 1~\text{m}$ that is discretized via an irregular triangular mesh wit $N_0=1647$ nodes, $N_1=4788$ edges and $N_2=3142$ triangles. Superparticles are injected randomly with uniform distribution at the bottom of the cavity in the region [$0.5-b_h,0.5+b_h$]. Here, $b_h=0.1$ m is the half-beam width. All four sides of the cavity are assumed to be perfect electric conductors (PEC). Superparticles are injected with initial velocity $v_0=5\times 10^6$ m/s along the positive $y$ direction at rate $10$ superparticles {($1~ \text{superparticle}\equiv 2\times 10^5$ electrons)} per timestep ($= 0.01$ ns ). The external voltage bias is set to $V_b=2\times 10^3$ V and external magnetic {flux} to $B_{ext}=B_0~\sin{(2\pi t/T_b)}~\mathbf{\hat{z}}$, where $B_0= 10^{-3}$ T and $T_b=20$ ns. Superparticles are absorbed as they hit the upper boundary. Time series data of degrees of freedom (DoF) of self-fields is stored at every $80^{th}$ timestep $(\Delta_n=80)$. The data set spans $n= 80$ to $n=80000$ (1000 datapoints).

\subsubsection{Self Electric Field Reconstruction}\label{result_stbeam_1}

\begin{figure} [hbt!]
    \centering
  \subfloat[ \label{fig:wavy_beam_ss_snap} ]{%
       \includegraphics[width=0.475\linewidth]{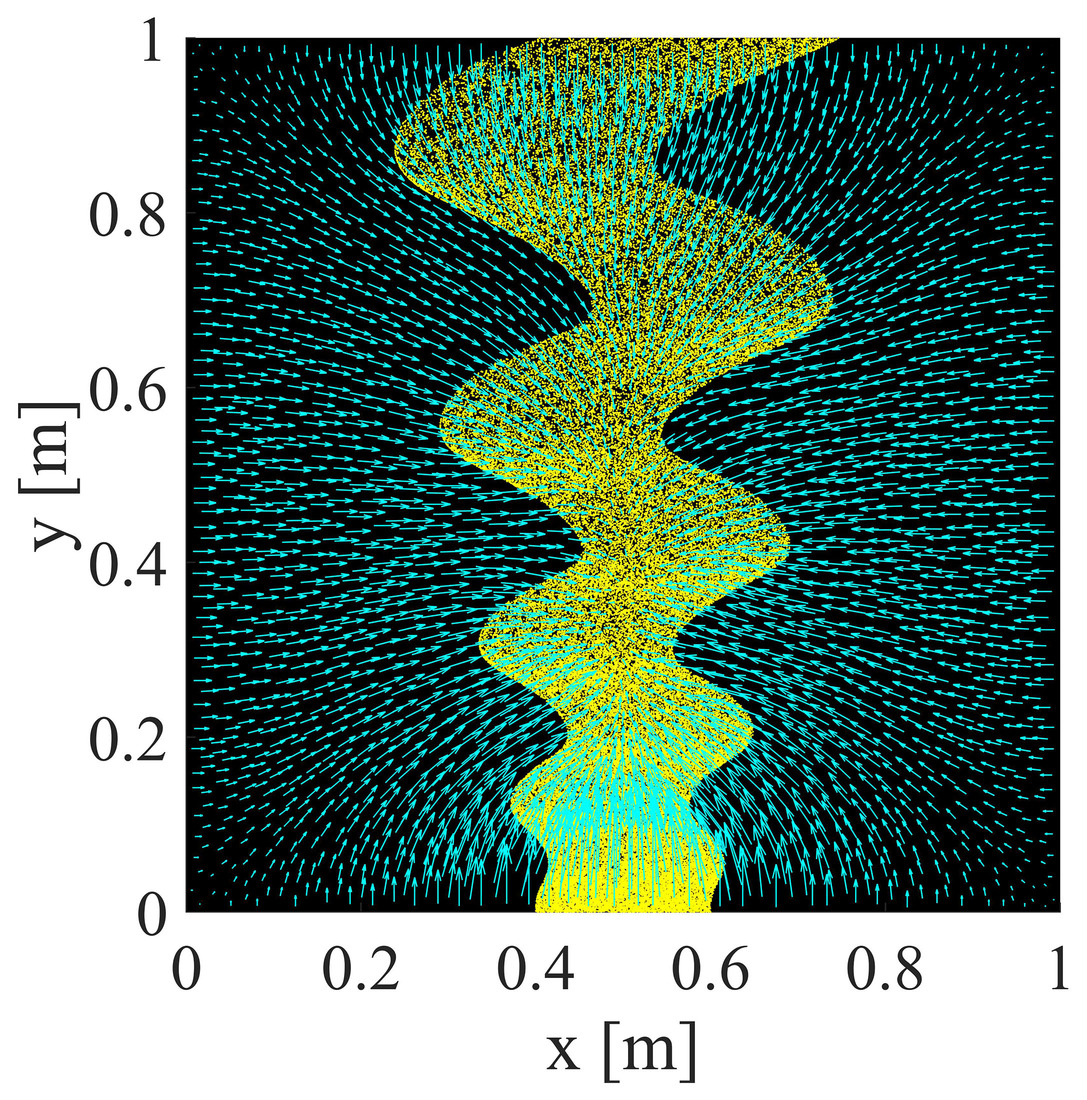}}
  \subfloat[ \label{fig:wavy_beam_ss_sing} ]{%
        \includegraphics[width=0.51\linewidth]{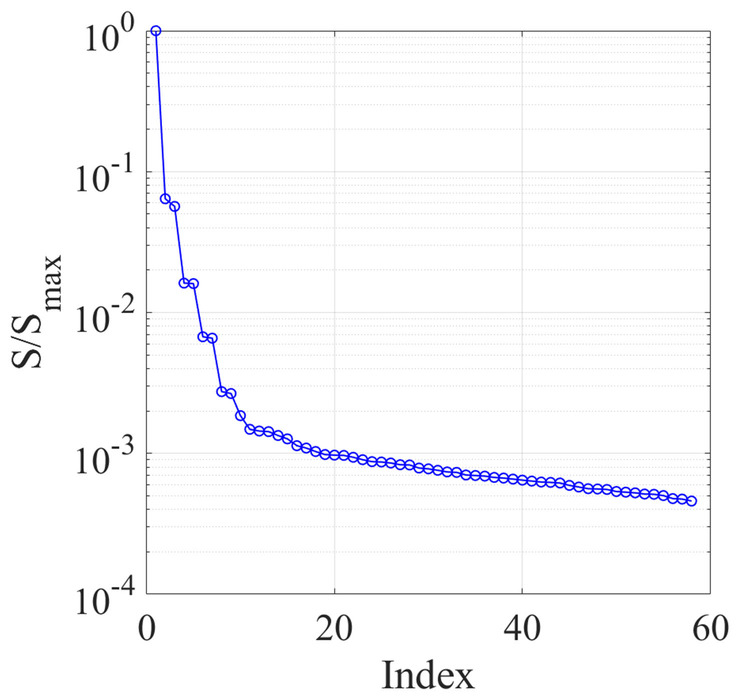}}
  \caption{\small{ (a) Snapshot of an oscillating 2-D electron beam at $n=64000$ in a square cavity, propagating along +ve $y$ direction. The cyan arrows show the self-electric field lines. (b) Normalized singular values from SVD of snapshot matrix in equilibrium state.}  \label{fig:wavy_beam_snap_sing}}
\end{figure}

\begin{figure} [H]
    \centering
  \subfloat[ \label{fig:wavy_ss_eigs} ]{%
        \includegraphics[width=0.31\linewidth]{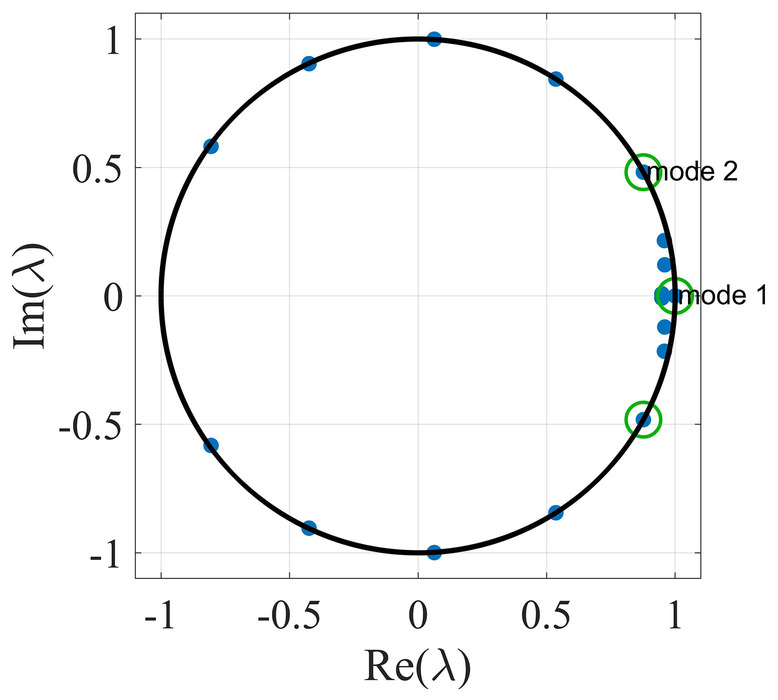}}
  \subfloat[$\Phi_1^{(eq)}$ \label{fig:wavy_ss_mode1} ]{%
       \includegraphics[width=0.34\linewidth]{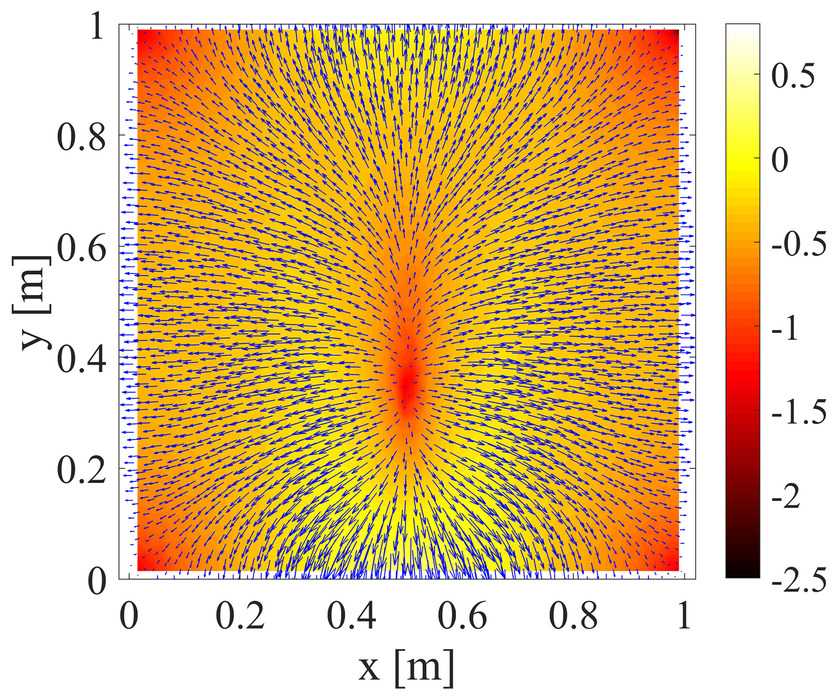}}
  \subfloat[$\Phi_2^{(eq)}$ \label{fig:wavy_ss_mode2} ]{%
        \includegraphics[width=0.34\linewidth]{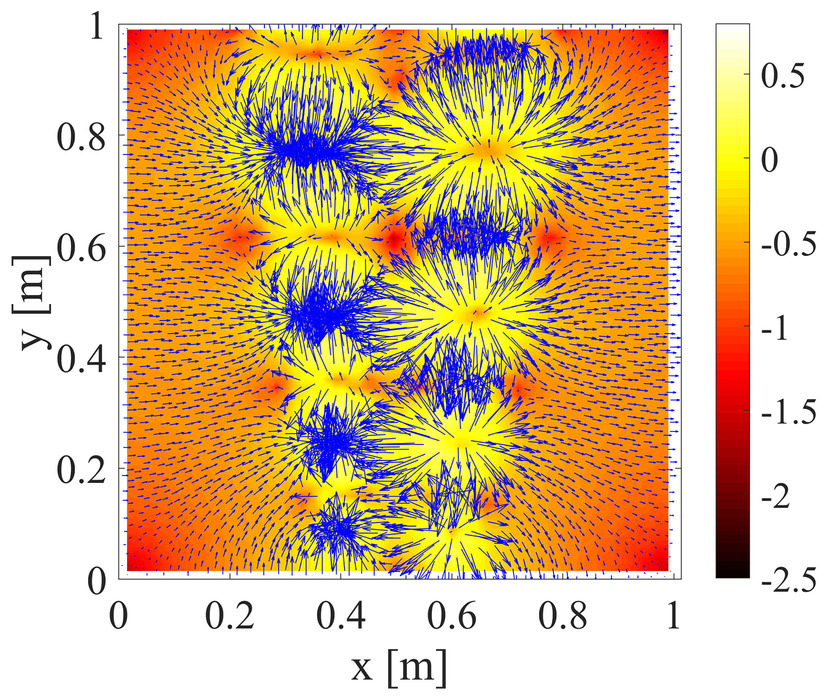}}
  \caption{\small{ (a) DMD eigenvalues in complex plane when DMD is performed on data from equilibrium. (b) First dominant mode. (c) Second dominant mode.}
  \label{fig:wavy_eigs}}
\end{figure}

Transience ends shortly after the beam reaches the upper boundary of the domain. The DMD window in equilibrium spans form $n=40080$ to $n=49600$, with consecutive samples $\Delta t=1.6$ ns apart. As seen in Fig.~\ref{fig:wavy_beam_ss_sing}, energy is primarily concentrated in the first few ($\sim 10$) modes, revealing existence of underlying low-dimensional coherent features. We truncate the SVD matrices at $r=17$, generating $9$ DMD modes, resulting in a reduced-order model with only $9$ degrees of freedom compared to $4788$ in the full-order finite element model. Fig.~\ref{fig:wavy_ss_eigs} with dominant eigenvalues marked by green circle indicates that DMD is able to successfully extract the stationary component $\Phi_1^{(eq)}$ and the oscillating component $\Phi_2^{(eq)}$ from the equilibrium state, with the oscillation frequency matching the frequency of oscillation of external magnetic flux. In equilibrium, these two modes contain more than $99\%$ of the energy. 

\begin{figure} [H]
\centering
  \subfloat[\label{fig:wavy_trans_self_corr} ]{%
       \includegraphics[width=0.5\linewidth]{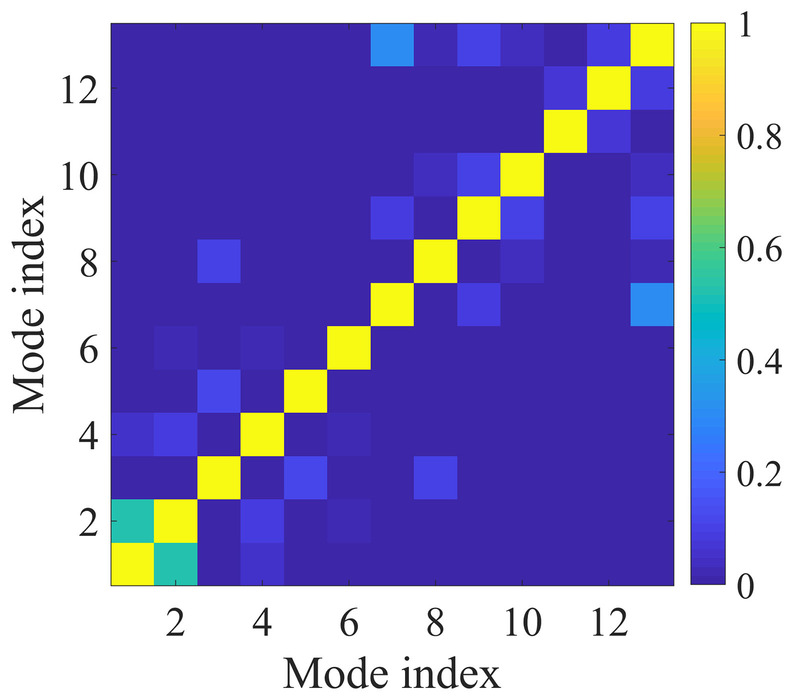}}
  \subfloat[\label{fig:wavy_ss_self_corr} ]{%
        \includegraphics[width=0.49\linewidth]{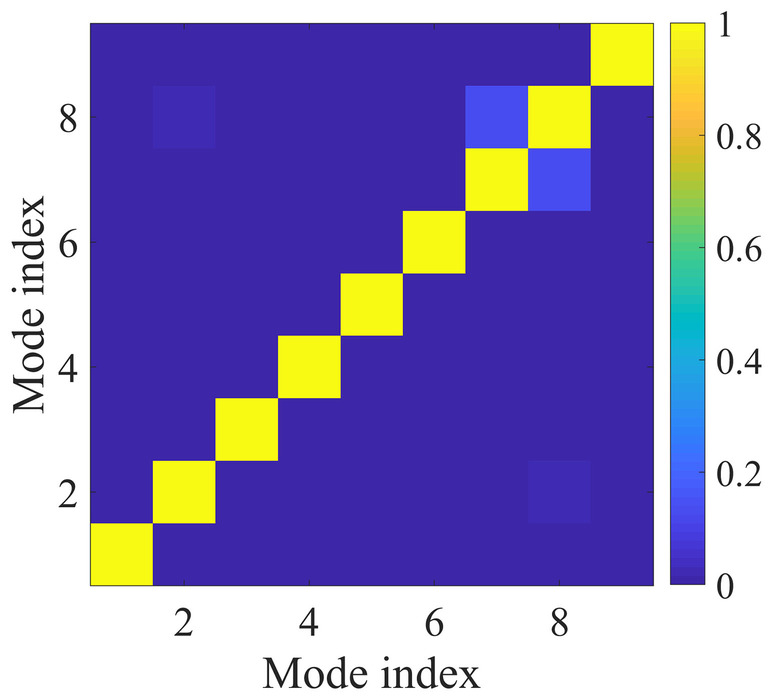}}
  \caption{\small{ (a) Coefficient $\rho$ between DMD modes from transient region. (b) Coefficient $\rho$ between DMD modes from equilibrium region.}
  \label{fig:wavy_self_corr}}
\end{figure}

\begin{figure} [H]
\centering
  \subfloat[\label{fig:wavy_trans_err} ]{%
       \includegraphics[width=0.5\linewidth]{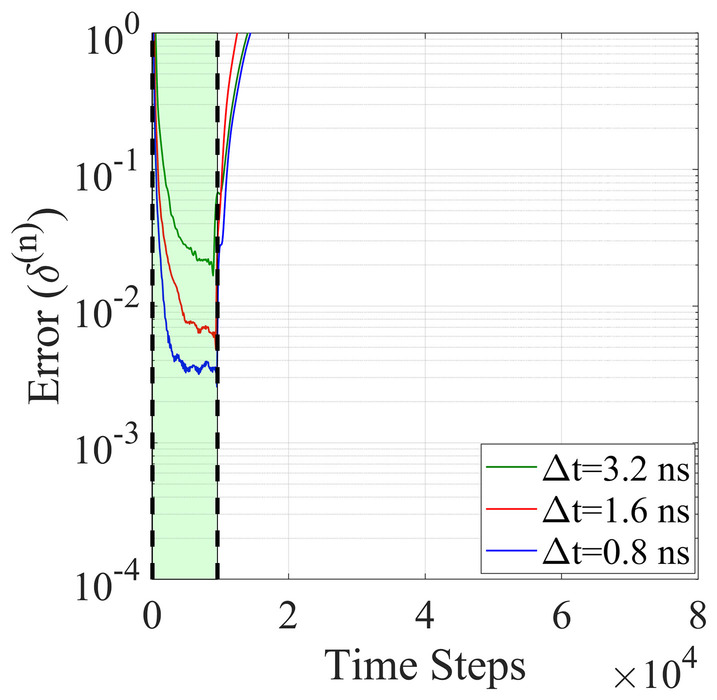}}
  \subfloat[\label{fig:wavy_ss_err} ]{%
        \includegraphics[width=0.5\linewidth]{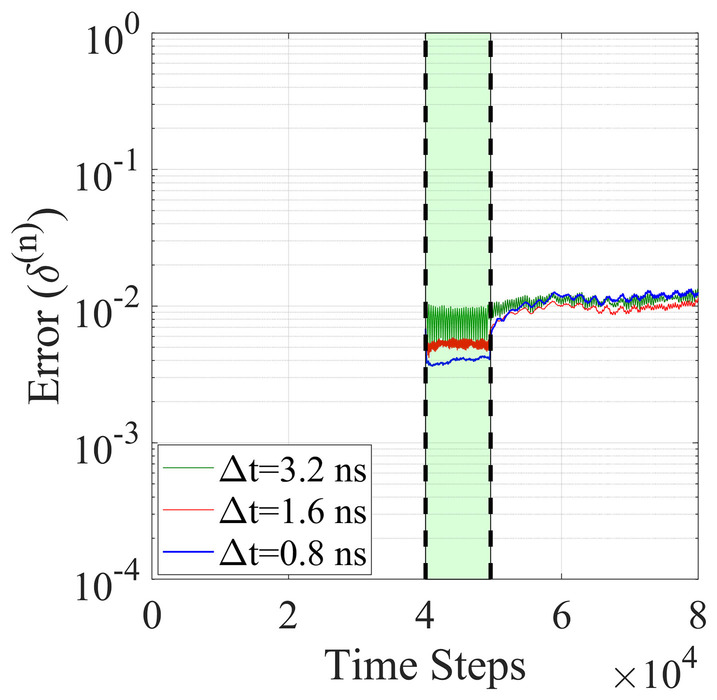}}
  \caption{\small{ Relative 2-norm error for reconstruction of self electric field, with green shaded area denoting the DMD window. (a) DMD window is in transient region. (b) DMD window is equilibrium region.}
  \label{fig:wavy_err}}
\end{figure}

As for the plasma ball example, we perform DMD during transience as well. This DMD window spans from $n=80$ to $n=9600$ with $\Delta t=1.6$ ns, $r=25$ and $13$ DMD modes. Fig. \ref{fig:wavy_self_corr} reveals a clear distinction in nature of correlation among equilibrium modes versus  correlation among transient modes: the former have greater separation while the latter have more overlap among each other. This phenomenon is similar to the plasma ball case. 
 Similar to the plasma ball case, the self-field reconstruction error stays within reasonable limits inside the interpolation region, but rapidly increases in the extrapolation region for transient DMD (Fig.  \ref{fig:wavy_trans_err}). However, for DMD in the equilibrium region (Fig. \ref{fig:wavy_ss_err}), the extrapolation error remains within acceptable bounds.
\begin{figure} [H]
\centering
  \subfloat[$\Phi_3^{(eq)}$ \label{fig:beam_trans_mode3} ]{%
       \includegraphics[width=0.33\linewidth]{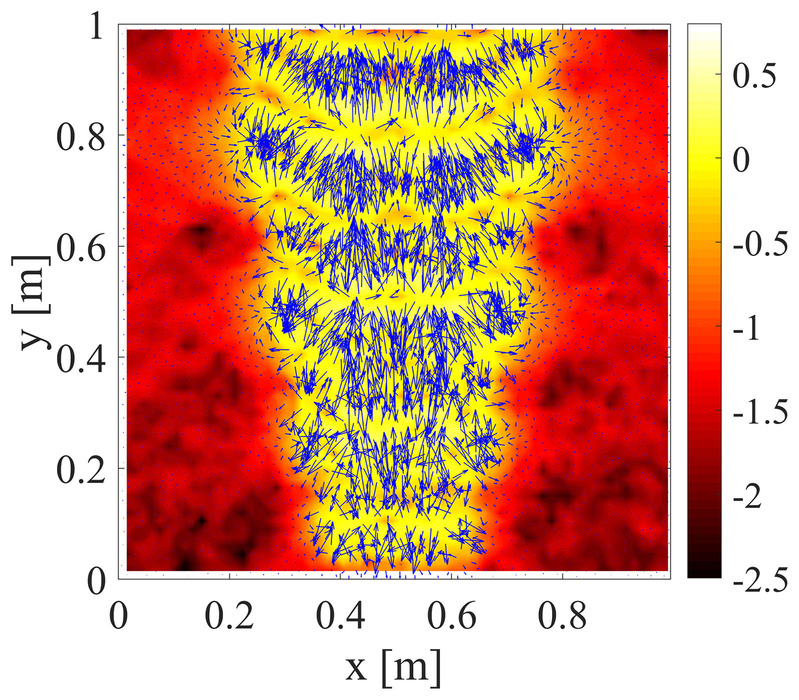}}
  \subfloat[$\Phi_4^{(eq)}$ \label{fig:beam_trans_mode4} ]{%
        \includegraphics[width=0.335\linewidth]{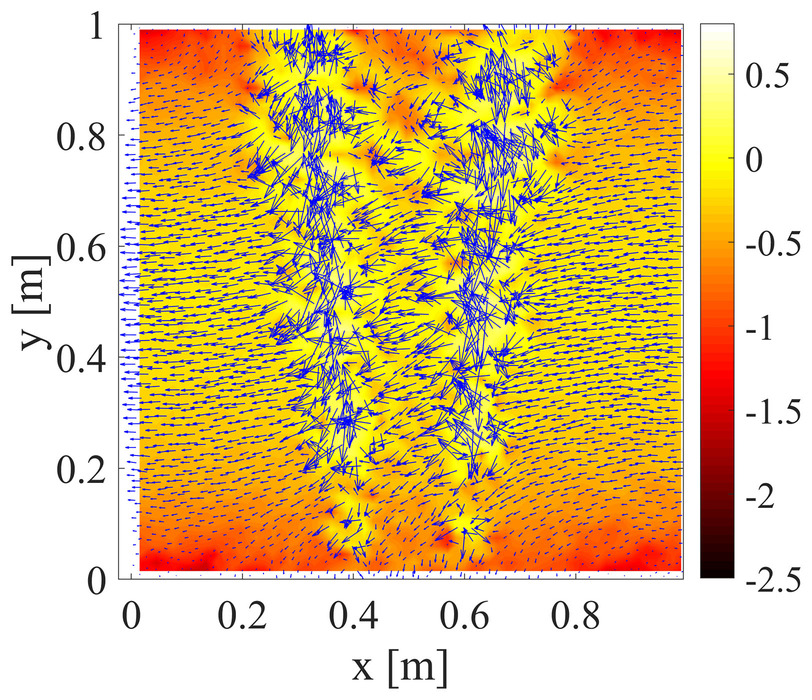}}
  \subfloat[$\Phi_5^{(eq)}$ \label{fig:beam_trans_mode5} ]{%
        \includegraphics[width=0.33\linewidth]{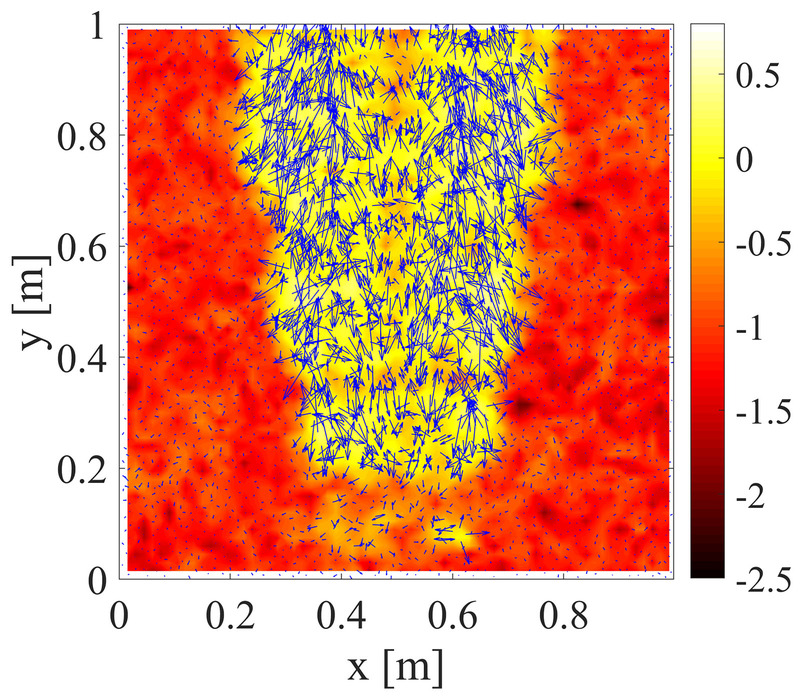}}\\
 \subfloat[$\Phi_6^{(eq)}$ \label{fig:beam_trans_mode6} ]{%
        \includegraphics[width=0.335\linewidth]{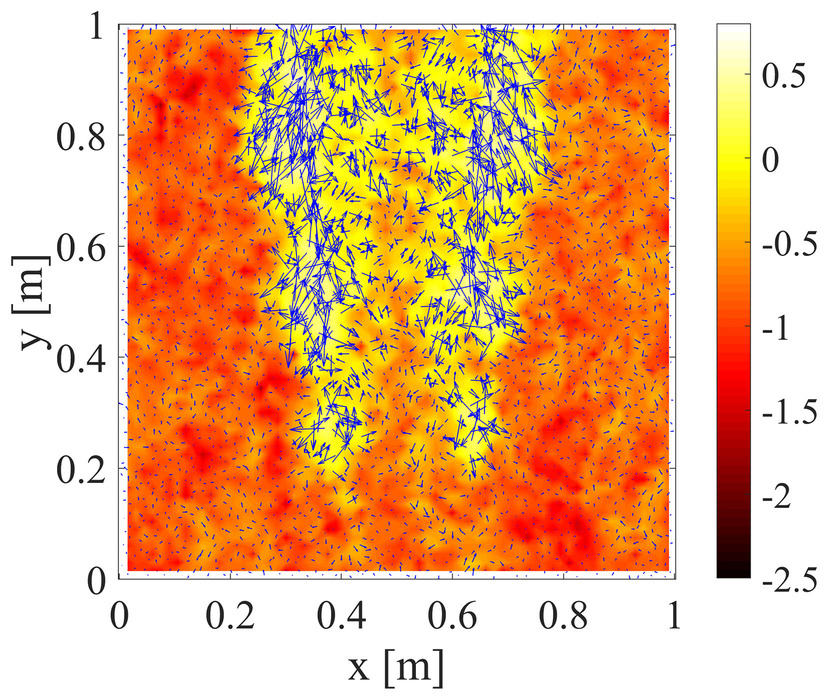}}
\subfloat[$\Phi_{7}^{(eq)}$ \label{fig:beam_trans_mode7} ]{%
        \includegraphics[width=0.335\linewidth]{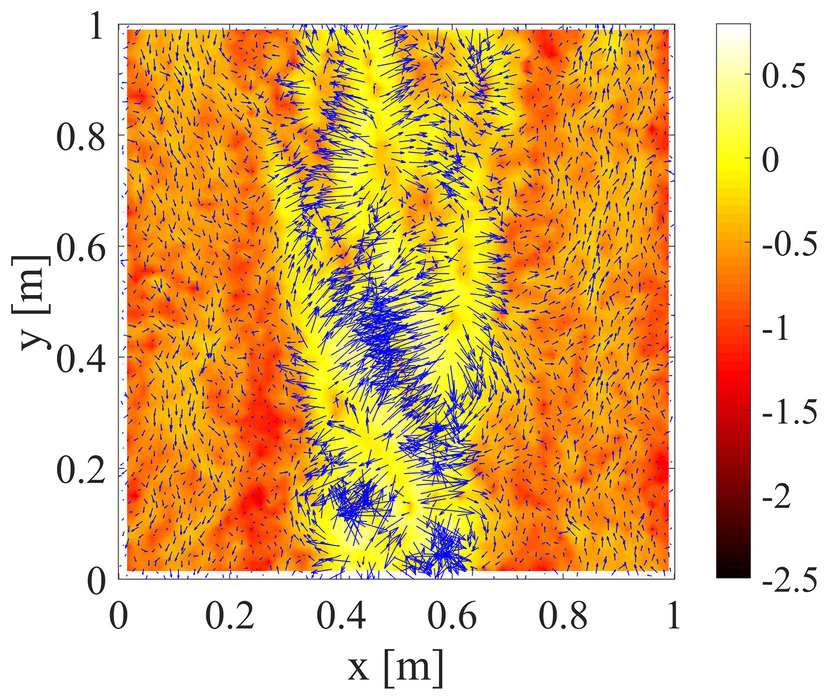}}
\subfloat[$\Phi_{8}^{(eq)}$ \label{fig:beam_trans_mode8} ]{%
        \includegraphics[width=0.33\linewidth]{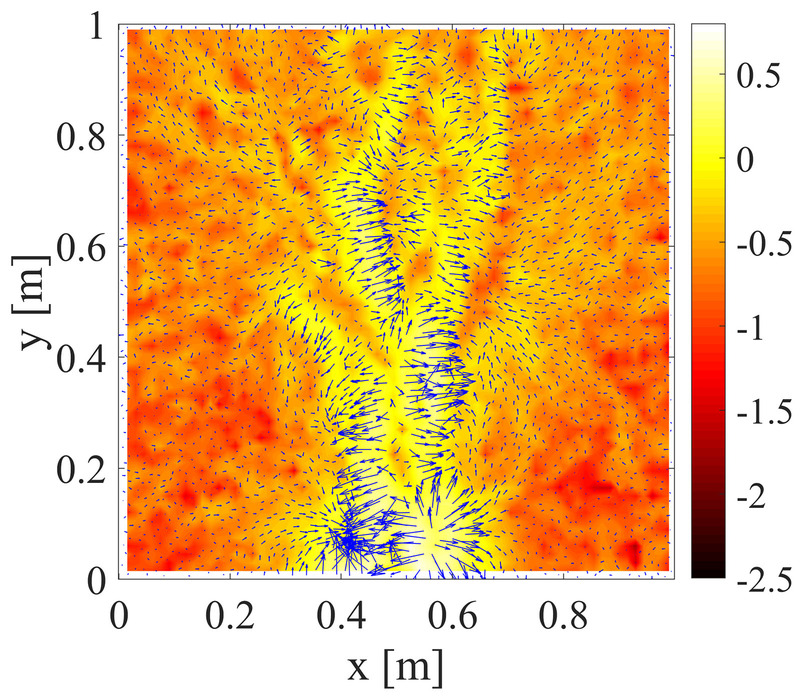}}
  \caption{\small{ First six recessive DMD modes extracted from equilibrium region of oscillating electron beam.}
  \label{fig:wavy_ss_rec_modes}}
\end{figure}

\subsubsection{Sliding-Window DMD}\label{result_stbeam_2}
We set $\beta_{thr}=0.01$ and $\Delta_k=3.2$ ns. Using prior knowledge about the oscillation period of the external magnetic flux ($T_b=20$ ns), we choose $T=56$ ns so that it covers multiple cycles of the forced oscillation. The resulting interval between successive snapshots is $\Delta t=1.6$ ns. 

\begin{figure} [hbt!]
\centering
  \subfloat[\label{fig:wavy_alpha_comp1} ]{%
       \includegraphics[width=0.5\linewidth]{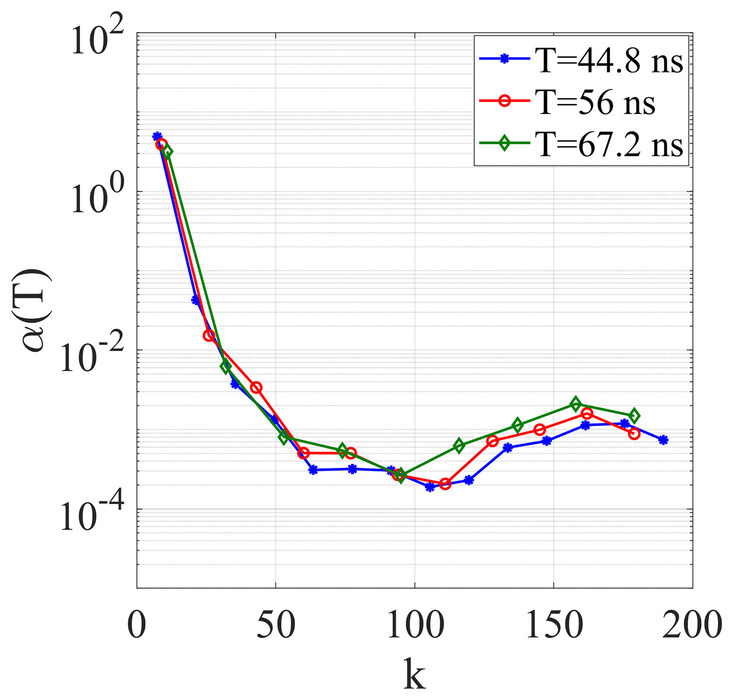}}
  \subfloat[\label{fig:wavy_alpha_comp2} ]{%
        \includegraphics[width=0.5\linewidth]{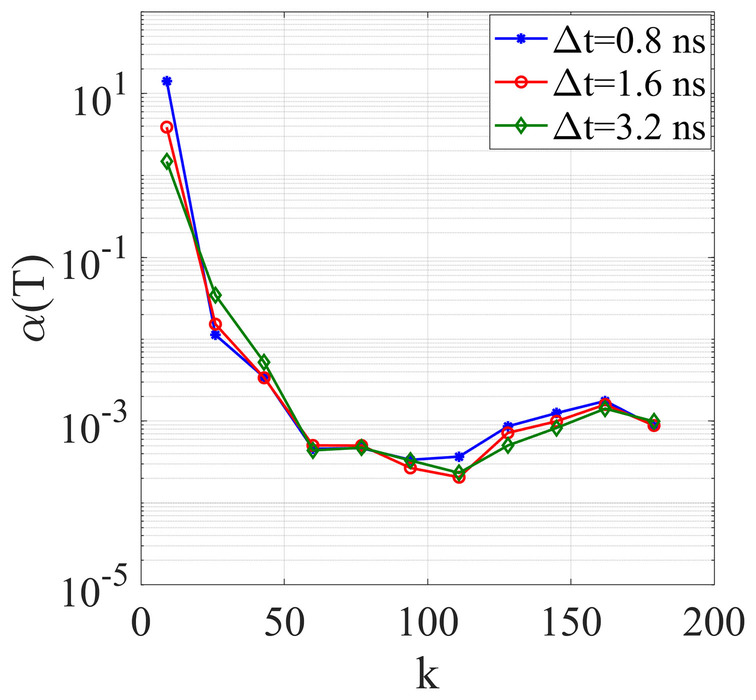}}
  \caption{\small{ (a) Sensitivity of Algorithm 2 towards window width $T~(\pm20\%)$, keeping fixed $\Delta t=1.6$ ns. (b) Sensitivity of Algorithm 2 towards sampling interval, keeping fixed $T=56$ ns.}
  \label{fig:wavy_alpha_comp}}
\end{figure}

\paragraph{Equilibrium Detection}
  The algorithm detects the steady-state at $k=136$ ($n_{st}(136)=43280$). The sensitivity of $\alpha(T)$ towards variation in $T$ and $\Delta t$ is shown in Fig.\ref{fig:wavy_alpha_comp}. 
\begin{figure}[H]
    \centering
	\includegraphics[width=0.99\linewidth]{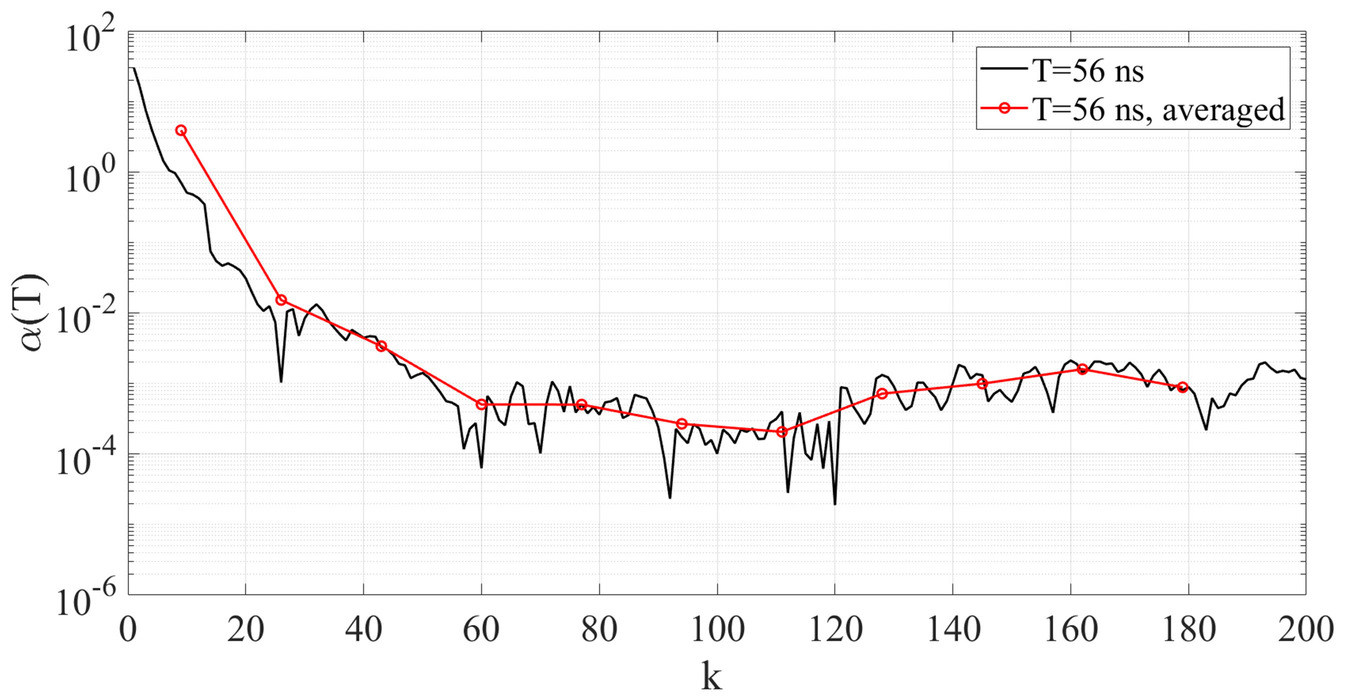}
    \caption{\small{Variation in $\alpha(T)$ for $\Delta t=1.6$ ns, as the DMD window slides towards equilibrium for oscillating electron beam. The red curve shows averaged $\alpha$ over 17 windows.}}
    \label{fig:wavy_alpha} 
\end{figure}
 
\begin{figure} [H]
\centering
  \subfloat[\label{fig:wavy_eig_move_mode1} ]{%
       \includegraphics[width=0.5\linewidth]{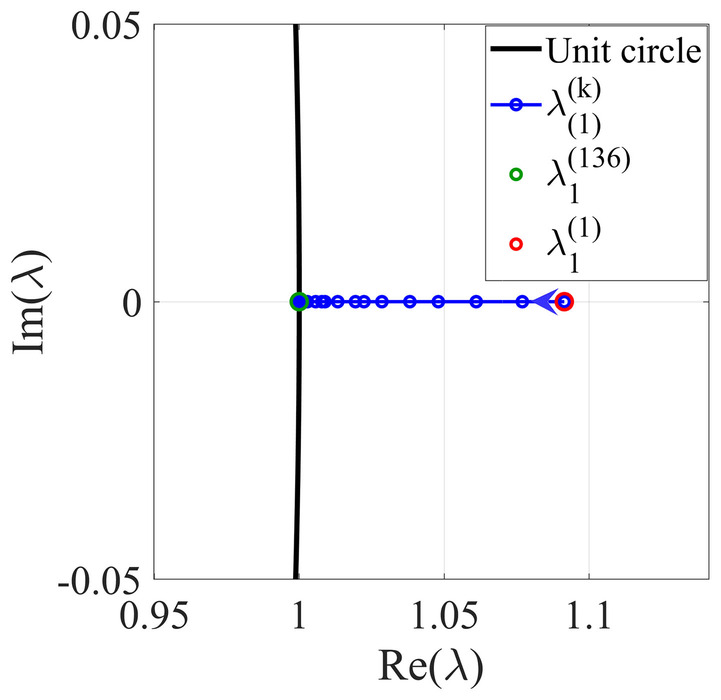}}
  \subfloat[\label{fig:wavy_eig_move_mode2} ]{%
        \includegraphics[width=0.5\linewidth]{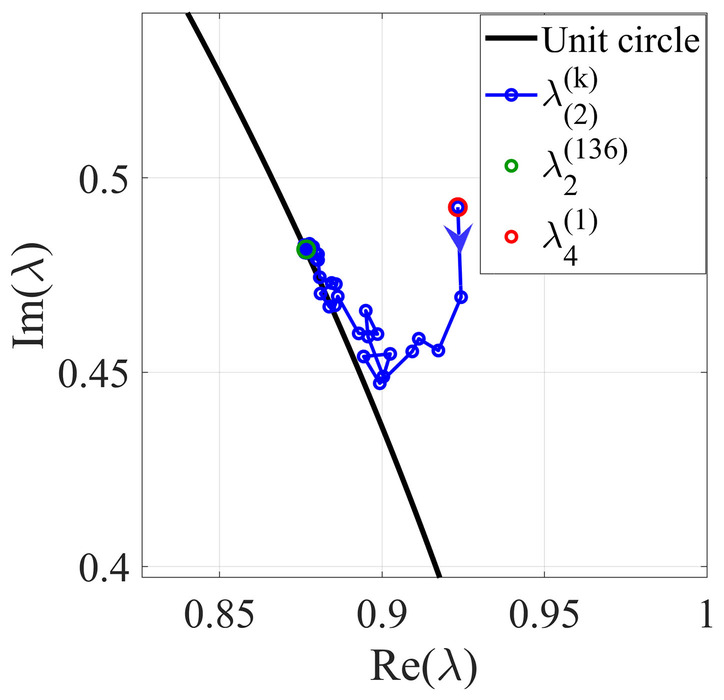}}
  \caption{\small{ (a) DMD eigenvalue movement corresponding to ($\lambda_1^{(136)},\Phi_1^{(136)}$). (b) DMD eigenvalue movement corresponding to ($\lambda_2^{(136)},\Phi_2^{(136)}$).}
  \label{fig:wavy_eig_move}}
\end{figure}

\begin{figure}[H]
    \centering
	\includegraphics[width=1\linewidth]{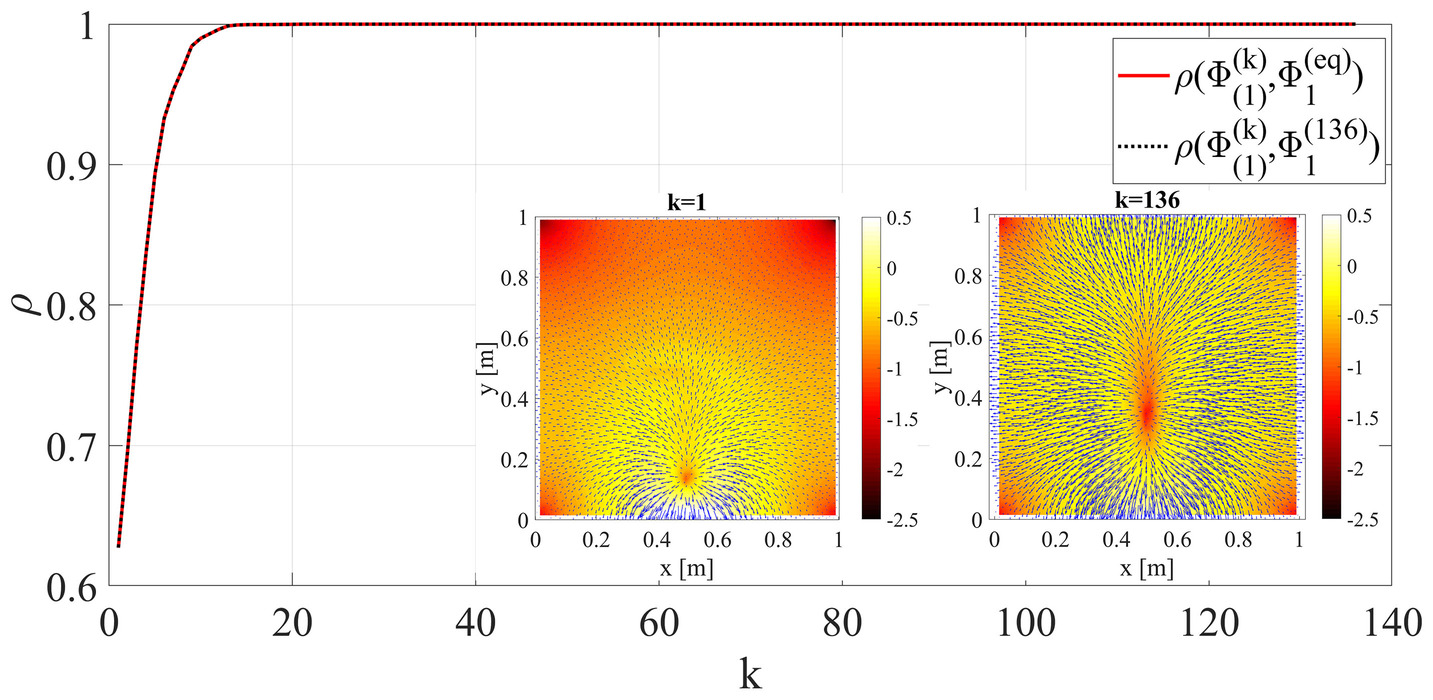}
    \caption{\small{Coefficient $\rho$ of $\Phi_1^{(136)}$ with its predecessors (black dotted curve). $\rho$ between $\Phi_1^{(eq)}$ and predecessors of $\Phi_1^{(136)}$ (red curve). Inset: $\Phi_1^{(136)}$ and its predecessor at $k=1$.}}
    \label{fig:wavy_mode1_conv} 
\end{figure}

\begin{figure}[H]
    \centering
	\includegraphics[width=1\linewidth]{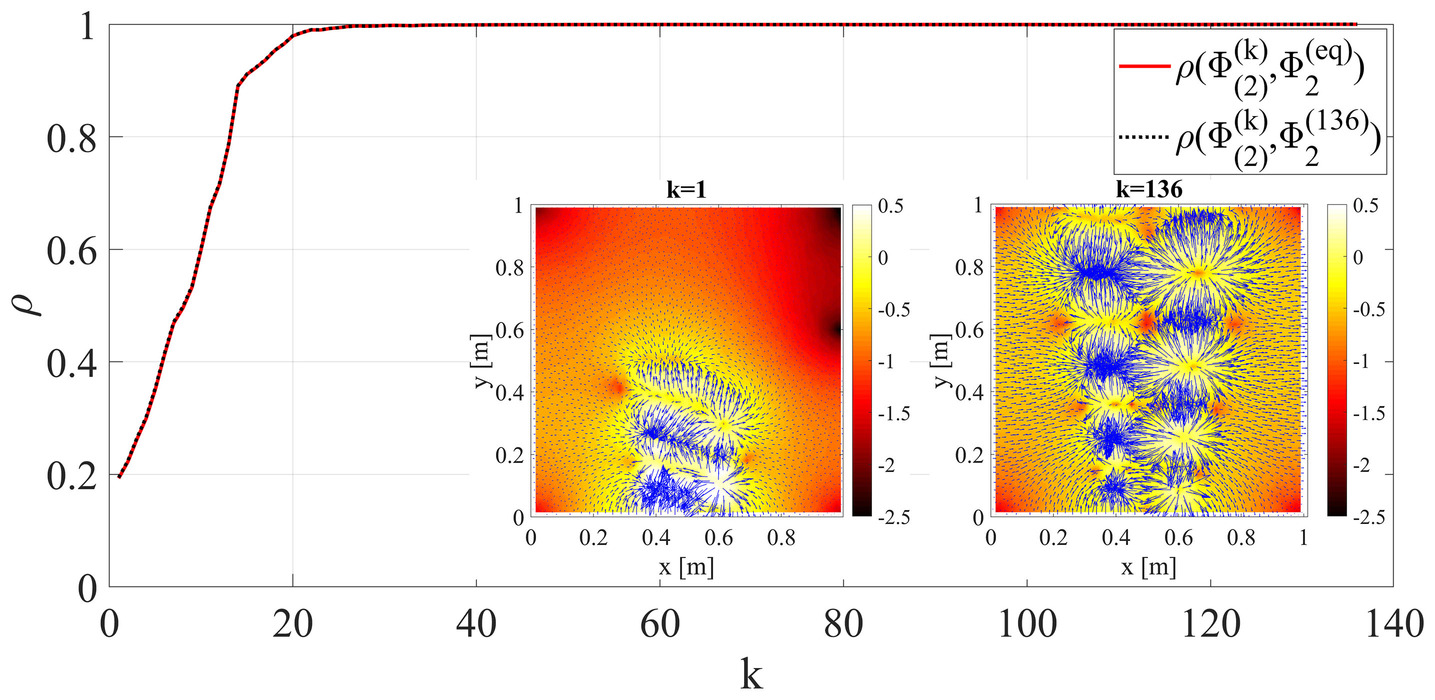}
    \caption{\small{Coefficient $\rho$ of $\Phi_2^{(136)}$ with its predecessors (black dotted curve). $\rho$ between $\Phi_2^{(eq)}$ and predecessors of $\Phi_2^{(136)}$ (red curve). Inset: $\Phi_2^{(136)}$ and its predecessor at $k=1$.}}
    \label{fig:wavy_mode2_conv} 
\end{figure}

\paragraph{Convergence of DMD Modes} There are two dominant DMD modes that describe equilibrium dynamics: the stationary mode and an oscillating mode corresponding to external magnetic flux oscillation frequency. It is of interest to track their
evolution to their final spatial configuration ($\Phi_1^{(136)}$ and $\Phi_2^{(136)}$) in equilibrium. The tracking algorithm reveals that the dominant stationary mode $\Phi_1^{(1)}$ in transient state ($k=1$) eventually evolves to the dominant stationary mode $\Phi_1^{(136)}$  in equilibrium ($k=136$). The inset in Fig. \ref{fig:wavy_mode1_conv} reveals that the mode shape $\Phi_1^{(1)}$ at $k=1$ is nothing but the self-field configuration of the straight beam (stationary component) emitting from the lower boundary of the mesh, whereas that of $\Phi_1^{(136)}$ suggests a full fledged straight electron  beam. Fig. \ref{fig:wavy_eig_move_mode1} shows convergent migration of the  DMD eigenvalue towards the unit circle. 
Similar behavior is observed for the eigenvalue corresponding to the oscillating mode $\Phi_2^{(136)}$ (Fig. \ref{fig:wavy_eig_move_mode2}), which is traced back to $\Phi_4^{(1)}$ in the first window. Interestingly, the fourth most energetic mode at $k=1$ evolves to become the second most energetic mode at $k=136$. Note that during transience, it is harder to separate modes in terms of energy due to complex dynamics and rapidly time varying amplitudes. {Tracking} evolution of the oscillating mode underscores how a relatively hidden feature in transience can become prominent in equilibrium. This gradual evolution in mode shape is captured by the continuous variation of parameter $\rho$ as seen in Figs. \ref{fig:wavy_mode1_conv} and \ref{fig:wavy_mode2_conv}.


\begin{figure}[H]
    \centering
	\includegraphics[width=1\linewidth]{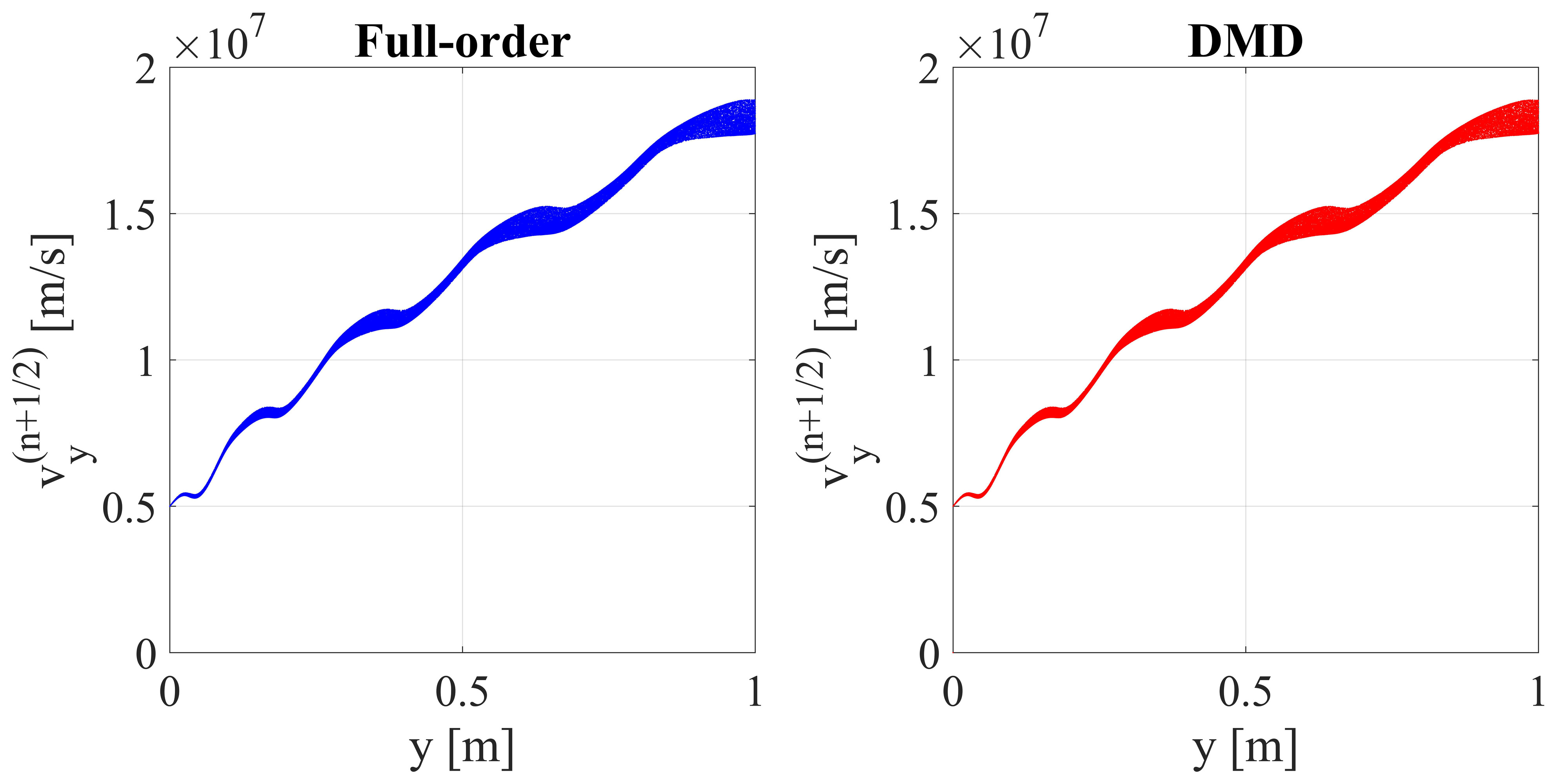}
    \caption{\small{The $y$-directional phase-space plot comparison between finite-element full-order EMPIC simulation (blue) and DMD (red) in extrapolation region ($n=76000$).}}
    \label{fig:wavy_ph_sp_y} 
\end{figure}

\begin{figure}[H]
    \centering
	\includegraphics[width=1\linewidth]{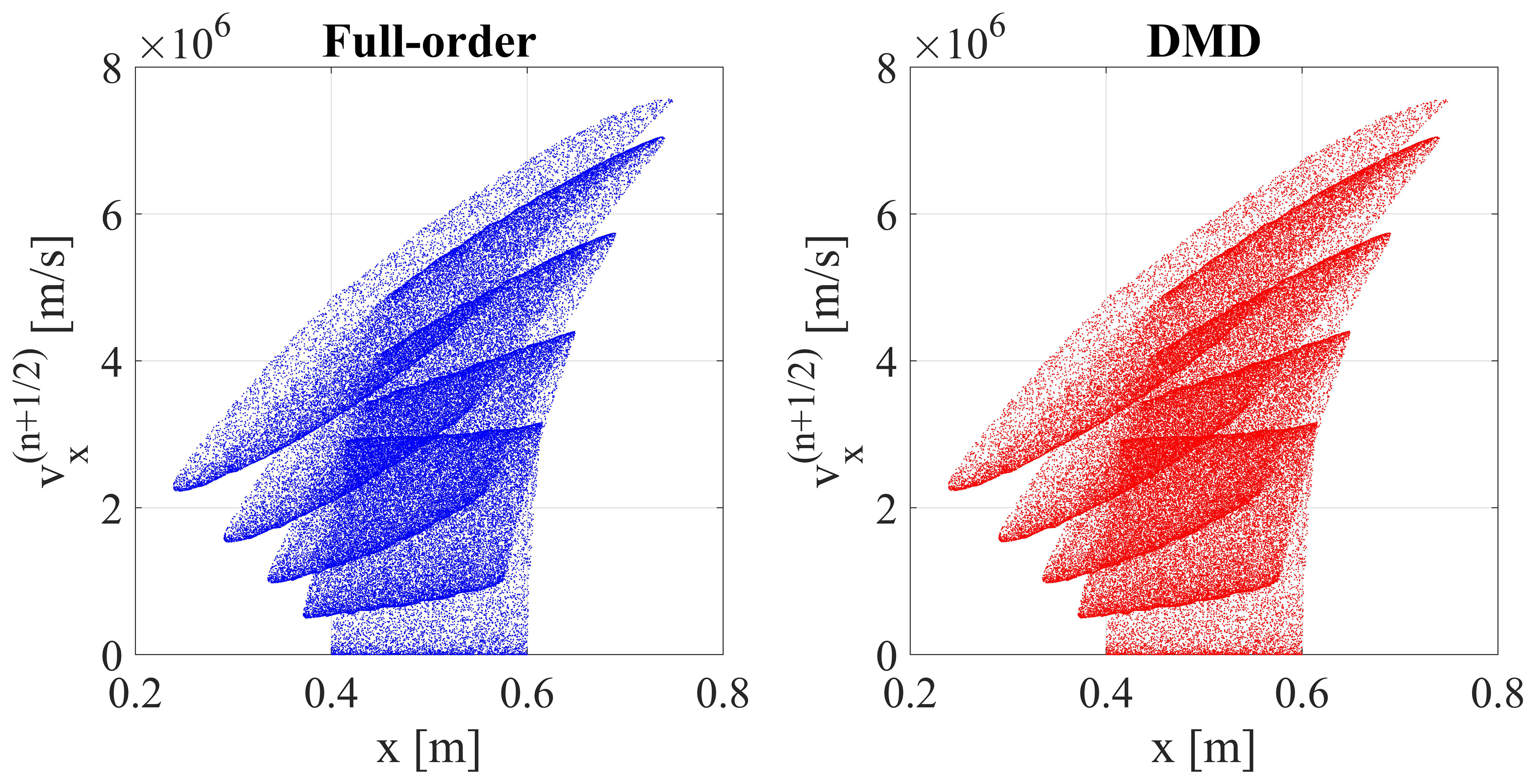}
    \caption{\small{The $x$-directional phase-space plot comparison between finite-element full-order EMPIC simulation (blue) and DMD (red) in extrapolation region ($n=76000$).}}
    \label{fig:wavy_ph_sp_x} 
\end{figure}

\begin{figure} [H]
    \centering
  \subfloat[ \label{fig:wavy_vel_y_err} ]{%
       \includegraphics[width=0.5\linewidth]{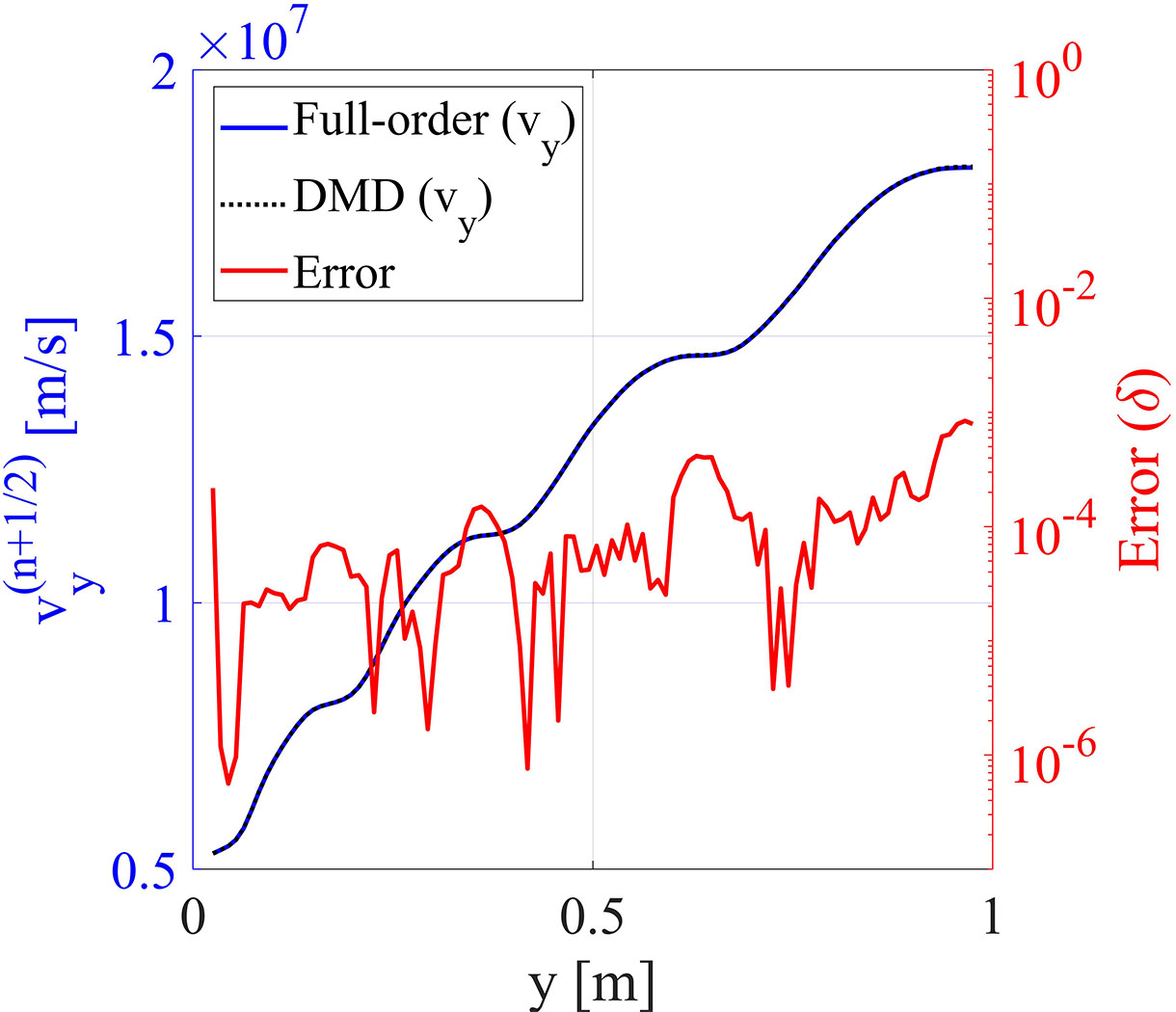}}
  \subfloat[ \label{fig:wavy_Np_y_err} ]{%
        \includegraphics[width=0.5\linewidth]{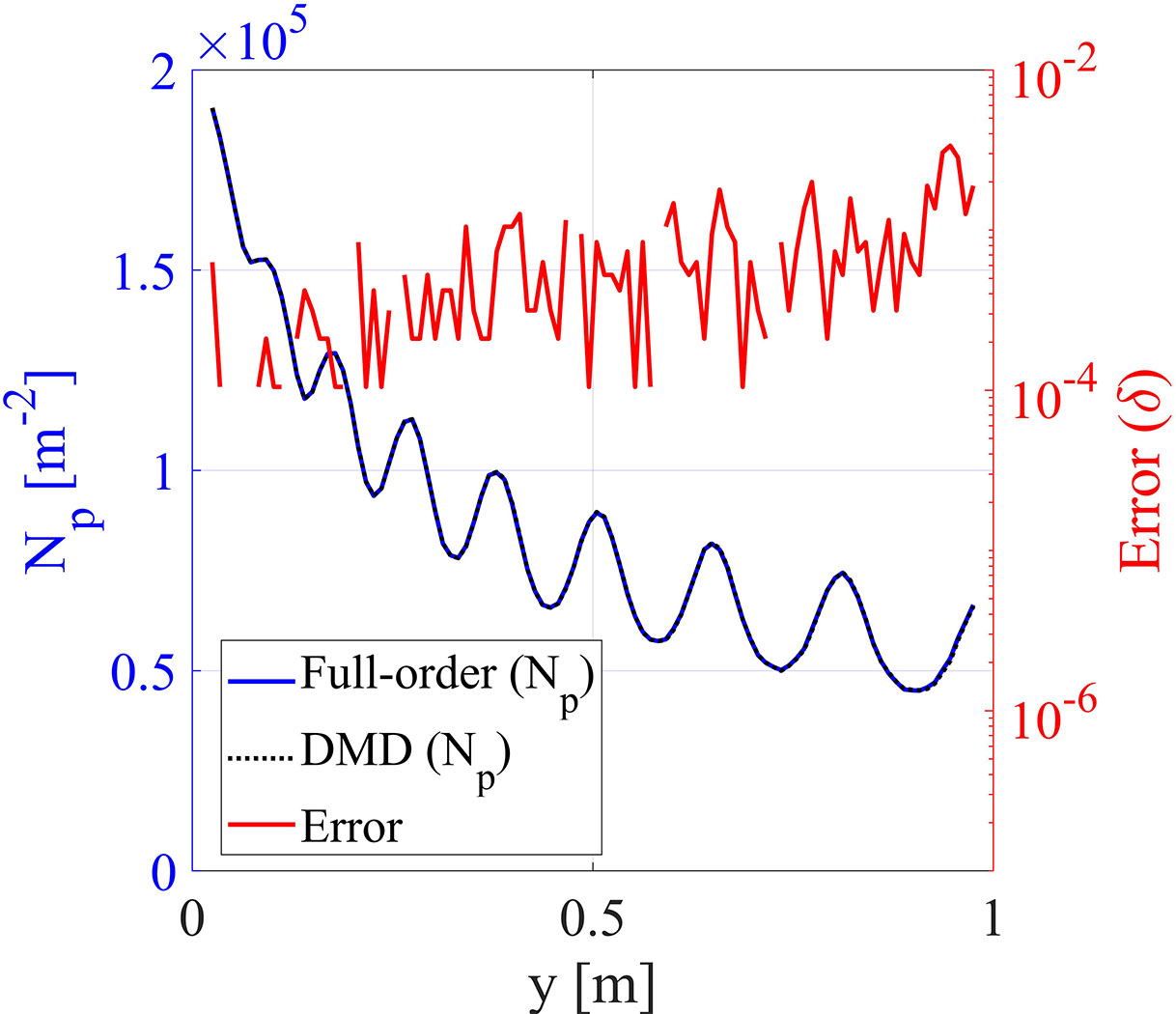}}
  \caption{\small{ Comparison between full-order and DMD predicted average velocity and particle density at $n=76000$ in the $y$-direction. Relative error for $\mathcal{X}(y)$ is defined as $\delta=|\hat{\mathcal{X}}(y)-\mathcal{X}(y)|/\max|\mathcal{X}(y)|$, where ``hat'' denotes the DMD approximation. (a) $y$-directional average velocity (left axis) and relative error (right axis) plot. (b) Particle density variation along the $y$-direction (left axis) and relative error plot (right axis). Few missing points in the error graph correspond to points where the error is below the log scale range shown.}}
  \label{fig:wavy_Np_vel_y_950}
\end{figure}

\begin{figure} [H]
    \centering
  \subfloat[ \label{fig:wavy_Np_vel_x} ]{%
       \includegraphics[width=0.49\linewidth]{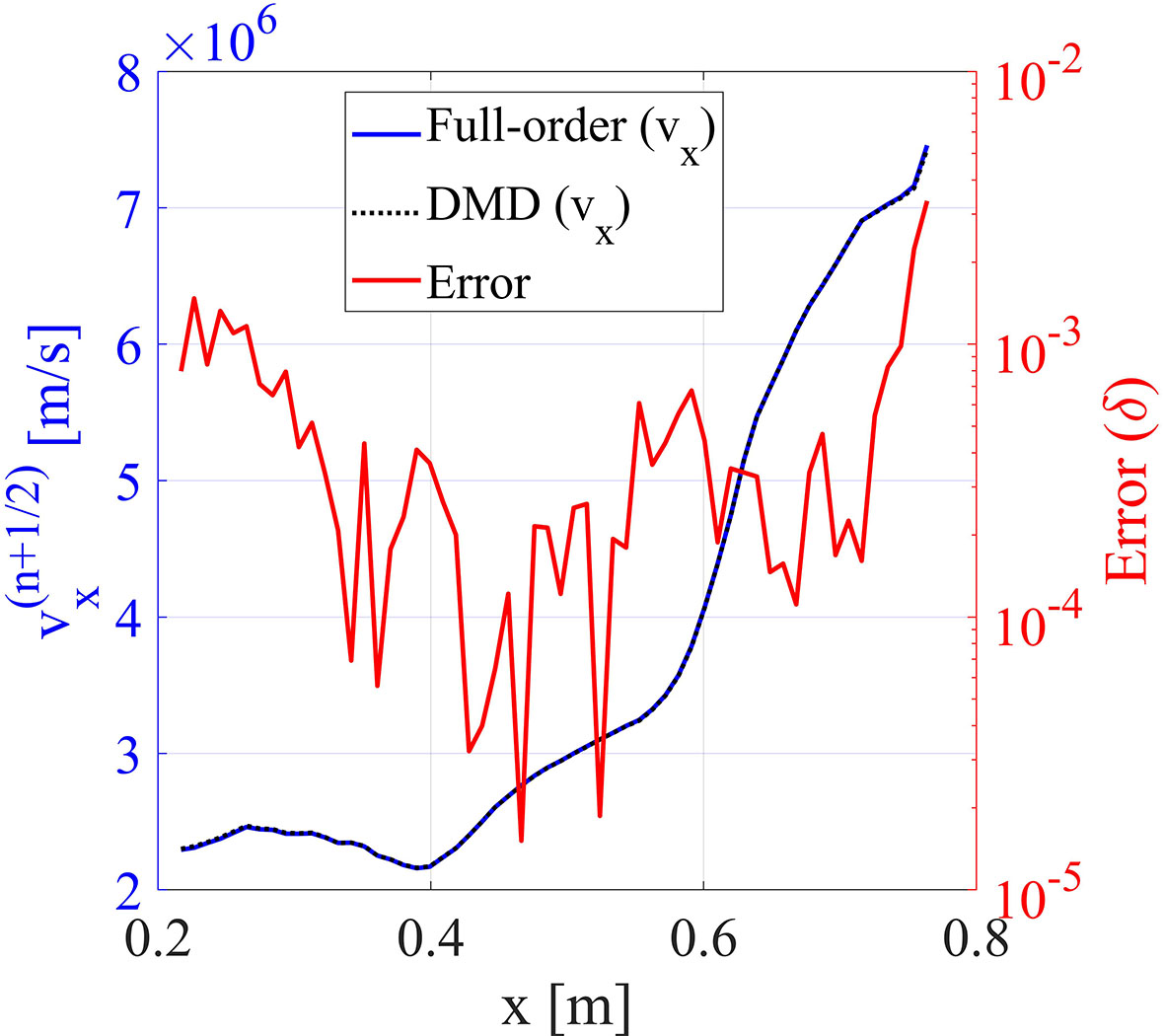}}
  \subfloat[ \label{fig:wavy_Np_vel_x_err} ]{%
        \includegraphics[width=0.5\linewidth]{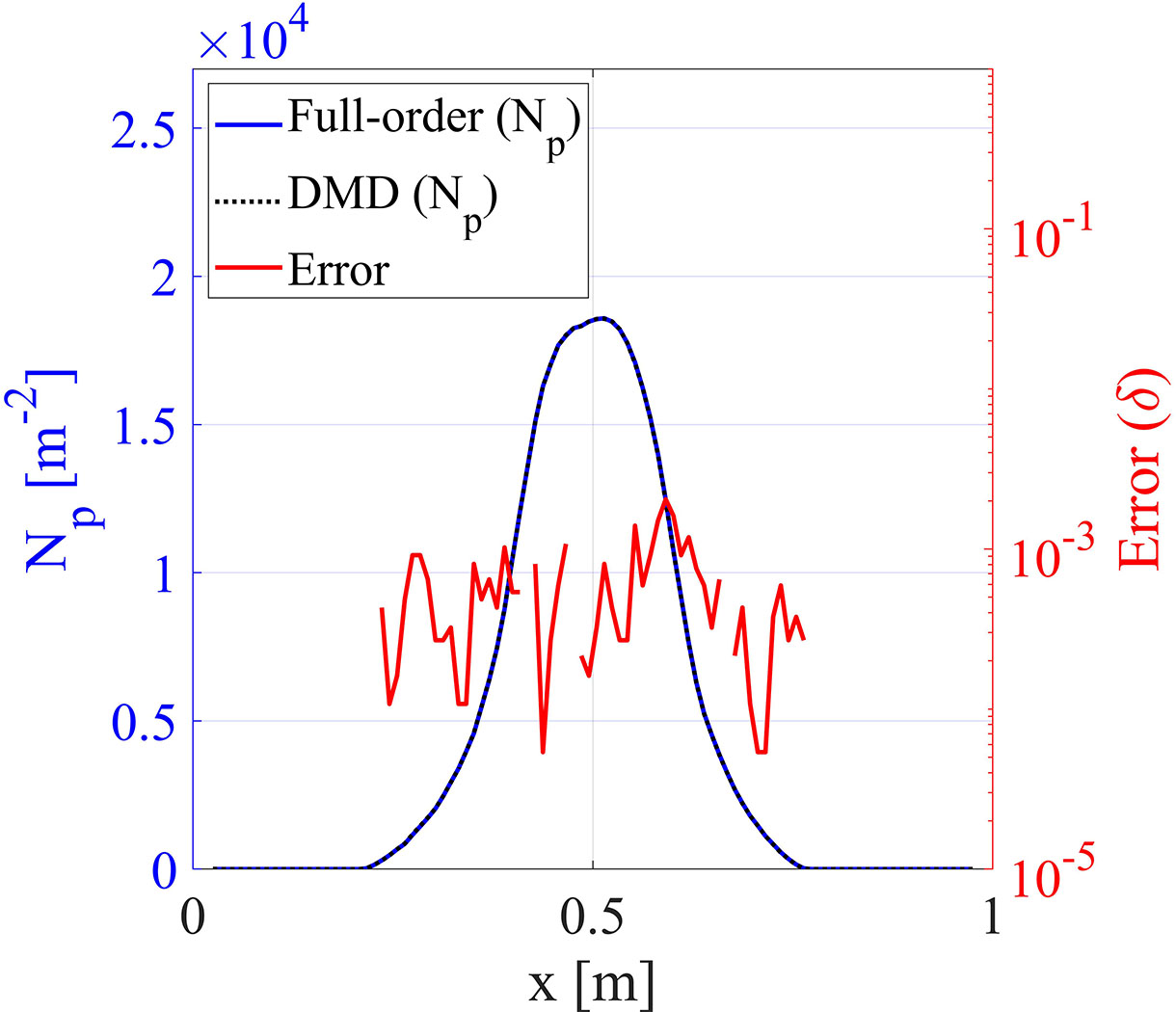}}
  \caption{\small{ Comparison between full-order and DMD predicted average velocity and particle density at $n=76000$ along the $x$-direction. Relative error is similarly defined as in Fig. \ref{fig:wavy_Np_vel_y_950}. (a) $x$-directional average velocity (left axis) and relative error (right axis) plot. (b) Particle density variation along $x$-direction (left axis) and relative error plot (right axis)}. The missing points in the error graph are below the log scale range shown.}
  \label{fig:wavy_Np_vel_x_950}
\end{figure}

\paragraph{Predicted Field and Particle Dynamics}
Recall that a key motivation for using a ROM such as DMD is to expedite the EMPIC simulation by predicting future self-fields and particle dynamics. The 2-norm relative error in predicted fields is close to $1\%$ after extrapolation from the window at $k = 136$. We compare the $x$ and $y$ directional phase-space plots {(Figs.~\ref{fig:wavy_ph_sp_y}-\ref{fig:wavy_ph_sp_x})} and the $x$ and $y$ directional average velocity and particle density (Figs.~\ref{fig:wavy_Np_vel_y_950}-\ref{fig:wavy_Np_vel_x_950}) at $n = 76000$, which extends well into the extrapolation region. 


\subsection{Electron Beam with Virtual Cathode Formation}\label{result_virt} 
\begin{figure} [hbt!]
    \centering
  \subfloat[ \label{fig:virt_snap} ]{%
       \includegraphics[width=0.48\linewidth]{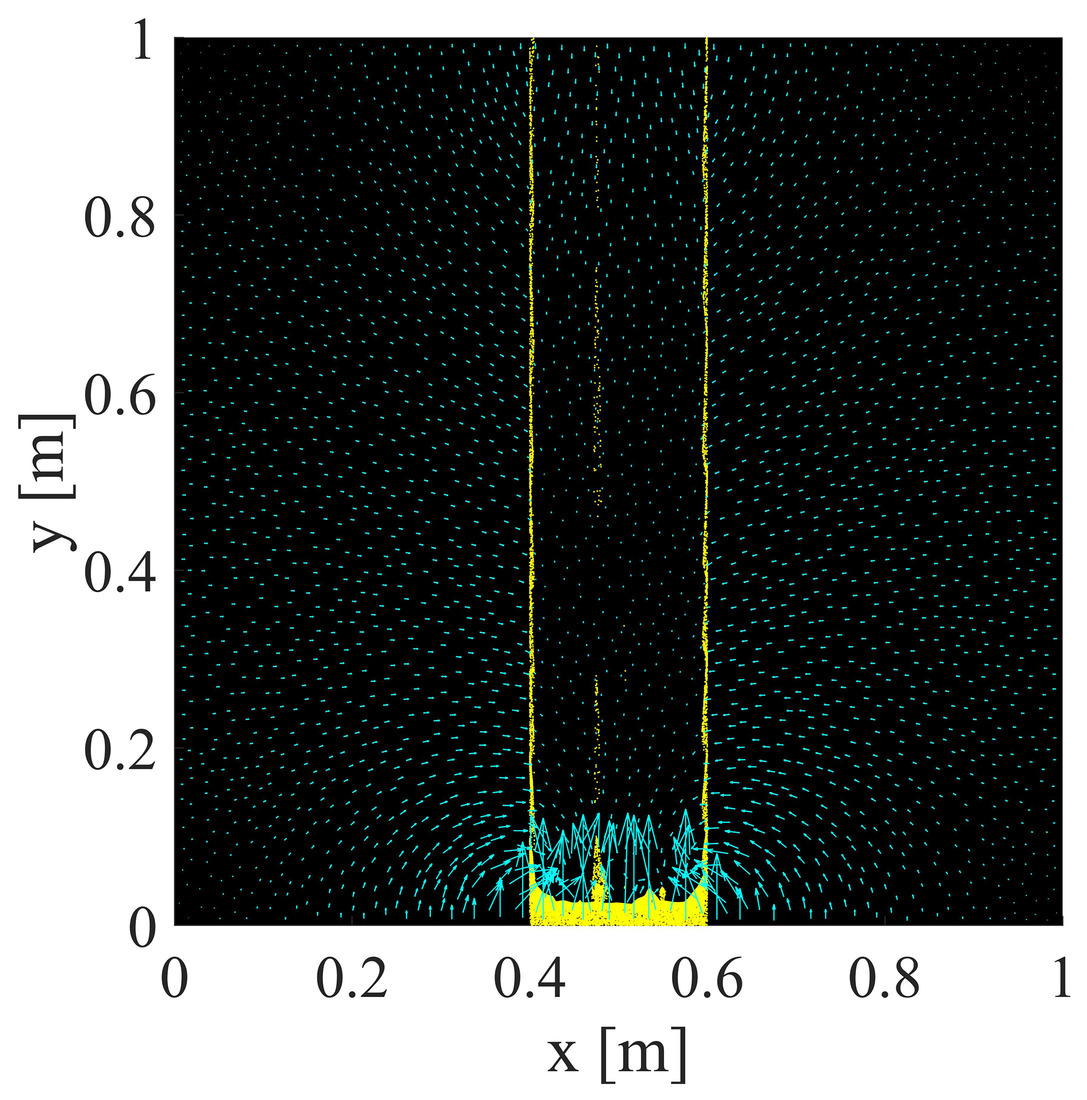}}
  \subfloat[ \label{fig:virt_ss_sing} ]{%
        \includegraphics[width=0.52\linewidth]{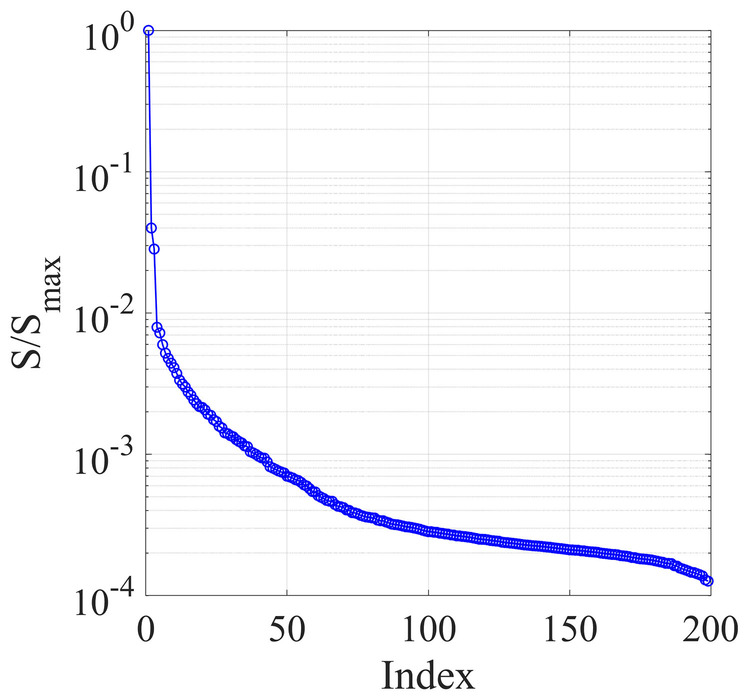}}
  \caption{\small{ (a) Snapshot of virtual cathode formation for 2-D electron beam at $n=80000$. The cyan arrows show the self electric field lines. (b) Normalized singular values for DMD in equilibrium region. }  \label{fig:virt_snap_sing}}
\end{figure}

A relatively complex example of interest is the reduced-order modelling of virtual cathode oscillations.
The setup of \ref{wavy_beam} is adopted with two major differences: (i.) the amount of injected current is increased $15$ times, and, (ii.) a $y$-directional non-oscillating confining magnetic {flux} is employed instead of a transverse oscillating magnetic flux. The superparticle ratio is increased to {$3\times 10^6$}, while holding the same  injection rate. The external voltage bias is turned off and a strong magnetic {flux}, $B = B_y\hat{y}$ is applied in the $y$ direction, with $B_y=100$ A/m. The increased current injection initiates virtual cathode formation, eventually leading to small oscillations near the root of the beam in the equilibrium state (Fig. \ref{fig:virt_snap}). The data set spans from timestep $n=80$ to $n=160000$, containing a total of $2000$ data points $(\Delta_n = 80)$, with $\Delta_t=0.02$ ns. Unlike the previous examples, we only discuss the key takeaways from DMD analysis for this problem.

\subsubsection{DMD in Equilibrium State} \label{result_virt_eq}

\begin{figure} [hbt!]
    \centering
  \subfloat[$\Phi_1^{(eq)}$ \label{fig:virt_ss_mode1} ]{%
       \includegraphics[width=0.5\linewidth]{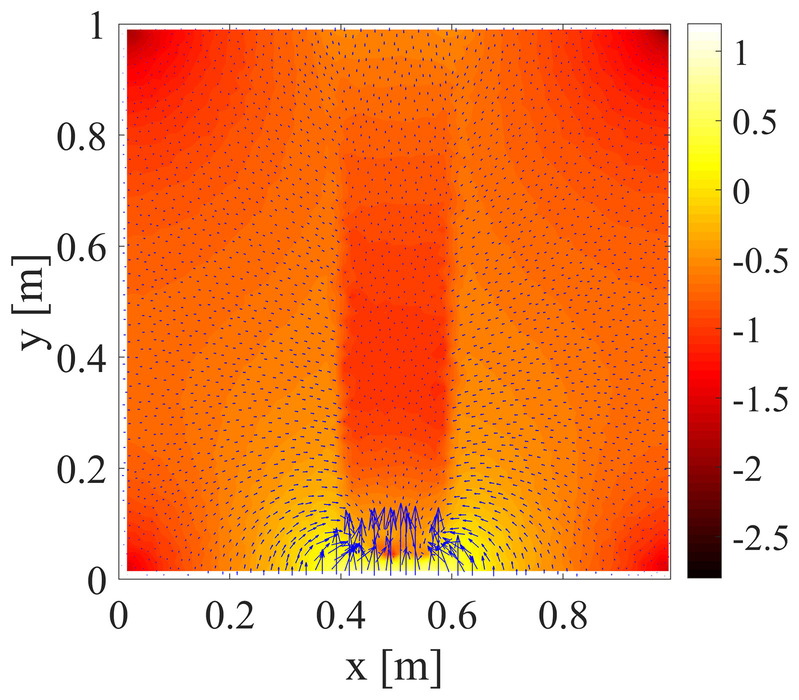}}
  \subfloat[$\Phi_2^{(eq)}$  \label{fig:virt_ss_mode2} ]{%
        \includegraphics[width=0.5\linewidth]{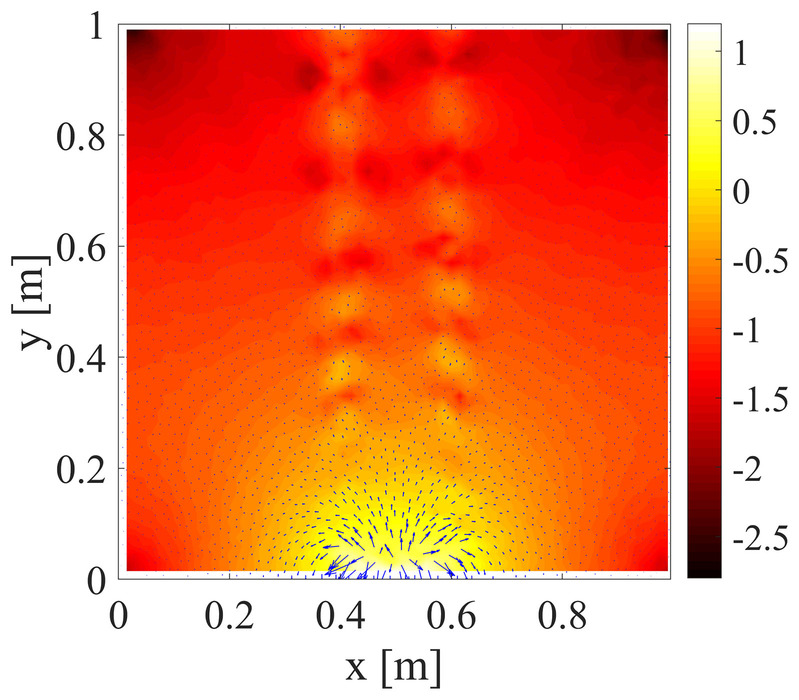}}
  \caption{\small{  Dominant modes extracted from equilibrium region of virtual cathode formation. }  \label{fig:virt_ss_modes}}
\end{figure}

The harvesting window spans from $n=88000$ to $n=120000$ with $r=71$ and $37$ DMD modes, although only two dominant modes capture more than $99\%$ of the total energy in equilibrium. Exponential decay in singular values {(Fig. \ref{fig:virt_ss_sing})} reveals the underlying low-dimensional structure in equilibrium dynamics. The stationary structure of the virtual cathode is represented by the mode $\Phi_1^{(eq)}$ and the small oscillations at the location of virtual cathode formation are captured by $\Phi_2^{(eq)}$.  The relative 2-norm error remains close to $1\%$ inside the harvesting window and oscillates around $5\%$ margin in the extrapolation region.

\subsubsection{Predicted Particle Dynamics}
\begin{figure}[H]
    \centering
	\includegraphics[width=1\linewidth]{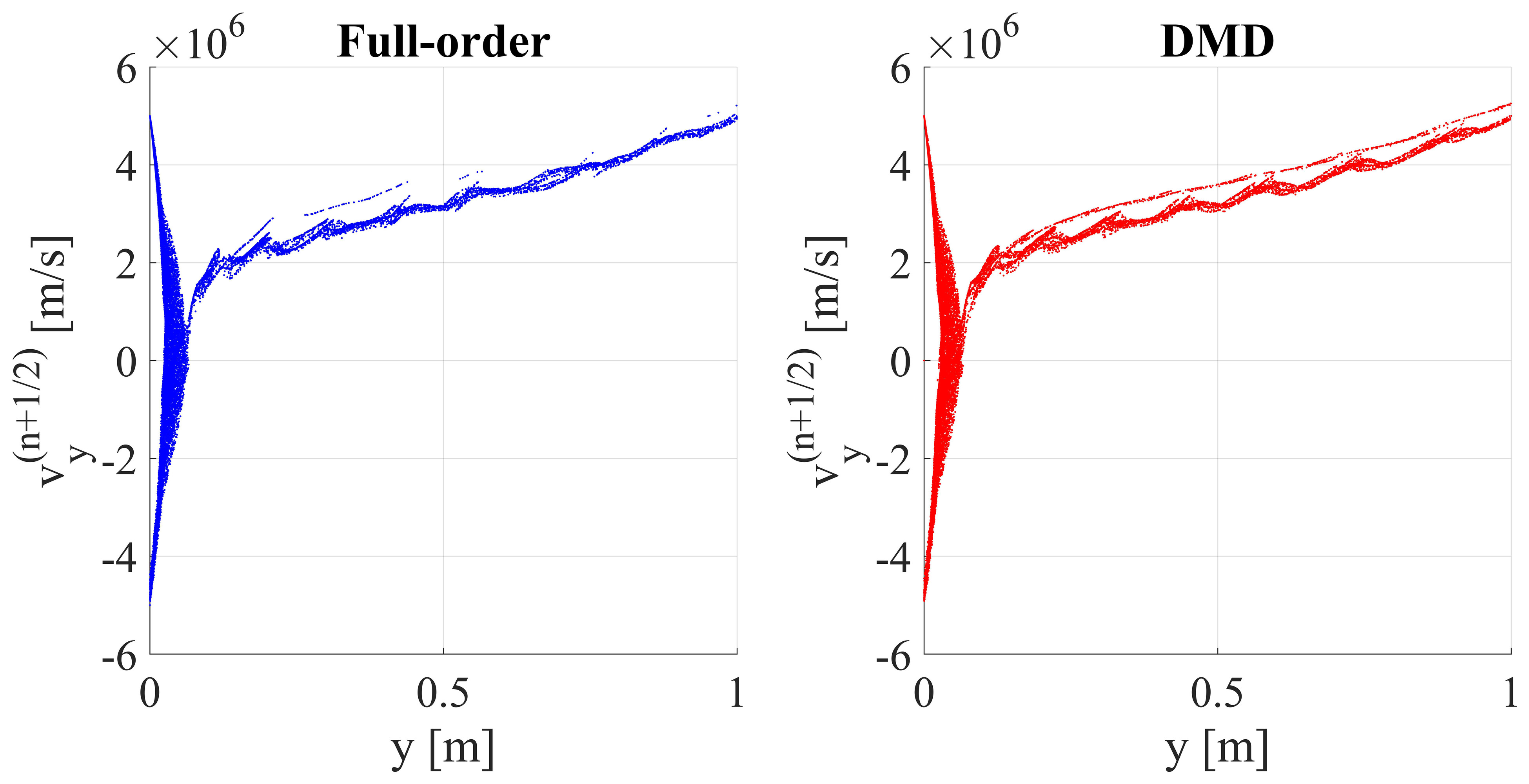}
    \caption{\small{The $y$-directional phase-space plot comparison between finite-element full-order EMPIC simulation (blue) and DMD (red) in extrapolation region ($n=128000$).}}
    \label{fig:virt_ph_sp_y} 
\end{figure}

\begin{figure}[H]
    \centering
	\includegraphics[width=1\linewidth]{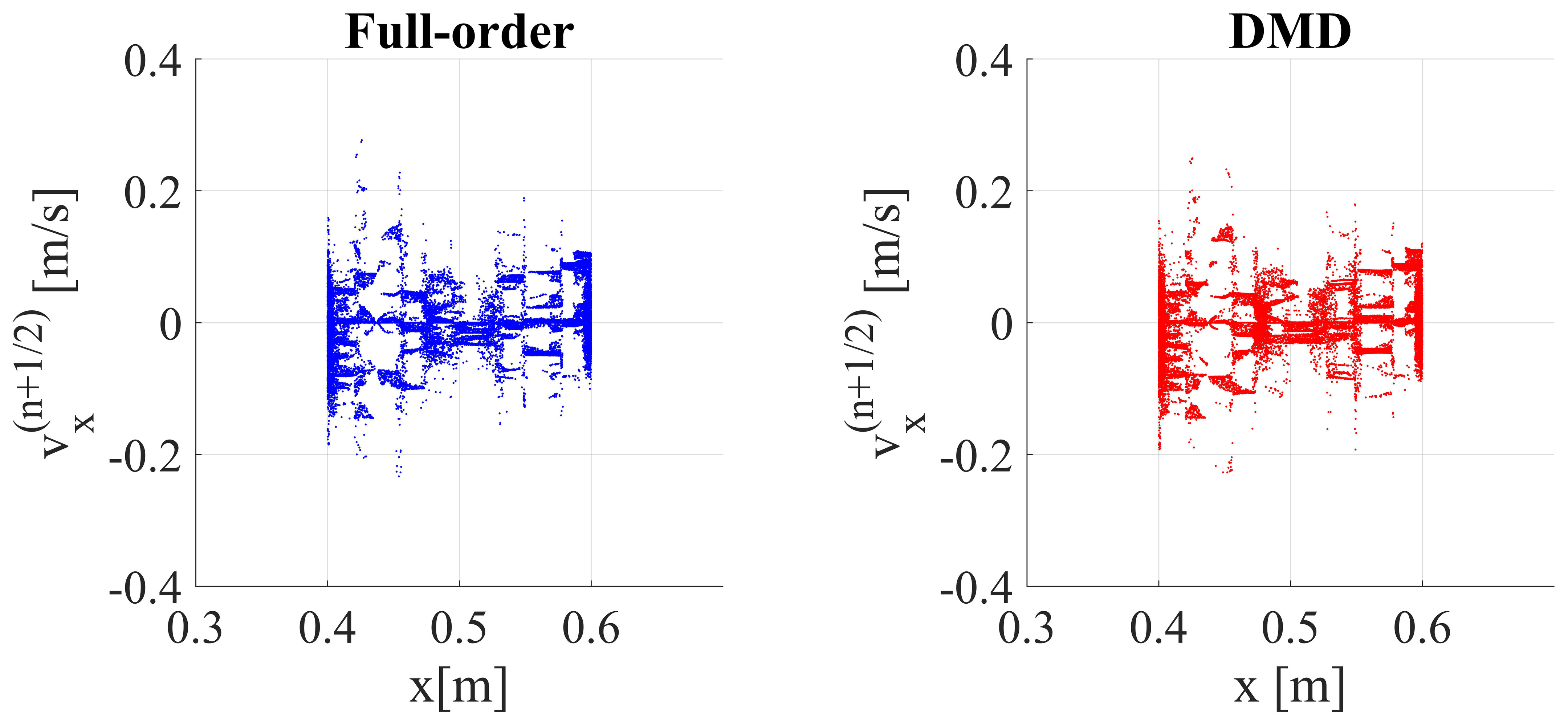}
    \caption{\small{The $x$-directional phase-space plot comparison between finite-element full-order EMPIC simulation (blue) and DMD (red) in extrapolation region ($n=128000$).}}
    \label{fig:virt_ph_sp_x} 
\end{figure}

\begin{figure} [H]
    \centering
  \subfloat[ \label{fig:virt_vel_y_err} ]{%
       \includegraphics[width=0.5\linewidth]{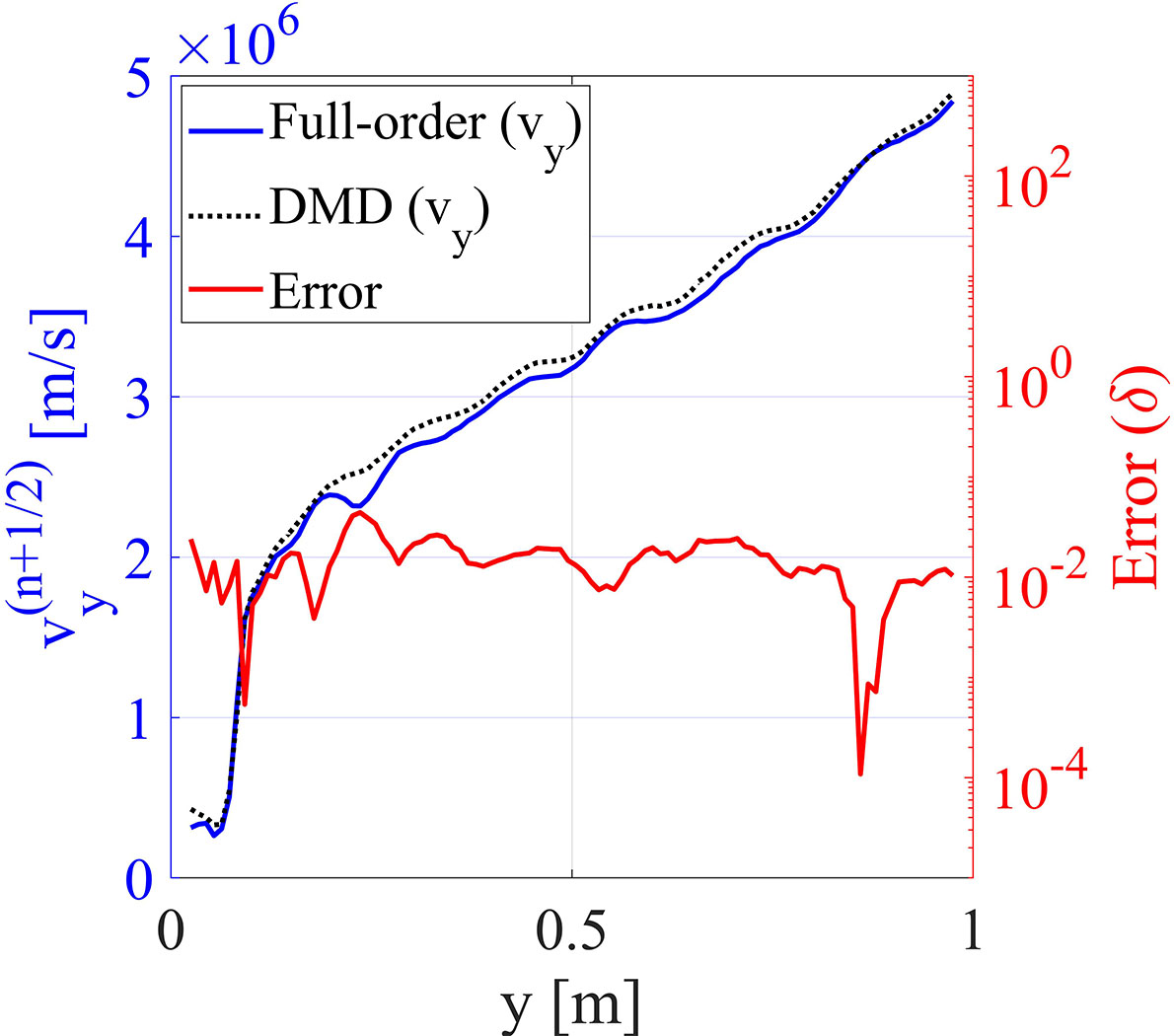}}
  \subfloat[ \label{fig:virt_Np_y_err} ]{%
        \includegraphics[width=0.5\linewidth]{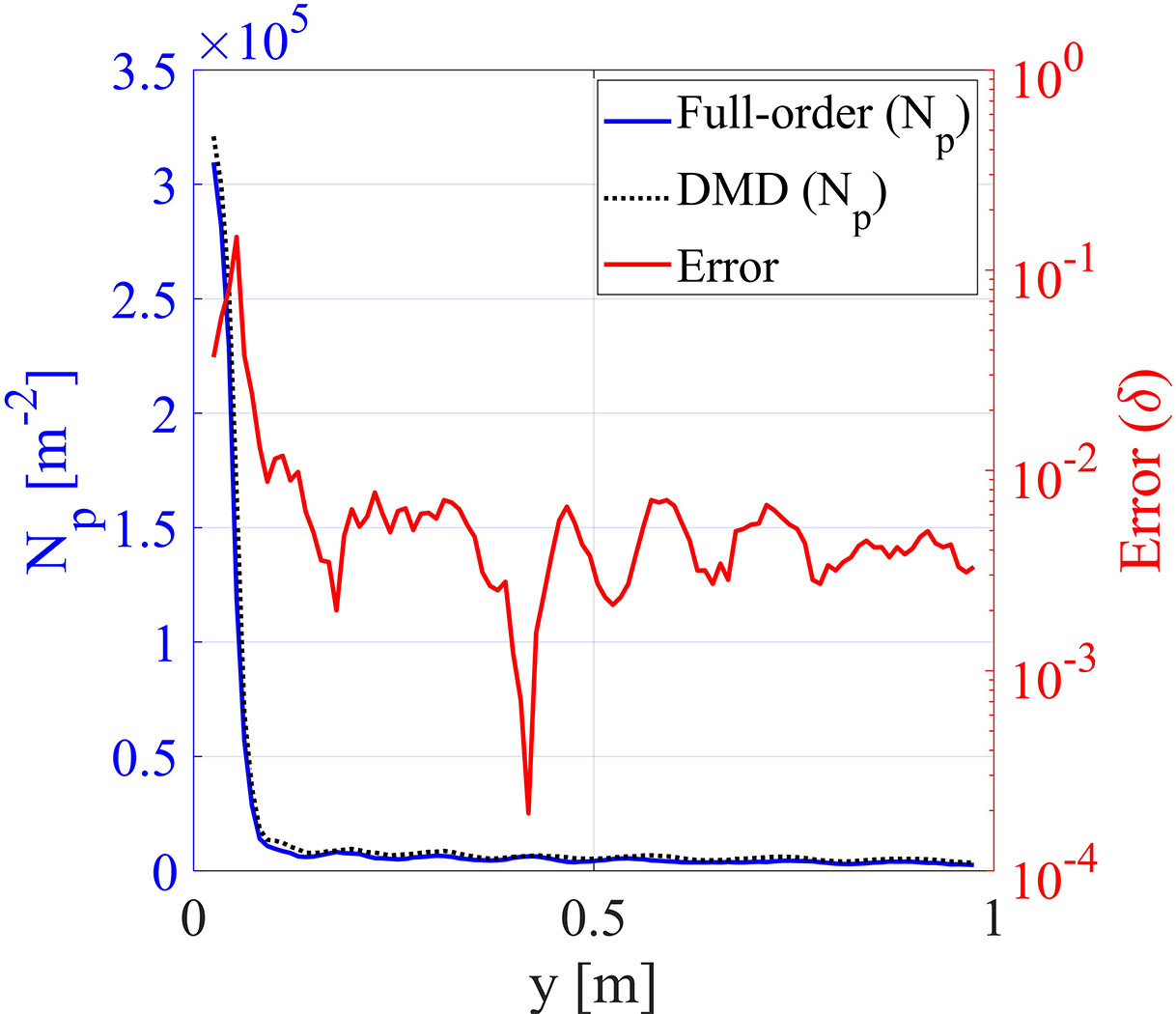}}
  \caption{\small{ Comparison between full-order and DMD predicted average velocity and particle density at $n=128000$ along the $y$-direction. Relative error in $v_y^{(n+1/2)}$ and $N_p$ are as defined in Fig. \ref{fig:wavy_Np_vel_y_950}. (a) $y$-directional average velocity plot and relative error. (b) Particle density variation along the $y$-direction and relative error.}}
  \label{fig:virt_Np_vel_y_1600}
\end{figure}

\begin{figure} [H]
    \centering
  \subfloat[ \label{fig:virt_vel_x_err} ]{%
       \includegraphics[width=0.5\linewidth]{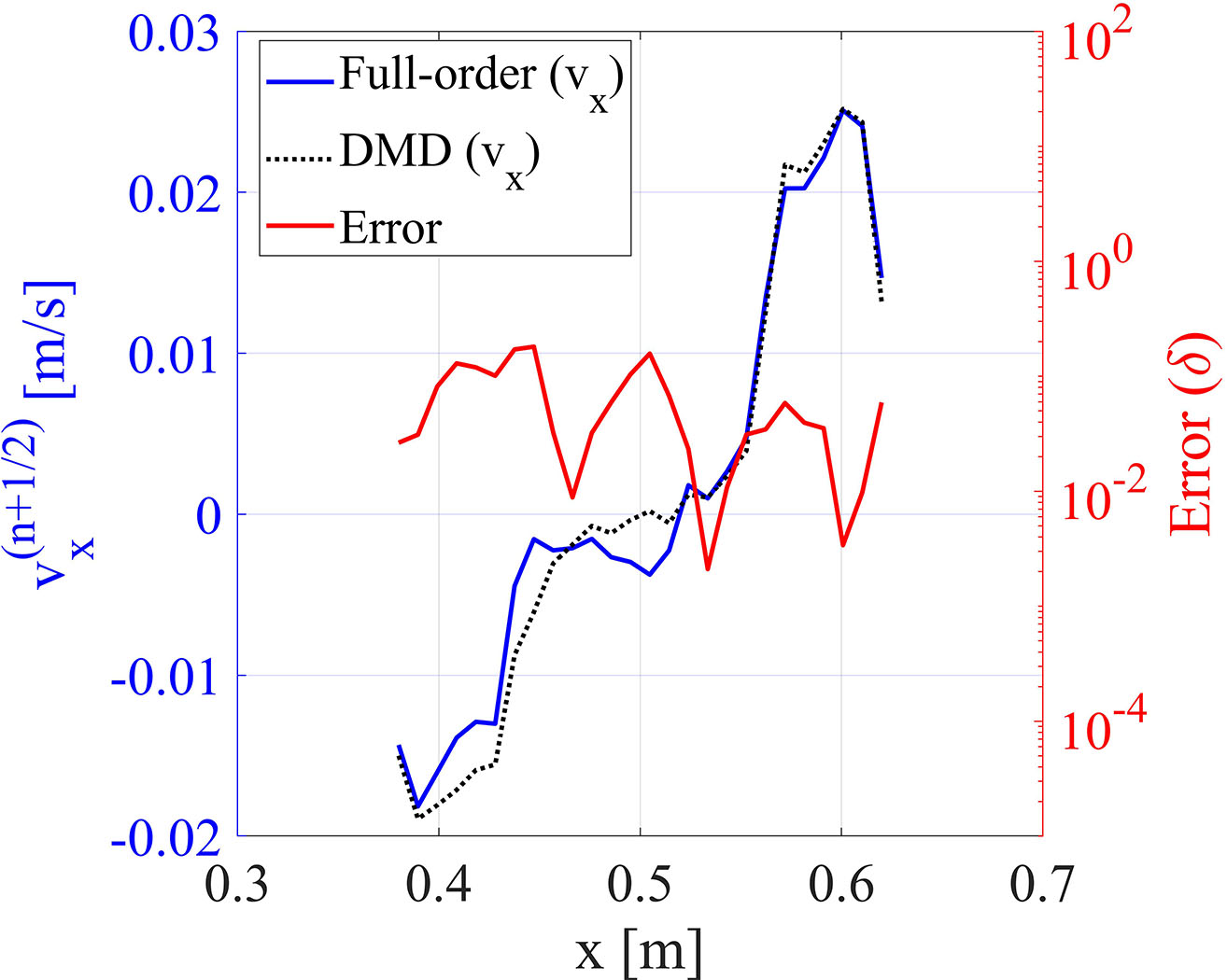}}
  \subfloat[ \label{fig:virt_Np_x_err} ]{%
        \includegraphics[width=0.49\linewidth]{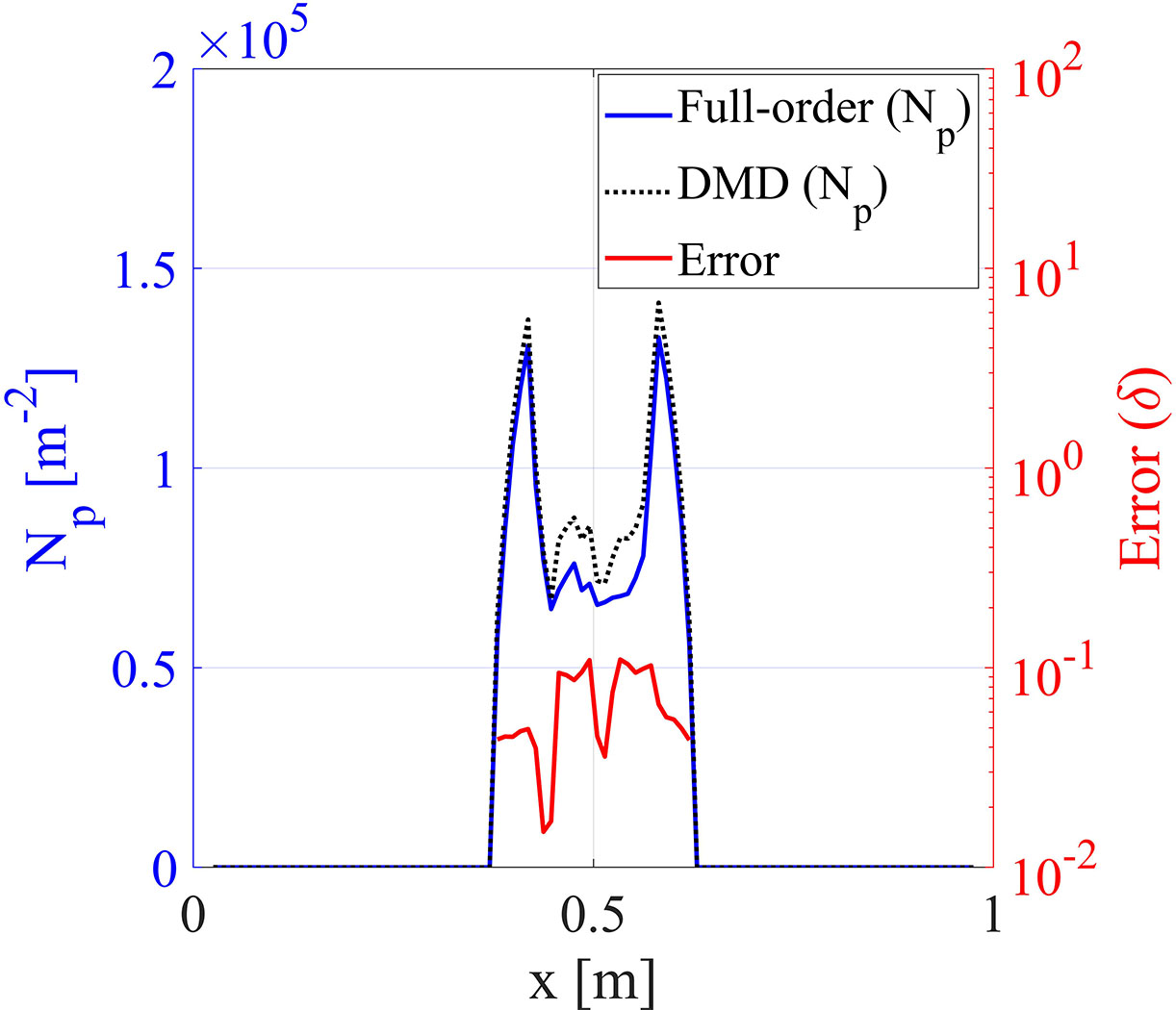}}
  \caption{\small{ Comparison between full-order and DMD predicted average velocity and particle density at $n=128000$ along the $x$-direction. (a) $x$-directional average velocity plot and relative error. (b) Particle density variation along the $x$-direction and relative error.}}
  \label{fig:virt_Np_vel_x_1600}
\end{figure}

We apply the sliding-window DMD method on self electric field data from the virtual cathode, with $\beta_{thr}=0.01$, $\Delta_k = 6.4$ ns, $T=160$ ns and $\Delta t=3.2$ ns. Equilibrium is detected at $k=180$ $(n_{en}(180)=65360)$, at which point the field-update is replaced with extrapolated self-field values from DMD. The predicted particle dynamics at $n=128000$ is shown in {Figs. \ref{fig:virt_ph_sp_y}-\ref{fig:virt_Np_vel_x_1600}}.

\section{Computational Complexity}\label{empic_speedup} 
The timestep complexity (runtime computational complexity  to evolve through one timestep) in our {\it explicit} particle-in-cell algorithm
is $\mathcal{O}(N_p + N)$ \citep{WOLF2016342} where $N_p$ is the number of particles and $N$ represents aggregate mesh dimension. More typically, for {\it implicit} field solvers the timestep complexity\footnote{The timestep complexity of our solver is reduced by employing a sparse approximate inverse of the finite-element mass matrix in the time stepping procedure~\cite{kim2011parallel,na2016local}. This strategy basically trades the reduction in timestep complexity for the one-time cost (incurred prior to time stepping) of computing the sparse approximate inverse.} is $\mathcal{O}(N_p+N^s)$, with $s\geq 1.5$.\par

 Usually, $N_p\gg N$ and therefore the field gather, particle push, and current scatter stages represent the main bottleneck, especially in serial computers. On the other hand, in parallel computers, one can exploit the fact that the particle steps are embarassingly parallelizable. Nevertheless, in large problems with millions of grid nodes and edges, the field update can also consume significant amount of time. DMD based reduced-order models for self-fields can reduce this cost for long term predictions. \par

In addition, EMPIC simulations are often run beyond the equilibrium onset (which is not known a priori). Let this post-equilibrium timestep index be denoted as $n_0$. The runtime of a typical EMPIC simulation up to timestep $n_0$ in a serial computer is then $\mathcal{O}(n_0N_p+n_0N^s)$. The runtime complexity of exact DMD is dominated by the SVD step, given by $\mathcal{O}(lN^2)$, where $l$ is the number of DMD snapshots. For the sliding-window  DMD method described in this paper, the equilibrium onset detection has a runtime complexity of $\mathcal{O}(lN^2~n_{eq}/\Delta n)$, assuming the sliding-window DMD terminates at timestep $n_{eq}$ with a typical window shift of one snapshot. Here $\Delta n$ represents the number of timesteps between two consecutive DMD snapshots. The resulting overall computational complexity of the sliding-window DMD is thus $\mathcal{O}(lN^2~n_{eq}/\Delta n+n_{eq}N_p+n_{eq}N^s)$. Consequently, the presented method is advantageous to determine self-fields for $N_p\gg N$ and/or $n_0\gg n_{eq}$.\par

If the particle dynamics at $n_0$ is also sought, then the reduced-order model for self-fields also provides some advantages given $n_0\gg n_{eq}$ since the field solver is obviated beyond $n_{eq}$. The overall computation complexity becomes $\mathcal{O}(lN^2~n_{eq}/\Delta n+n_0N_p+n_{eq}N^s)$ compared to the original cost of $\mathcal{O}(n_0N_p+n_0N^s)$. If $N_p\gg N$, it turns out that the computation advantage is insignificant. However, if $N_p$ and $N$ are comparable, then the sliding-window DMD model is advantageous for $n_0\gg n_{eq}$. \par

For simplicity, the above estimates assume a serial implementation. As noted, in parallel computers, one can readily exploit the fact that all particle steps (gather, pusher, and scatter) are embarrassingly parallelizable. In that case, the runtime estimates would of course depend on other factors such as the number of available processors.

\section{Concluding Remarks} 
This work introduced a DMD approach for the reduced-order modeling of kinetic plasmas. Data is harvested from high-fidelity EMPIC simulations and used to extract key (low-dimensional) features as well as to predict/ extrapolate the problem dynamics to later times. Extraction of key features/modes is shown to be instrumental in providing physical insight into the problem and can facilitate the application of model predictive control methods. Accurate prediction of nonlinear limit-cycle behavior can be non-trivial, especially in simulations based on large meshes with many elements. The sliding-window DMD approach correctly identifies the onset of  limit-cycle behavior, which enables accurate prediction of the self-field and particle dynamics beyond the equilibrium detection point, and thus has the potential to speed-up EMPIC simulations for long term prediction. These methods were demonstrated on plasma ball and electron beam examples. Future work will involve improving the algorithm for in-line detection of knee/elbow region for the $\alpha$ variation and implementing model order reduction directly on the particle dynamics.

\section{Acknowledgment}\label{ack}
This work was partially supported by the Defense Threat Reduction Agency  under Grant {HDTRA1-18-1-0050}, the Air Force Office of Scientific Research under Grant No. FA9550-20-1-0083 and the Ohio Supercomputer Center under Grant {PAS-0061}.

\clearpage
\appendix

\section{\label{lorenz_96}Equilibrium Detection for Lorenz'96 Model}

\begin{figure} [tbh!]
\centering
  \subfloat[$n_{beg}=100$ \label{fig:lorenz_snap100} ]{%
       \includegraphics[width=0.33\linewidth]{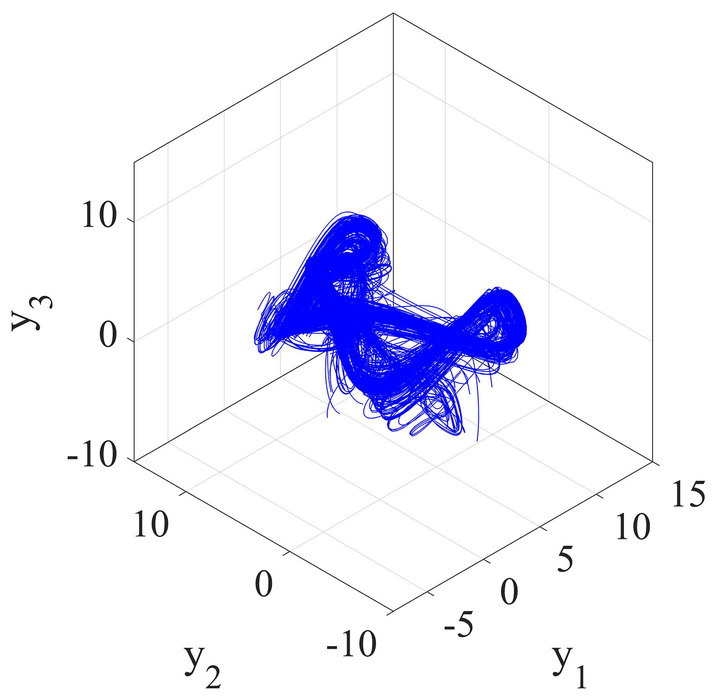}}
  \subfloat[$n_{beg}=300$ \label{fig:lorenz_snap300} ]{%
        \includegraphics[width=0.33\linewidth]{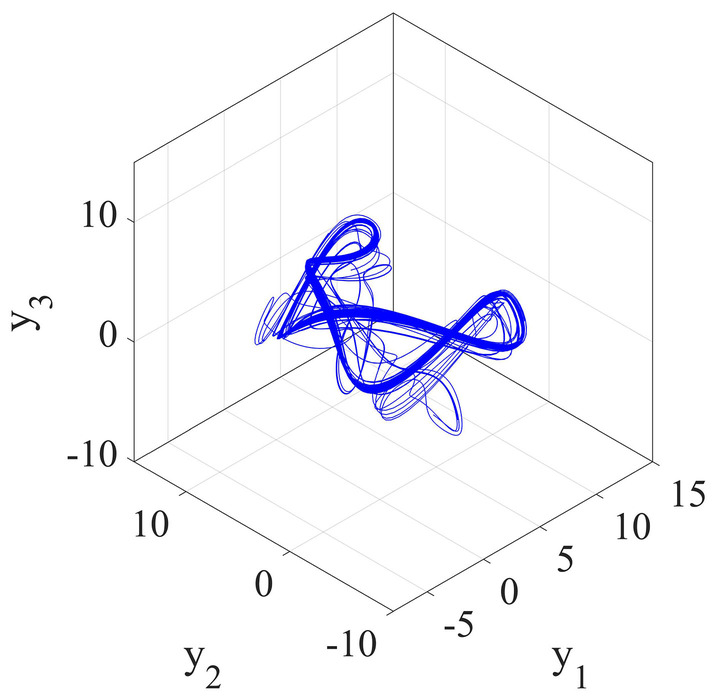}}
  \subfloat[$n_{beg}=500$ \label{fig:lorenz_snap500} ]{%
        \includegraphics[width=0.33\linewidth]{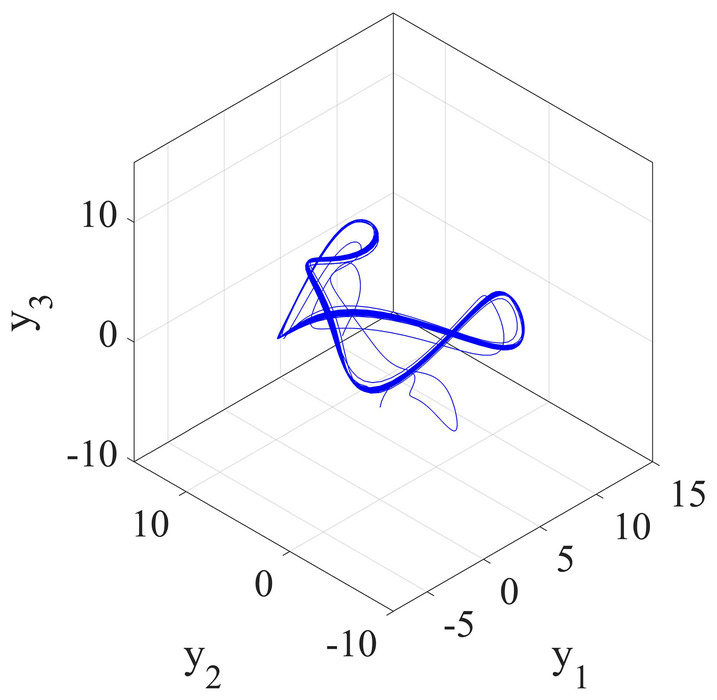}}\\
 \subfloat[$n_{beg}=700$ \label{fig:lorenz_snap700} ]{%
        \includegraphics[width=0.33\linewidth]{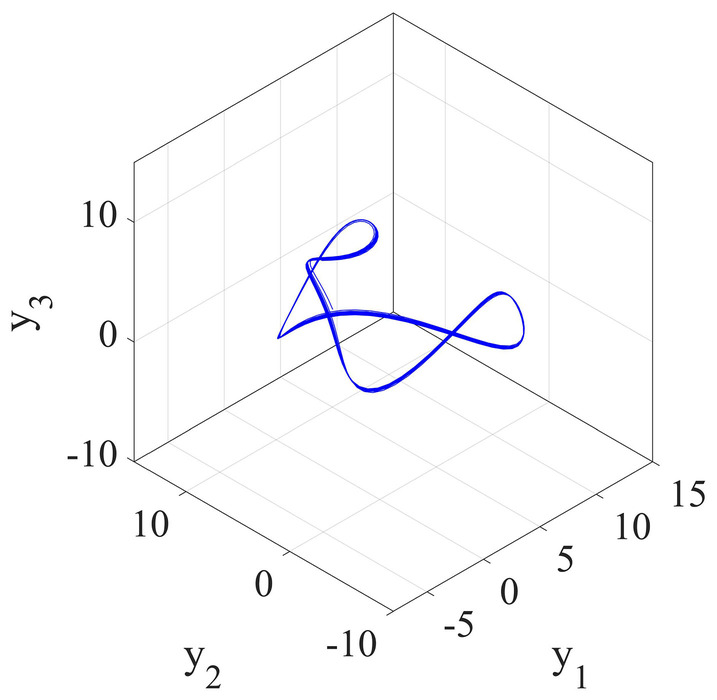}}
\subfloat[$n_{beg}=900$ \label{fig:lorenz_snap900} ]{%
        \includegraphics[width=0.33\linewidth]{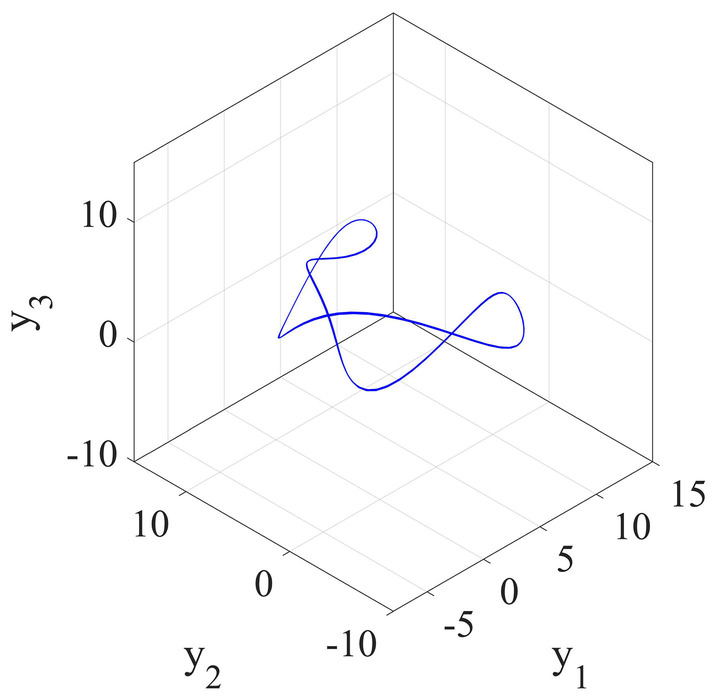}}
 \subfloat[$n_{beg}=1100$ \label{fig:lorenz_snap1100} ]{%
        \includegraphics[width=0.33\linewidth]{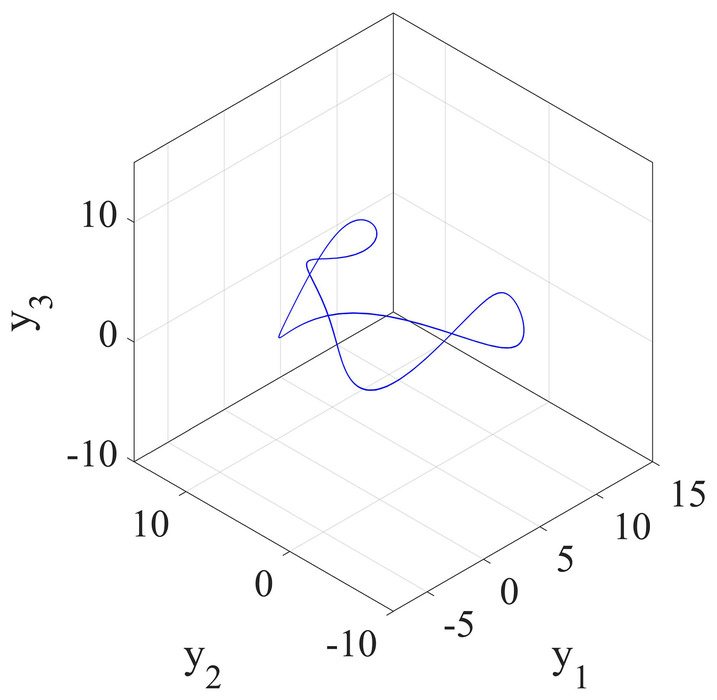}}
  \caption{\small{ Trajectory of state-space ($y_1,y_2,y_3$) for $200$ different initial conditions, from $n=n_{beg}$ to $n=6001$. The noisy trajectories for early values of $n_{beg}$ suggests that the states are yet not evolving on a periodic orbit, thus still in transience. From $n_{beg}=900$ we see a clear periodic orbit, indicating onset of equilibrium.  }
  \label{fig:lorenz_snaps}}
\end{figure}

The Lorenz'96 model was introduced by Edward Lorenz in 1996 \cite{lorenz1996predictability} as a simplified model for predicting atmospheric phenomena. It has been widely used with data assimilation and ensemble forecasting techniques \cite{lorenz1998optimal, kerin2020lorenz, lucarini2011statistical, gallavotti2014equivalence}. Here, we use the Lorenz'96 model to demonstrate the effectiveness of  Algorithm 2 in detecting the onset of the equilibrium state in a perhaps more familiar setting. For $N$ states, the equations governing the dynamics are 
\begin{align}
    &\dot{y_i}=(y_{i+1}-y_{i-2})y_{i-1}-y_i+F_e\\
     & y_{-1}=y_{N-1}\\
   &y_0=y_N\\
   &y_{N+1}=y_1
\end{align}
where, $y_i$ ($i=1,2,~.~.~.,N$) represents the state of the dynamical system and $F_e$ the external force.  We set $N=4$ and $F_e=10$, leading to stable limit cycle behavior {(Fig. \ref{fig:lorenz_snap1100})}. 200 random realizations of the initial state are achieved in the following form: $(y_1^{(0)}, y_2^{(0)}, y_3^{(0)}, y_4^{(0)}) = (F_e+\hat{\delta}_j, F_e, F_e, F_e)$, $j=1, 2, \ldots,200$, where $\hat{\delta}_j$ is randomly generated with uniform distribution in $[-2,2]$. The system is solved until $t = 120$ units, with a total of $n = 6001$ timesteps. Each ``snapshot'' is a $800 \times 1$ vector formed by stacking the instantaneous values of each of the 200 realizations of the four states.
\begin{figure}[t]
    \centering
	\includegraphics[width=0.9\linewidth]{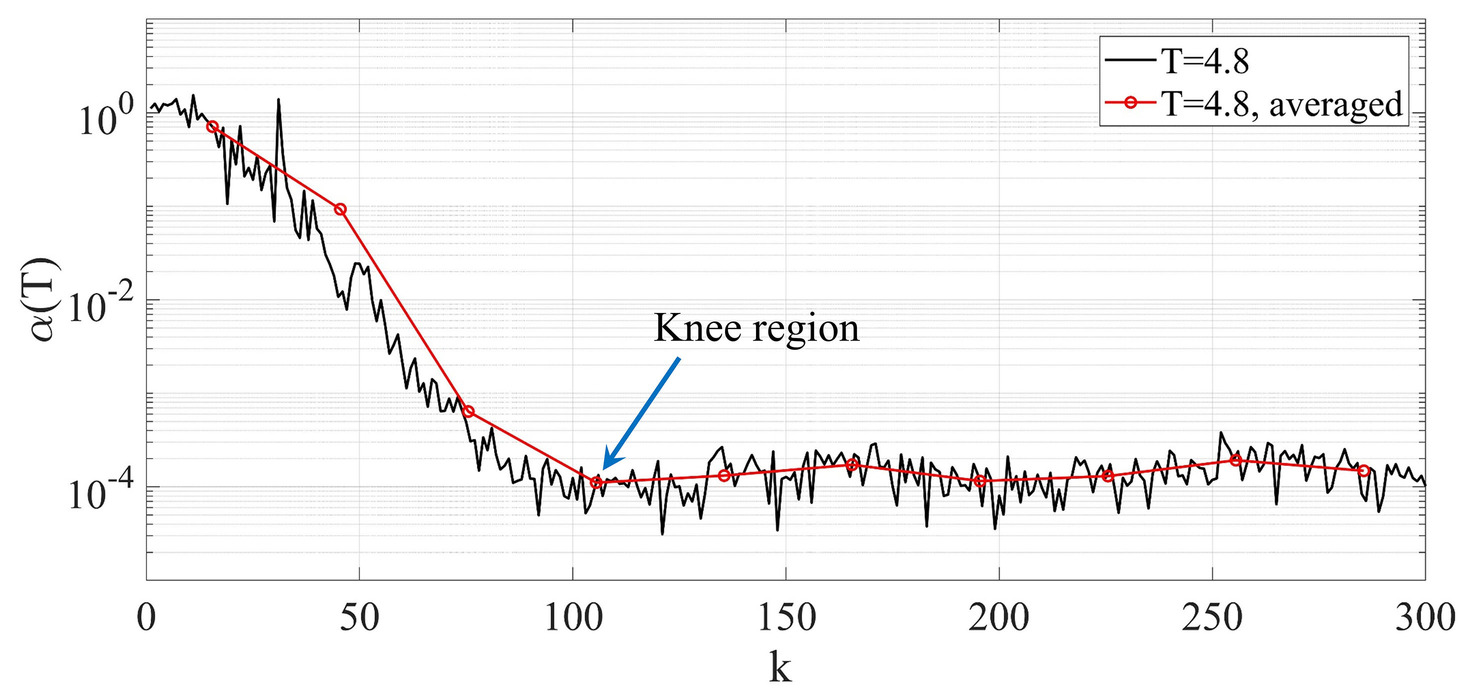}
    \caption{\small{Variation in $\alpha(T)$ as the window slides towards equilibrium for $\Delta t=0.08$. The knee region of the averaged (red) graph is at around $k=105$.}}
    \label{fig:lorenz_alpha_var} 
\end{figure}\par

Algorithm 2 is used to identify the onset of equilibrium with $\beta_{thr}=0.01$ and $\Delta_k = 0.16$. Using $T=4.8$ allows it to cover multiple cycles of the limit cycle oscillations with interval between two consecutive snapshots being $\Delta t=0.08$. It is seen that $\alpha(T)$ decreases with increasing $k$ initially but eventually converges {(Fig. \ref{fig:lorenz_alpha_var})}, with the knee/elbow region marking the transition from transient to steady-state. In this figure, the rolling average of $\alpha(T)$ over $\floor{T/\Delta_k}$ data points is shown (in this case, $30$ datapoints).
The knee/elbow region for the averaged graph is clearly visible around $k=105$, indicating state transition around $n_{st}(105) = 833$ to $n_{en}(105) = 1073$, where $n_{st}(k)$ and $n_{en}(k)$ are respectively the starting and ending timestep of $k^{th}$ window.

\begin{figure} [t]
    \centering
  \subfloat[ \label{fig:lorenz_alpha_comp1} ]{%
       \includegraphics[width=0.5\linewidth]{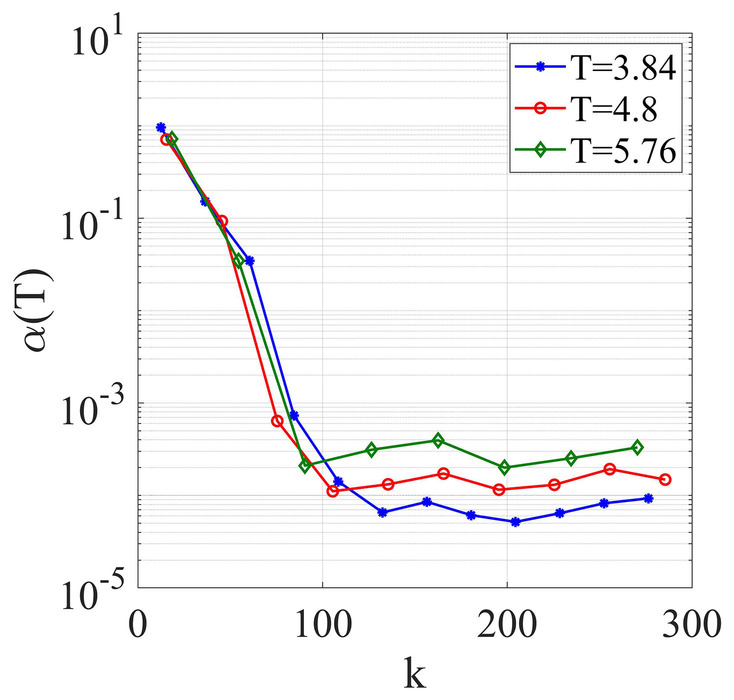}}
  \subfloat[ \label{fig:lorenz_alpha_comp2}]{%
        \includegraphics[width=0.5\linewidth]{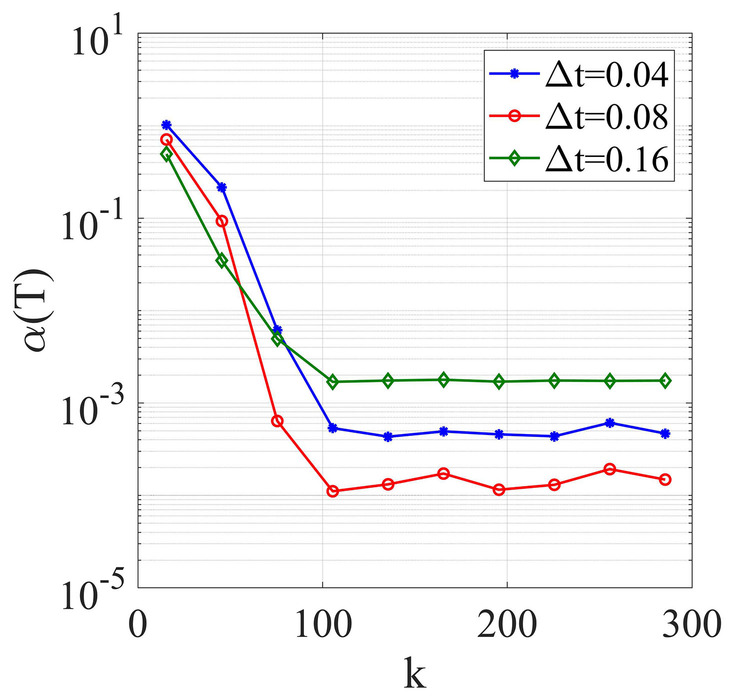}}
  \caption{\small{ (a) Sensitivity of Algorithm 2 towards window width $T~(\pm20\%)$, keeping fixed $\Delta t=0.08$. Equilibrium detected at $k=168,150$ and $144$ for $T=3.84,4.8$ and $5.76$ respectively. (b) Sensitivity of Algorithm 2 towards sampling interval $\Delta t$ , keeping fixed $T=4.8$. Equilibrium detected at $k=180,150$ and $150$ for $\Delta t=0.04,0.08$ and $0.16$ respectively.}
  \label{fig:lorenz_alpha_comp}}
\end{figure}

This result is validated by plotting the state-space trajectories, see {Fig.~\ref{fig:lorenz_snaps}}. For  $n_{beg} = 900$, we observe formation of a periodic orbit {(Fig. \ref{fig:lorenz_snap900})} indicating the onset of equilibrium.
This is in agreement with Algorithm 2, which suggests state transition in the range $n = 833$ to $n = 1073$. The sensitivity of $\alpha(T)$ towards variation in $T$ and $\Delta t$ is shown in {Fig. \ref{fig:lorenz_alpha_comp}} ($\Delta_k$ equivalent to 8 timesteps for each case). The early detection of equilibrium region for large window widths {(Fig. \ref{fig:lorenz_alpha_comp1})} can be attributed to the ``look-ahead" artifact due to the finite width of DMD time window. As in other examples in this work, in-line detection of the knee is adversely impacted by the use of the non-negative slope criterion. As seen in {Fig. \ref{fig:lorenz_alpha_comp2}}, the knee region appears first, visually speaking, around $k=105$, but the first non-negative slope is encountered much later, around $k=135$ for $\Delta t=0.04$. 
The algorithm detects it at $k=180$ and terminates. The delayed detection is not necessarily a drawback in the sense that it provides a conservative estimate. When the graph of $\alpha(T)$ is ``noisy'', there is potential for false positives in an in-line approach. The non-negative slope criterion provides robustness against such false detection, but at the cost of precision since detection of equilibrium is delayed. As mentioned before, delayed equilibrium detection does not affect prediction accuracy, but it does affect its computational efficiency. More work is needed for building better methods for in-line detection of the knee region in noisy datasets. Discussion of additional textbook example can be found in {\cite{nayak2021detecting}}.

\bibliography{mybibfile}

\end{document}